\documentclass[a4paper,11pt]{article}
\pdfoutput=1 
\usepackage{jcappub} 

\usepackage{epsfig}
\usepackage{graphics}
\usepackage{bm}
\usepackage{color}
\usepackage{colortbl}
\usepackage{dcolumn}   
\usepackage{bm}     
\usepackage{bbm}       
\usepackage{amssymb}  
\usepackage{amsmath}
\usepackage{latexsym}
\usepackage{float}
\usepackage{ifthen}
\usepackage{caption,subfig}
\usepackage{enumerate}
\usepackage{url}
\usepackage{caption,subfig}
\usepackage{amsopn}
\usepackage{hyperref}
\usepackage{comment}
\usepackage{animate}

\usepackage{amsfonts}
\usepackage{multirow}
\usepackage{array}
\usepackage{booktabs}
\usepackage{rotating}

\usepackage{ulem}
\def\clap#1{\hbox to 0pt{\hss#1\hss}}

\def\bea{\begin{eqnarray}}
\def\eea{\end{eqnarray}}
\def\be{\begin{equation}}
\def\ee{\end{equation}}

\def\d{\mathrm{d}}
\newcommand{\bpm}{\begin{pmatrix}}
\newcommand{\epm}{\end{pmatrix}}

\newcommand{\lp}{\left(}
\newcommand{\rp}{\right)}
\newcommand{\lb}{\left[}
\newcommand{\rb}{\right]}

\newcommand{\vPhi}{\vec{\Phi}}

\renewcommand{\geq}{\geqslant}
\renewcommand{\leq}{\leqslant}

\newcommand{\Tcoh}{T_\mathrm{coh}}
\newcommand{\Lcoh}{L_\mathrm{coh}}
\newcommand{\zcoh}{z_\mathrm{coh}}

\newcommand{\zmix}{z_\mathrm{mix}}

\newcommand{\zbroad}{z_\mathrm{broad}}

\newcommand{\tr}[1]{{\mathrm{Tr}\lb  #1\rb}}
\newcommand{\nM}{\hat{\nu}}
\newcommand{\cM}{\hat{C}}
\newcommand{\pM}{\hat{\Pi}}
\newcommand{\mM}{\hat{M}}
\newcommand{\eM}{\hat{E}}
\newcommand{\tM}{\hat{\theta}}
\newcommand{\iM}{\hat{I}}

\newcommand{\tg}{\Theta_g}

\newcommand{\dLgw}{d^\text{GW}_{L}}

\newcommand{\dc}{\Delta c}
\newcommand{\dm}{\Delta m}
\newcommand{\dnu}{\Delta \nu}
\newcommand{\dmu}{\Delta \mu}
\newcommand{\onu}{\omega_\nu}

\definecolor{orange}{rgb}{1,0.5,0}
\definecolor{mygray}{gray}{0.9}

\definecolor{darkgreen}{rgb}{0.33, 0.42, 0.18}

\author[a]{Jose Maria Ezquiaga}
\emailAdd{ezquiaga@uchicago.edu}
\affiliation[a]{Kavli Institute for Cosmological Physics and Enrico Fermi Institute, The University of Chicago, Chicago, IL 60637, USA.}
\affiliation[b]{Department of Astronomy \& Astrophysics, The University of Chicago, Chicago, IL 60637, USA}
\affiliation[c]{Department of Physics and Astronomy, Columbia University, New York, NY 10027, USA}

\author[a,b]{, Wayne Hu}
 \emailAdd{whu@background.uchicago.edu}
\author[c]{, Macarena Lagos}
\emailAdd{m.lagos@columbia.edu}
\author[a,b]{and Meng-Xiang Lin}
\emailAdd{mxlin@uchicago.edu}

\title{Gravitational wave propagation beyond general relativity: \\ waveform distortions and echoes}

\abstract{We study the cosmological propagation of gravitational waves (GWs) beyond general relativity (GR) across homogeneous and isotropic backgrounds.  We consider scenarios in which GWs interact with an additional tensor field and use a parametrized phenomenological approach that generically describes their coupled equations of motion.  We analyze four distinct classes of derivative and non-derivative interactions: mass, friction, velocity, and chiral. We apply the WKB formalism to account for the cosmological evolution and obtain analytical solutions to these equations. We corroborate these results by analyzing numerically the propagation of a toy GW signal. We then proceed to use the analytical results to study the modified propagation of realistic GWs from merging compact binaries, assuming that the GW signal emitted is the same as in GR. 
We generically find that tensor interactions lead to copies of the originally emitted GW signal, each one with its own possibly modified dispersion relation. 
These copies can travel coherently and interfere with each other leading to a scrambled GW signal, or propagate decoherently and lead to echoes arriving at different times at the observer that could be misidentified as independent GW events. 
Depending on the type of tensor interaction, the detected GW signal may exhibit amplitude and phase distortions with respect to a GW waveform in GR, as well as birefringence effects. 
We discuss observational probes of these tensor interactions with both individual GW events, as well as population studies for both ground- and space-based detectors. 
}
\date{\today}

\keywords{gravitational waves, tests of general relativity, dark energy, modified gravity.}
\begin{document}

\maketitle

\section{Introduction}
The detection of gravitational waves (GWs) by the LIGO--Virgo collaboration \cite{LIGOScientific:2018mvr, Abbott:2020niy} has launched the development of novel tests of general relativity (GR), and the presence of exotic components in the Universe. Generally, deviations from GR may appear both during emission due to high-energy new physics (see e.g.\ \cite{Yunes:2009ke, Yunes:2011aa, Tahura:2018zuq, Zhang:2019iim}), and during propagation due to low-energy new physics, but in this paper we focus on the latter and assume that the emitted signal is the same as that expected in GR. 
Previous analyses on the propagation of GWs on a homogeneous and isotropic universe have shown that the presence of additional fields interacting non-trivially with gravity can lead to deviations from GR in the GW propagation speed, luminosity distance and phase evolution of the GW signal (see e.g.\ reviews in \cite{Ezquiaga:2018btd, Belgacem:2019pkk}), and some of these effects have already been constrained by the LIGO/Virgo collaboration with current observations \cite{LIGOScientific:2019fpa}. However, other effects of non-trivial cosmological scenarios still remain to be better understood and modeled so that they can be tested in the future, such as those appearing when GWs interact with another tensor-like cosmological field.

In this paper, we study the cosmological propagation of GWs in a homogeneous and isotropic universe, in the case where gravity is coupled to another cosmological field beyond the concordance $\Lambda$CDM model. Since GWs do not couple to vector or scalar fields at linear order in perturbation theory, we assume the additional field to be described by another rank-2 tensor. Note that in the case of inhomogeneous backgrounds, vector and scalar fields can couple to GWs \cite{Garoffolo:2019mna}, such as in the case of lensing beyond GR \cite{Ezquiaga:2020dao,Tasinato:2021wol}, but these scenarios are not considered here. 
In particular, we follow \cite{Jim_nez_2020} and adopt a model-independent approach considering parametrized deviations from GR in the equations of motion (EoM) for linear tensor perturbations, which allow us to describe a wide variety of possible interactions between GWs and the additional tensor field. We focus our attention on four particular interactions: \emph{velocity} (which appear as second-order spatial derivatives of the fields), \emph{friction} (which appear as first-order time derivative of the fields), \emph{mass} (non-derivative contributions of the fields), and \emph{chiral} (first-order spatial derivatives of the fields that break parity symmetry). 
Specific choices for the free parameters of this framework will correspond to cosmological scenarios coming from different modified gravity theories.
Examples of theories described by this approach include massive bigravity \cite{deRham:2010kj,Hassan:2011zd}, Yang-Mills theories \cite{Cervero:1978db,Galtsov:1991un,Darian:1996mb}, Abelian multi-gauge fields in a gaugid configuration \cite{Piazza:2017bsd} and multi-Proca theories \cite{ArmendarizPicon:2004pm,Hull:2014bga,Allys:2016kbq,Jimenez:2016upj,GallegoCadavid:2020dho} (see also \cite{Jim_nez_2020} for a survey of the theory landscape). Note that these free parameters are not necessarily constants, but instead are allowed to evolve on cosmological timescales with the expansion of the universe. As a consequence, we will have two coupled EoM (for GWs and the extra tensor field) with time-dependent parameters, which typically cannot be solved analytically. 
Nevertheless, one can obtain simple analytical solutions using the Wentzel–Kramers–Brillouin (WKB) approximation, in cases where the period of the GWs is much smaller than the Hubble timescale (which is valid for current and next-generation planned GWs detectors). 

The approach used in this paper was originally applied in \cite{Jim_nez_2020} to understand the deviations from GR caused by the additional tensor field, and here we refine the formalism and further study its  phenomenological implications. More specifically, we extend \cite{Jim_nez_2020} by describing in detail all the deviations from GR expected in the chirping GW signal of a coalescence of binary compact objects at different stages during its propagation to the observer.
We also extend and connect to previous investigations that have been performed for the specific models of massive bigravity \cite{Max:2017kdc,Max:2017flc} and gauge field dark energy \cite{Caldwell:2016sut,Caldwell:2018feo}. As has been discussed in these previous works, due to the interactions
between the metric and the additional tensor field, GWs are composed by a superposition of wavepackets---the propagating eigenstates---that can individually exhibit deviations from GR in their amplitude and dispersion relation. Because of these modified dispersion relations (MDRs), each eigenstate may propagate at a different physical speed. We clarify that it is the group velocity that determines the
arrival time of wavepacket components, as opposed to alternative quantities such as the particle velocity suggested in \cite{Will:1997bb,Mirshekari:2011yq}. 
These eigenstates that compose the net GW signal are emitted at the same time but, depending on the propagation time, they may eventually arrive at different times if they have a different propagation speed, leading to the detection of separately-identifiable GW signals, or echoes. When this happens it is said that the eigenstates have decohered. Before decoherence, on top of the modifications of individual eigenstates, the total detected GW signal may differ from that of GR because these eigenstates may interfere with each other. This interference may lead to a time or frequency-dependent oscillation of the amplitude, as discussed in \cite{Max:2017kdc,Belgacem:2019pkk,Jim_nez_2020}, but we crucially find that phase distortions of the GWs will also be present in theories like massive bigravity. 

We first corroborate our analytical findings by numerically solving the coupled tensor EoM for a Gaussian wavepacket as a toy localized GW signal and, for simplicity, ignoring the cosmological expansion. We analyze separately the distortions of the Gaussian signal in case where the metric has mass, friction, velocity and chiral interactions with the extra tensor field. In particular, in each case we discuss how the amplitude, shape and polarization content of the Gaussian changes during propagation, and how/when GW echoes will be detected. In the case of chiral interactions, we show in detail how the detected signal can change its degree of circular polarization (or amplitude birefringence) and also suffer a polarization rotation (called here phase birefringence, but also known as velocity birefringence \cite{Zhao:2019xmm}) with respect to the emitted signal.
These toy Gaussian examples help build intuition on the various propagation regimes of the GWs, and the expected deviations from GR. 

We then apply the WKB formalism and our analytical results to the propagation of realistic chirping GW signals from a binary black hole coalescence observed by LIGO-type detectors (although generalizations to next-generation GW detectors are discussed as well). In addition to the modifications present for the Gaussian wavepacket, these realistic examples allow us to identify how the phase of the GW signal is distorted during propagation. Whereas some of the examples we show exhibit both distortions in the phase and amplitude of the GW signal, we also discuss examples in which individual detected GW events may still have a GR-like waveform but with different amplitude, phase or polarization (a summary of the possible waveform modifications is given in Table \ref{table:summary_time2} and Fig.\ \ref{fig:summary_waveform_distortions}). 
In these cases, we discuss how a bias in the reconstructed source parameters such as its luminosity distance, coalescence phase, inclination and orientation would happen if GR was assumed during the parameter estimation analysis of the detected signal. Nevertheless, a population analysis may help distinguish these GR-like signals from true GR signals, but detailed forecasts in model constraints are left for future work.

This paper is organized as follows. In Section \ref{sec:wavepacket} we introduce the general parametrized EoM for GWs coupled to another tensor field, and describe the analytical solutions that are obtained in the case of constant parameters as well as the case of time-dependent parameters with the use of the WKB formalism. The GW solution is a superposition of propagating eigenstates, each one having its own MDR. General properties of these MDRs are discussed in App.\ \ref{App:WKB}.
In subsec.\ \ref{sec:GW-propag} we define three relevant timescales---mixing, decoherence and broadening---describing different physical regimes in the propagation of the net GW signal and the behavior of the individual eigenstates. Here we highlight the role of the group velocity in the propagation of the GWs.
In Section \ref{sec:examples_mixing}, we consider a toy example of an emitted Gaussian plane wavepacket of GWs, and analyze its propagation numerically in the case of time-independent free parameters in the EoM. We confirm our analytical findings and illustrate how this toy signal gets distorted in the presence of different types of mixing: velocity (subsec.\ \ref{sec:velmixing}), mass (subsec.\ \ref{sec:massmixing}), friction (subsec.\ \ref{sec:friction_mixing}), and chiral (subsec.\ \ref{sec:chiral}). Readers that are familiar with the physics of tensor interactions can skip Sec.\ \ref{sec:examples_mixing}, and go directly to Section \ref{sec:implications} where we use our analytical results from the WKB formalism to analyze the propagation of realistic waveforms from binary black hole mergers, assuming time-dependent parameters in the EoM. 
Since the WKB formalism gives the GW solution in spatial-momentum space, we discuss how to obtain the GW solution in frequency space instead as this latter gives the appropriate description of a temporally-varying GW signal detected at a given location (additional details on connecting momentum and frequency space are also given in App.\ \ref{App:mono_k_omega}). 
We start by summarizing the main observational effects that can be found in subsec.\ \ref{main_effects}, and exemplifying them for velocity (subsec.\ \ref{sec4:vel}), mass (subsec.\ \ref{sec4:mass}), friction (subsec.\ \ref{sec4:friction}) and chiral (subsec.\ \ref{sec4:chiral}) interactions. For each case, we discuss the three relevant timescales of mixing, decoherence and broadening and compare them to each other. Here we show how the total duration of the GW event that would be detected could shorten or lengthen due to phase distortions. These changes in duration can also cause decoherence to be never achieved even if the eigenstates propagate at different speeds if these individual signals get considerably stretched in time (so as to never separate from each other and lead to the individual echoes). In addition, we show how GWs suffer from phase and amplitude distortions, which for mass interactions appear always together. 
In the case of chiral interactions, we also show how the polarization of the signal changes during propagation (characterized by two parameters described in App.\ \ref{App:polarization}) and exhibits amplitude and phase birefringence.  
We study when individual signals may still look like GR waveforms but with different source parameters, focusing mainly on how the GW luminosity distance can be biased if GR is assumed. We furthermore discuss how populations of GW events may help distinguish these scenarios from events coming truly from GR. 
Finally, in Section \ref{sec:discussion} we summarize our findings and discuss their implications and future prospects.

Throughout this paper we use natural units with $c=1$ and $\hbar=1$, unless explicitly mentioned otherwise.
\section{Cosmological GW propagation beyond general relativity}
\label{sec:wavepacket}

Let us consider a perfectly homogeneous and isotropic universe, where the space-time is described by a metric $g_{\mu\nu}=\bar{g}_{\mu\nu}+h_{\mu\nu}$, with a background Friedmann-Robertson-Walker (FRW) metric $\bar{g}_{\mu\nu}$ whose line element in conformal time $\eta$ is given by:
\begin{equation}
    ds^2=a^2(\eta) [-d\eta^2+d\vec{x}^2]
\end{equation}
and linear perturbations $h_{\mu\nu}$. Around homogeneous and isotropic backgrounds, linear perturbations can be decomposed into scalar, vector, and tensor components, according to how they transform under 3-dimensional spatial rotations. Each one of these types of perturbations will evolve independently and can be studied separately. GWs will be described by the tensor modes, which carry two degrees of freedom in GR corresponding to the two possible polarizations of the massless spin-2 graviton. These tensor modes will be described by the transverse traceless spatial components of $h_{\mu\nu}$: $\partial^i h_{ij}=0$ and $\bar{g}^{ij}h_{ij}=0$ for the background FRW metric $\bar{g}_{\mu\nu}$. The two polarizations are usually labelled  $+$ and $\times$ for the linear states, or in terms of right and left-handed circular states $L$ and $R$. In GR, these two polarizations evolve in the same way and hence satisfy the same EoM. In this case, we will use $h$ to describe the amplitude of GWs of either polarizations.  

Even over homogeneous and isotropic backgrounds, GWs could mix with tensor perturbations of additional fields, which could arise in modified gravity theories and non-trivial realizations of the cosmological principle (see introduction of \cite{Jim_nez_2020}). In general, the presence of additional cosmological fields would both change the background and introduce couplings with $h$. In the following we focus on the latter and analyze how the mixing can affect the wavepacket propagation in the presence of one additional field, generically described by $s$. We will begin with a phenomenological approach solving the evolution of GWs in momentum space 
\begin{equation}\label{FourierTransform}
    h(\eta,k)=\frac{1}{\sqrt{2\pi}}\int dx e^{-ikx}\tilde{h}(\eta,x)\,,
\end{equation}
and similarly for $s$. Here $x$ is the spatial coordinate along the direction of the wavevector. Because in position space the waves $\tilde{h}(\eta,x)$ and $\tilde s(\eta,x)$ are real, we see that $h(\eta,-k)=h^*(\eta,k)$ and $s(\eta,-k)=s^*(\eta,k)$. 
Therefore, without loss of generality, we assume for the rest of the paper that $k>0$. 
We analyze a generic propagation equation for second-order derivative models that can be parametrized as:
\be \label{eq:generalequation}
\lb\hat{I}\frac{\d^2}{\d\eta^2} + \nM(\eta)\frac{\d}{\d\eta}+\cM(\eta) k^2 + \pM(\eta) k +\mM(\eta) \rb\bpm h \\ s \epm=0\,,
\ee
where $\nM$, $\cM$, $\pM$ and $\mM$ are the real-valued dimensionful friction, velocity, chiral and mass mixing matrices, respectively, which may depend on time but not on $k$. Here, $\hat{I}$ is the identity matrix, and thus we have assumed that there are no second-order temporal derivative interactions. While this parametrized equation of motion allows for any elements in the mixing matrices, additional conditions on these elements may be required to avoid unstable perturbations, as we will see in the detailed examples considered in Sec.\ \ref{sec:examples_mixing} and \ref{sec:implications}. In addition, if these mixing matrices come from specific non-linear gravity theories (e.g.\ massive bigravity, Yang-Mills, etc.), further restrictions on the matrix elements and their time evolutions will be required in order to ensure consistency with the background evolution and properties predicted by these theories.

We note that in Eq.\ (\ref{eq:generalequation}) we have ignored the possible presence of terms of the form $kd/d\eta$ (and other parity-violating terms \cite{Zhao:2019xmm}) that can appear in theories like Chern-Simons gravity \cite{Alexander:2009tp, Yunes:2010yf}. Similarly, we could have also considered higher spatial derivative terms, as they do not automatically bring instabilities into the system, unlike higher temporal derivative terms that bring Ostrogradski instabilities \cite{Ostrogradsky:1850fid, Woodard:2015zca}. Nevertheless, the WKB approach used in this paper to solve these coupled EoM can be straightforwardly generalized to include these other types of interactions, and to even include more than 2 fields interacting with each other. 

We assume that these EoM come from an action principle, and therefore by renormalizing appropriately the fields $h$ and $s$, one can always bring the matrices $\cM$, $\pM$ and $\mM$ into symmetric forms, whereas $\nM$ can always be brought into a matrix such that $\nM_{12}=-\nM_{21}$. In this paper, we will assume this is the case. 
Note that in this equation we are implicitly assuming that $s$ also carries two tensor polarizations that couple to $h$, but as long as each polarization of $h$ propagates equally there is no need for a special subscript for polarization. When we consider the possibility of polarization-dependent interactions, we will introduce a subscript for polarization. In this case, each one of the polarizations will still satisfy an equation of the form Eq.~(\ref{eq:generalequation}) but the coefficients in the equation may differ. 

In the case in which the interactions between the tensor fields are switched-off, the formalism discussed in this paper allows us to generically describe the modified propagation of GWs over cosmological backgrounds:
\begin{align} \label{eq:prop_single}
   & h'' + \nu(\eta) h' + \omega^2(\eta,k)h=0\,, 
\end{align}
where primes denote derivatives with respect to conformal time and we have defined the dispersion relation
\begin{align} \label{eq:dispersion_single}
   & \omega^2 = c_h^2(\eta)k^2 + \pi(\eta) k + m_g^2(\eta)\,,
\end{align}
where all the parameters are allowed to be time dependent. 
The modified propagation of GWs in light of multi-messenger GW astronomy has been extensively studied in the literature \cite{Lombriser:2015sxa,Nishizawa:2016kba,Bettoni:2016mij,Belgacem:2017ihm, Baker:2017hug,Ezquiaga:2017ekz, Haiman:2017szj,Lagos:2019kds, Bonilla:2019mbm, Mastrogiovanni:2020gua,Mukherjee:2020mha} (see also \cite{Ezquiaga:2018btd,CANTATA:2021ktz} for reviews). Having an electromagnetic (EM) counterpart allows one to directly constrain the propagation speed $c_h$ and the additional friction term $\nu$. 
However, the GW signal alone can also be used to constrain modified gravity, due to the effects that the terms $\pi$ and $m_g$ in the MDR generate in the GW phase evolution. This has also been studied in the past \cite{Will:1997bb, Mirshekari:2011yq, Yunes:2016jcc, Abbott:2017vtc, LIGOScientific:2019fpa} for single fields, but in this paper we clarify how to properly account for the induced waveform distortions.

\subsection{GW solutions with mixing}

In order to illustrate the general behavior of the GWs solutions due to the mixing with $s$, we start by discussing the case when all the matrices in Eq.~(\ref{eq:generalequation}) are constants in time, and obtain the solution for plane waves. In the simple case where there is no friction, $\nM=0$, then the EoM can be simply written as:
\begin{equation}
    \left[ \hat{I} \frac{\d^2}{\d\eta^2}+  \hat{W} \right]\vPhi=0,
\end{equation}
where we have defined $\hat{W}\equiv\cM k^2+\pM k+\mM$ and $\vPhi\equiv(h,s)$. Since $\hat{I}$ and $\hat{W}$ commute, we can diagonalize them simultaneously by changing basis, and hence completely decouple the EoM. This means that $h$ and $s$ will generically be described by linear combinations of two modes $H_{1,2}$ that evolve independently:
\begin{equation}
    \bpm H_1 \\ H_2 \epm = \hat{U}^{-1}\bpm h \\ s \epm\,
\end{equation}
with $\hat{U}$ being the unitary basis transformation matrix. The \emph{propagation eigenstates} $H_A$ (where $A \in \{ 1,2 \}$) can be solved in terms of plane waves $H\propto  e^{i{\omega_A}\eta}$, 
with the eigenfrequencies determined by
\be \label{eq:determinant1}
\det\lb \hat{W}-\iM {\omega_A^2} \rb=0\,.
\ee
This equation will always have four constant solutions of the form $\pm \omega_1$ and $\pm \omega_2$ arising from
\be
\omega^2_{A}=\frac{1}{2}\lp\tr{\hat{W}}\pm\sqrt{4\hat{W}_{12}\hat{W}_{21}+\lp\hat{W}_{11}-\hat{W}_{22}\rp^2}\rp\, \quad \text{if} \quad \hat{W}_{11}>\hat{W}_{22},\label{omegaA_nofriction}
\ee 
where we choose the convention that $\omega_1^2=\hat{W}_{11}$ in the no-mixing limit (i.e.\ when $\hat{W}_{12}=\hat{W}_{21}=0$), so that $\omega_1$ always describes the propagation of the field $h$ in this limit. Therefore, since we define the square root as $\sqrt{x^2}=|x|$ for a real number $x$, $\omega_1^2$ is defined as (\ref{omegaA_nofriction}) with a plus sign in front of the square root if $\hat{W}_{11}>\hat{W}_{22}$, otherwise $\omega_1^2$ is defined with a minus sign. In addition, we will use the overall sign convention that $\omega_A>0$.
Therefore, the independent eigenmodes are given by:
 \begin{equation}\label{EingeModesH}
    H_A=H_{0+,A}e^{i\omega_{A}\Delta\eta} + H_{0-,A}e^{-i\omega_{A}\Delta\eta}\,,
\end{equation} 
where $H_{0\pm,A}$ are constants  determined by the initial conditions at $\eta_0$ and $\Delta\eta=\eta-\eta_0$. 
Notice that since $k>0$ and $\omega_A>0$ by definition, ``$+$" represents a wave propagating in the $-x$ direction whereas ``$-$" represents a wave propagating in the $+x$ direction.

We can transform back to the interaction basis using the matrix of eigenvectors:
\be \label{eq:eM1}
\hat{U}=\frac{\hat{E}}{\sqrt{|\det (\hat{E})|}}; \quad \hat{E}=\bpm 1 & -\frac{\hat{W}_{12}}{\hat{W}_{11}-\omega_{2}^2} \\ -\frac{\hat{W}_{21}}{\hat{W}_{22}-\omega_{1}^2} & 1\epm\,.
\ee
Notice that $\eM_{21}=-\eM_{12}$ so this mixing matrix takes the form of a rotation matrix, up to an arbitrary normalization that we have chosen such that the diagonal terms are unity.
Therefore the unitary mixing matrix $\hat{U}$ involves a renormalization by $\sqrt{|\det (\hat{E})|}$.
The result of the mixing is thus fully parametrized by a \emph{mixing angle}, which is chosen to be such that 
$\tan\Theta_g= \hat E_{12}$ (i.e.~$\hat{E}$ describes a $\vec{\Phi}$ basis that is rotated by $\Theta_g$ clockwise from that of $\vec{H}$), and the frequency difference
$\Delta\omega=\omega_2-\omega_1$, as is familiar from the case of mass mixing of highly relativistic neutrinos. In this paper, we have assumed that $\omega_{A}$ approaches the value $\hat{W}_{AA}$ in the limit of $\hat{W}_{12},\hat{W}_{21}\rightarrow 0$ (see Eq.\ (\ref{omegaA_nofriction})), so that the off-diagonal terms of the mixing matrix $\hat{E}$ vanish in the no-mixing limit (and hence $\Theta_g$ vanishes). In this case, the frequency $\omega_{1}$ ($\omega_2$) determines the propagation of the field $h$ ($s$) in the no mixing limit. We emphasize though that the choice of labels $1,2$ in the eigenfrequencies is a convention that does not affect the final physical result. 

The final solution can be then written as:
\be\label{eq:Phisol1}
\vPhi=\hat{U}\,\vec{H}=\hat{U} \left[ \hat{P}_+ \vec{H}_{0+}  + \hat{P}_- \vec{H}_{0-}\right] 
=\hat{U} \left[ \hat{P}^* \vec{H}_{0+}  + \hat{P} \vec{H}_{0-}\right] 
\,,
\ee
where we have introduced a diagonal propagation matrix for the eigenstates
\be  \hat{P} = \hat P_- = \hat P_+^* 
=\bpm e^{-i\omega_{1}\Delta\eta} & 0 \\ 0 & e^{-i\omega_{2}\Delta\eta} \epm\,.
\ee

If we now allow for a friction matrix, $\nM\not=0$, the EoM are not exactly diagonalizable as $\nM$ will not commute with $\hat{W}.$ Due to the linear derivative in time in the EoM, waves propagating in a given direction are not the same as the time reversal of waves propagating in the opposite direction. Therefore, there is no single matrix $\hat{U}$ that diagonalizes the system, but we can instead obtain two separate $\hat{U}_{+}$ and $\hat{U}_{-}$ matrices that diagonalize the EoM of waves propagating in the two different directions. 
We start by changing to the propagation basis and obtaining the complex eigenfrequencies $\theta_A$, which will satisfy the following equation:
\be \label{eq:determinant}
\det\lb \hat{W}-\iM\theta_A^2+i\nM\theta_A\rb=0\,.
\ee
This equation will generically have four complex constant solutions of the form (see Appendix \ref{App:WKB})
\begin{equation} \label{eq:freq_real_imag}
    \theta_{A\pm}= \pm  \omega_{A} +i \Gamma_A \,,
\end{equation}
where we assume that in the no mixing limit (i.e.\ all diagonal matrices), $ \theta_{A\pm}$ are such that $\Gamma_A=\hat{\nu}_{AA}/2$, and thus the eigenfrequencies 1 and 2 describe the propagation of $h$ and $s$, respectively. Note that there may be models in which the eigenfrequency solutions with mixing are not smoothly connected to those of the no-mixing limit. In those cases, we will choose arbitrarily what eigenfrequency will be 1 and 2, although they will still have the form on Eq.\ (\ref{eq:freq_real_imag}).
This result implies that, in the propagation basis, each eigenstate will follow an independent evolution determined by 
\be
\lb \frac{\d^2}{\d\eta^2} +2\Gamma_{A}\frac{\d}{\d\eta} + \left( \omega_{A}^2 + \Gamma_A^2 \right) \rb H_A = 0\,,
\ee
whose solution is $H_A \propto e^{\pm i \omega_A\eta -\Gamma_A\eta}$. 
Thus $\omega_A$ corresponds to the oscillatory piece of the wave and
determines the phase and group velocity of the waves.  Hence $\pm$ again represents the direction of propagation. $\Gamma_A$ determines the damping or growth of the wave with time and therefore is the same for waves propagating in either direction.

The associated eigenvector matrix that transforms the states back to the interaction basis is given by:
\be \label{eq:eM}
\hat{U}_\pm=\frac{\hat{E}_\pm}{\sqrt{|\det (\hat{E}_\pm)|}}; \quad \hat{E}_\pm=\bpm 1 & -\frac{\hat{W}_{12}+i\nM_{12}\theta_{2\pm}}{\hat{W}_{11}-\theta_{2\pm}^2+i\nM_{11}\theta_{2\pm}} \\ -\frac{\hat{W}_{21}+i\nM_{21}\theta_{1\pm}}{\hat{W}_{22}-\theta_{1\pm}^2+i\nM_{22}\theta_{1\pm}} & 1\epm\,.
\ee
Note that $\pm$ are again related by conjugation as must be the case for a real wavepacket to remain real after propagation:
\begin{equation}
    \eM_\pm = \eM_R \pm i\eM_I\,.
\end{equation}
The final solutions for $h$ and $s$ can then be written as:
\bea
\vPhi &=& \hat{U}_+ \hat{P}_+ \vec{H}_{0,+} + \hat{U}_{-} \hat{P}_{-} \vec{H}_{0,-}\, \nonumber\\
&=& \hat{U}^* \hat{P}^* \vec{H}_{0,+} + \hat{U} \hat{P} \vec{H}_{0,-}\,,
\label{vPhiFriction}
\eea
where $\hat{U}=\hat{U}_-$ and is not necessarily unitary when including friction.  The propagation matrix is now
\be 
\hat P= \hat P_- = \hat P_+^* 
=\bpm e^{- i\omega_{1}\Delta\eta-\Gamma_1\Delta\eta} & 0 \\ 0 & e^{- i\omega_{2}\Delta\eta-\Gamma_2\Delta\eta} \epm\,.
\ee    
From here we explicitly see that friction leads to an exponential suppression in the modes, and that now there is no global transformation matrix that allows us to express the $\vPhi$ solutions as Eq.~(\ref{eq:Phisol1}), without first determining the direction of propagation.
On the other hand, by first separating out the components that propagate in opposite directions, we can write down two separate equations analogous to Eq.~(\ref{eq:Phisol1}).

To clarify the role of mixing and propagation, 
let us consider a wave propagating in the $+x$
direction only and specified initially by
$\vec{\Phi}_0=(h_0,s_0)$ at $\eta_0$.  After propagating 
for $\Delta \eta$ 
\begin{eqnarray}
\vPhi &=&
\hat U \hat P \hat U^{-1} \vPhi_0
\label{eq:propsoln}
\\
&=&
\frac{1}{1 -\hat E_{12}\hat E_{21}}
\bpm 
(h_0-\hat E_{12} s_0) e^{-i\omega_1 \Delta\eta - \Gamma_1\Delta\eta}+ \hat E_{12}(s_0 - \hat E_{21} h_0) 
e^{-i\omega_2 \Delta\eta - \Gamma_2\Delta\eta}
\\ 
(s_0 - \hat E_{21} h_0 )e^{-i\omega_2 \Delta\eta - \Gamma_2\Delta\eta} - \hat E_{21}(h_0 -\hat  E_{12} s_0)e^{-i\omega_1 \Delta\eta -\Gamma_1\Delta\eta}
\epm \,.
\nonumber
\end{eqnarray}
In general, the mixing is described by 4 parameters: the real and imaginary parts of
$E_{12}$ and $E_{21}$ respectively. 

In the special case that we consider below where the initial state is purely $h_0$, then the detected $h$ after propagation depends only on the combination $\eM_{12} \eM_{21}$: 
\begin{eqnarray}\label{hsolconstparam}
h(\eta,k)  
&=& \frac{h_0(k)}{{1 -\hat E_{12}\hat E_{21}}}
\left(
e^{-i\omega_1\Delta\eta - \Gamma_1\Delta \eta} - \hat E_{12}\hat E_{21} 
e^{-i\omega_2 \Delta\eta - \Gamma_2\Delta\eta}\right),
\end{eqnarray}
whereas $s$ depends separately on $\eM_{12}$ and $\eM_{21}$.
Therefore it is convenient to define the mixing  angle as:\footnote{Note that a change in the normalization of $\hat{E}$ will not change the mixing angle but will change the corresponding $\vec{H}$ vector that satisfies the EoM. 
Although Eq.~(\ref{eq:mixing_angle}) is not manifestly invariant under rescalings of $\hat{E}$, more generally one generically has to divide by $\eM_{11}\eM_{22}$ the right hand side of this equation. As previously mentioned, the sign of $\eM$ can be fixed by the convention that $\Theta_g$ describes a clockwise rotation from $\vec{\Phi}$ to $\vec{H}$.} 
\be
\tan^2\Theta_g(\eta,k)= | \eM_{12}\eM_{21} |.  \label{eq:mixing_angle}
\ee
We can also define an associated phase to the mixing, which is given by:
\begin{equation}\label{mixphase}
   \phi_g(\eta,k)=-i\ln\lp\frac{\eM_{12}\eM_{21}}{\vert\eM_{12}\eM_{21}\vert}\rp\,.
\end{equation}
Notice that in the absence of friction $\hat{E}$ is a purely real rotation matrix and hence $\phi_g=\pi$ and $\Theta_g$ becomes the usual mixing angle that fully determines the mixing matrix (and therefore only one parameter determines both $h$ and $s$). The same will hold when we only have friction mixing with off-diagonal terms, since in that case $\hat{E}_{12}$ and $\hat{E}_{21}$ are purely complex.
The amplitude of $h$ can be conveniently expressed as:
\begin{equation}\label{eq:amplitude_const}
\vert h(\eta,k)\vert^2 =  \vert h_0(k)\vert^2\left[ |f_1(\eta,k)|^2 + |f_2(\eta,k)|^2  - 2 |f_1(\eta,k)||f_2(\eta,k)|\cos\lp\Delta\omega\Delta\eta - \phi_g\rp\right].
\end{equation}
Here, we have introduced the functions $f_1$ and $f_2$ such that they describe the contributions from the eigenstates $H_{1,2}$ to the detected signal $h$:
\footnote{One should note that in order to separate the time variation of the amplitude from the high frequency oscillation of the phase, we need $\Delta \omega\ll \omega_A$. }
\begin{equation}
    h(\eta,k)=h_0(k) \left[ f_1(\eta,k)e^{-i\omega_1\Delta \eta} +f_2(\eta,k)e^{-i\omega_2\Delta \eta}\right]
\end{equation}
where
\begin{equation}
 f_1(\eta,k)=  \frac{e^{-\Gamma_{1}\Delta\eta}}{1 - \tan^2\tg e^{i\phi_g}}\,  \quad ; \quad f_2(\eta,k)= -f_1(\eta,k)\tan^2\tg\,e^{-\Delta\Gamma\Delta\eta +i\phi_g}\,,
\end{equation}
and we have defined the damping difference $\Delta\Gamma=(\Gamma_{2}-\Gamma_{1})$.
Notice that the mixing phase $\phi_g$ not only shifts the phase of the mixing but also changes the overall amplitude of $|f_1|$ and $|f_2|$ together whereas $|f_2|/|f_1| = \tan^2\Theta_g e^{-\Delta \Gamma \Delta \eta}$.

Therefore for the case of interest, the change to the propagation of $h$ is
determined by four parameters: $\Delta \omega$, which determines the mixing frequency and the difference in phase velocities; $\Delta\Gamma$ which determines the relative damping of these components; $\Theta_g$ which controls the amplitude of mixing; $\phi_g$ which controls the phase of mixing. For a monochromatic wave, the mixing term creates an amplitude modulation  that oscillates
between $(|f_1|\pm |f_2|)|h_0|$, and as we shall see, for a wavepacket with a spread in frequencies, this modulation can eventually decohere into two wavepackets with amplitudes $|f_1h_0|$ and $|f_2h_0|$ respectively.

We can perform a similar construction for the $-x$ propagating components which are given by $\omega_A \rightarrow -\omega_A$ and
$\phi \rightarrow -\phi$. 
For more general initial or detection conditions there would be two additional mixing parameters describing the
 mixing from and into the $s$ state.  
 
Finally to construct a wavepacket after mixing due to propagation in the $+x$ direction
out of the Fourier components of a real initial 
wavepacket, we have:
\begin{equation} \label{eq:wavepacket_prop_solution}
    \vPhi(\eta,x) = \frac{1}{\sqrt{2\pi}} \int_0^\infty dk \left[
    e^{i k x}\hat U \hat P \hat U^{-1}
\vPhi_0(k) +
    e^{-i k x}\hat U^* \hat P^* \hat U^{-1*}
\vPhi_0^*(k)\right] .
\end{equation}
 Notice that an initially real wavepacket remains real.  A similar relation follows for the component propagating in the $-x$ direction with  $\hat U \hat P \hat U^{-1}$ terms replaced by their conjugate.  

\subsubsection{WKB approximation}\label{WKBsubsection}
If we allow the coefficients in the EoM to vary in time, in the limit of a large GW frequency compared to the time variation of the matrix elements, which will always be the case if they evolve on cosmological time scales, we can solve the propagation in momentum space with a WKB approximation \cite{Jim_nez_2020}. 
In essence, what the WKB approximation does is to provide a framework to obtain iteratively the change of basis and the propagation eigenstates, which will be characterized by a time dependent frequency and amplitude. The WKB solution is obtained solving the propagation equation (\ref{eq:generalequation}) order by order in terms of a small dimensionless parameter $\epsilon$, 
which corrects for the  evolution of the amplitude and phase, whose rate of change is taken to be slow compared with the frequency of the wave.
Following \cite{Jim_nez_2020}, for modes propagating in a given direction (either $+x$ or $-x$), it is convenient to introduce the following ansatz for plane waves that iteratively corrects the constant coefficient solution (\ref{eq:propsoln})
\begin{eqnarray} \label{eq:wkb_form}
\vPhi(\eta)  &\equiv&\hat U (\eta) \hat P(\eta,\eta_0) \hat U^{-1}(\eta_0) \vPhi_0 \nonumber\\ 
&=&
\hat U (\eta)\,\hat{P}_0(\eta,\eta_0) \,[\hat{Q}_0(\eta,\eta_0) + \epsilon \hat{Q}_1(\eta,\eta_0)+ \cdots] \hat{U}^{-1}(\eta_0) \vPhi_0\,.
\end{eqnarray}
Note that we can again interpret this solution as one where we have the mixing matrix $\hat{U}$ transforming the propagation eigenmodes $\vec{H}$ to the observed modes $\vec{\Phi}$, such that $\vec{\Phi}(\eta)=\hat{U}(\eta)\vec{H}(\eta)$. The way these eigenmodes propagate will be described by the propagation matrix $\hat{P}$, where $\vec{H}(\eta)=\hat{P}(\eta)\vec{H}_0$ with $\vec{H}_0=\hat{U}^{-1}(\eta_0)\vec{\Phi}_0$. We will see that these eigenmodes propagate effectively independently from each other in the WKB approximation and, as before, they have eigenfrequencies given by $\hat{\theta}$. Also, the full propagation matrix $\hat P$ will contain the matrices $\hat{Q}_i$, which correct its constant-coefficient form:
\be
\hat P_0(\eta,\eta_0) = e^{+\frac{i}{\epsilon}\int_{\eta_0}^{\eta}\hat\theta(\eta') d\eta'}\,.
\ee
Here we have introduced an $\epsilon^{-1}$ scaling since we assume that the phase matrix $\hat{\theta}$ evolves much more quickly than the amplitude matrices (that are of order $\epsilon^n$ with $n\geq 0$). 
When the coefficients are constants, we have at lowest order that $\hat{Q}_0$ is the identity matrix, in which case we recover the results of the previous section. 

When applying the WKB approximation, one can see that the leading order EoM determine the eigenfrequencies $\tM$, which are given by the roots of the quartic equation (\ref{eq:determinant}). Thus, they have the same structural form as the constant coefficient case but now each parameter from the propagation equation could be a function of time. As a consequence, we obtain four solutions of the form (\ref{eq:freq_real_imag}), which can again be separated into real and complex parts as:
\be
\tM_\pm(\eta)=\bpm \theta_{1 \pm}(\eta) & 0 \\ 0 & \theta_{2 \pm}(\eta)\epm=\pm\hat{\omega}(\eta)+i\hat{\Gamma}(\eta)\,,
\ee
which describes $+x$ propagating modes for $\hat{\theta}_{-}$, and $-x$ propagating modes for $\hat{\theta}_{+}$.
Similarly, the leading order WKB equation tell us that $\hat U(\eta)$ will correspond to the same eigenvector matrix as in Eq.~(\ref{eq:eM}) at time $\eta$, for each $\hat{\theta}_{\pm}$ solution. Thus, the first elements of the WKB expansion look like the constant solution substituting the constant coefficients by time dependent functions. 
In other words, at leading order the WKB eigenstates follow adiabatically the local changes in the background.

Next, we analyze the first-order corrections from the WKB expansion to the constant coefficient solution. This WKB order gives a first-order differential equation for the propagation eigenmodes upon expressing  $\vec{H}(\eta)=\hat{P}_0(\eta,\eta_0) \vec{H}_Q(\eta)$, with $\vec{H}_Q(\eta)=\hat{Q}_0(\eta,\eta_0)\vec{H}_0$:
\begin{equation}\label{EqnH0Q}
    (2\hat{\theta} + i \hat{U}^{-1}\hat{\nu}\hat{U})\hat{P}\vec{H}_Q' + (\hat{\theta}' +2\hat{U}^{-1}\hat{U}'\hat{\theta} +i \hat{U}^{-1}\hat{\nu}\hat{U}')\vec{H}_Q=0.
\end{equation}
This equation tells us how $\hat{Q}_0$ evolves in time. In particular, we can rewrite (\ref{EqnH0Q}) as\footnote{This equation is equivalent to equation (2.11) of \cite{Jim_nez_2020}.}:
\begin{align}
&\hat{Q}'_{0}=-\hat{A}_\text{WKB} \hat{Q}_{0};\quad \hat{Q}_0=e^{-\int_{\eta_0}^{\eta}  \hat{A}_\text{WKB}(\eta')d\eta'}\,,
\label{eq:WKBeps1}
\end{align} 
where
\begin{equation}
    \hat{A}_\text{WKB}=\hat P_0^{-1}\lp 2\hat{U}\tM+i\nM\hat{U}\rp^{-1}\lp\hat{U}\tM'+2\hat{U}'\tM+i\nM\hat{U}'\rp \hat P_0.
\end{equation}
Note that the solution in Eq.~(\ref{eq:WKBeps1}) has free integration constants that we have fixed so that
$Q_0(\eta_0,\eta_0)=\hat I$ since $\vPhi(\eta_0)=\vPhi_0$.
From here we obtain that, at leading order in $\epsilon$, the most general solution for $\vec{\Phi}$ will have the following form:
\begin{equation}\label{WKBSolPhi}
    \vec{\Phi}(\eta)= \sum_{\pm}\hat{U}_{\pm}(\eta) \hat P_{0\pm}(\eta) 
    e^{-\int_{\eta_0}^\eta \hat{A}_{\text{WKB}\pm}(\eta') d\eta'}\hat{U}^{-1}_{\pm}(\eta_0) \vPhi_{0\pm}.
\end{equation}
This WKB result can be generalized iteratively going to higher orders in the expansion $\epsilon$. 
Note that the $-x$ propagation matrices $\hat{U}_{+}$, $\hat{P}_{0+}$ and $\hat{A}_{\text{WKB}+}$ are related by conjugation to those of the $+x$ propagation, as described in the previous section. From now on, we focus only on the $+x$ propagating modes and drop the corresponding subscript $-$ from all the matrices.

In the case of time varying coefficients, we see that generically $\hat{A}_\text{WKB}$ (and hence $\hat{Q}_0$) is a time varying non-diagonal matrix that corrects the leading-order propagation of the system determined by $\hat{U}$ and $\hat{P}_0$. 
The off-diagonal terms of $\hat{A}_\text{WKB}$ mean that the two eigenmodes $\vec{H}$ found in the previous section for constant coefficients (that can be interpreted as ``instantaneous" eigenmodes) no longer propagate independently. Indeed, in Eq.~(\ref{EqnH0Q}) we can see that the matrix coefficients of $\vec{H}_Q'$ and $\vec{H}_Q$ are not always diagonal, and thus the two propagation eiegnmodes are generically expected to mix with each other. 
Nevertheless, there is a simplifying adiabatic regime, in which these eigenmodes do propagate nearly independently. This happens when the change in the local eigenmodes
from the change of the mixing coefficients is slowly varying compared with the mixing time itself. 
Schematically, we can factor out of $\hat{A}_\text{WKB}$ the terms that come from $\hat P_0$ and $\hat P_0^{-1}$ and write its elements  as:
\begin{equation}
    \hat{A}_{\text{WKB},\,AB} = e^{i\int (\theta_A-\theta_B)d\eta} \mathcal{A}_{AB}\, ,
\end{equation}
where $\mathcal{A}_{AB}$ (for $A,B={1,2}$) is a slowly-varying linear combination of $\hat{\theta}'$ and $\hat{U}'$.
For the diagonal terms $A=B$ this exponential prefactor is unity, but for the off-diagonal
terms $A\ne B$ it oscillates with the mixing frequency 
$|\theta_1-\theta_2|$.  Therefore the off-diagonal term in 
$\ln Q_0$ is the integral of the product of a slowly varying and oscillating term, which is expected to average out when these oscillations are fast enough as they should be for sufficiently high $k$. 
On the other hand, the diagonal terms in $\ln Q_0$ are integrals over terms that are proportional to $\theta'$ and $U'$ and therefore reflect the net change in the properties of the local eigenstates
 across
the propagation distance.   
The relative suppression of the off-diagonal terms means that two local eigenstates propagate nearly 
independently.\footnote{Note that even when the two  eigenmodes do propagate nearly independently, one cannot directly neglect their interactions in Eq.~(\ref{EqnH0Q}). If one were to do that, Eq.~(\ref{EqnH0Q}) would lead to two independent solutions for the components of $\vec{H}_Q$, which would have the structure $H_{Q,A}\propto \exp\{\int d\eta\, (\tilde{\theta}'_A)/(2\theta_A+i\tilde{\nu}_A)\}$, where $\theta_A=\hat{\theta}_{AA}$, $\tilde{\nu}_A=(U^{-1}\hat{\nu}\hat{U})_{AA}$ and $\tilde{\theta}'_A=(\hat{\theta}' +2\hat{U}^{-1}\hat{U}'\hat{\theta} +i \hat{U}^{-1}\hat{\nu}\hat{U}')_{AA}$. However, this solution does not yield exactly the same result as when neglecting the off-diagonal terms after obtaining the full solution $\vec{H}_Q=\exp\{\int d\eta\, \hat{A}_\text{WKB}\}\vec{H}_0$.}

As an example, let us assume no friction: $\hat{\nu}=0$. In this case, we find that $\hat{A}_\text{WKB}$ is given by:
\begin{equation}
\hat{A}_\text{WKB}=\begin{bmatrix} 
 \frac{1}{2}\frac{\theta_1'}{\theta_1} &  0 \\
0 & \frac{1}{2}\frac{\theta_2'}{\theta_2}.
\end{bmatrix}
+ \Theta_g'\begin{bmatrix} 
 0 &  \frac{\theta_2}{\theta_1}e^{-i\int (\theta_1-\theta_2)d\eta} \\
-\frac{\theta_1}{\theta_2}e^{i\int (\theta_1-\theta_2)d\eta} & 0
\end{bmatrix}.
\end{equation}
Here we explicitly see that the off-diagonal terms oscillate at the mixing frequency. Since to obtain $\vec{\Phi}$ we integrate over $\hat{A}_\text{WKB}$, we expect these quickly oscillating off-diagonal terms to average out as long as the WKB time variations are on a much longer scale than the mixing time.
Note that the diagonal terms in $\hat A_\text{WKB}$ can be explicitly integrated because they form total derivatives. Assuming the adiabatic approximation in which we neglect the contribution of the off-diagonal terms, the solution for $\vec{\Phi}$ simplifies to:
\begin{equation}
\vec{\Phi}=\hat{U}(\eta)
\hat P_0 
\begin{bmatrix} 
 \sqrt{ \frac{\theta_1(\eta_0)}{\theta_1(\eta)}} &  0 \\
0 & \sqrt{ \frac{\theta_2(\eta_0)}{\theta_2(\eta)}}
\end{bmatrix} \hat U^{-1}(\eta_0) \vPhi_0, 
\end{equation}
which trivially generalizes the usual WKB form for the amplitude and phase evolution. In general cases with friction, the matrix $\hat{A}_{\rm WKB}$ takes a more complicated form. However, in the large-$k$ limit and for small mixing angles (i.e.~$|E_{21}|,|E_{12}|\ll 1$), one expects\footnote{This can be confirmed by noticing that for theories with an EoM with a maximum of two derivatives, then $\hat{W}$ is of order $\mathcal{O}(k^2)$ in the large-$k$ limit, and the eigenfrequencies in $\hat{\theta}$ are $\mathcal{O}(k)$. Therefore, the mixing matrix $\hat{U}$ has elements that are at most of order $\mathcal{O}(k^0)$. Then, the matrix $\hat{U}^{-1}\hat{\nu}\hat{U}$ has elements $\mathcal{O}(k^0)$ and they will be subdominant compared to $\hat{\theta}$. Similarly, we will have that $|\hat{U}^{-1}\hat{\nu}\hat{U}'|\sim \mathcal{O}(k^0)$ and can be neglected. However, $|\hat{U}^{-1}\hat{U}'\hat{\theta}|\sim\mathcal{O}(k^1)$ and can be comparable to $\hat{\theta}'$. However, in the small mixing angle regime (i.e.~small deviations from GR) we can also neglect $|\hat{U}^{-1}\hat{U}'\hat{\theta}|$.} 
$|\hat{\theta}| \gg |\hat{U}^{-1}\hat{\nu}\hat{U}|$  and  $|\hat{\theta}'|\gg |\hat{U}^{-1}\hat{U}'\hat{\theta} + i \hat{U}^{-1}\hat{\nu}\hat{U}'|$
so that $\hat{A}_{\rm WKB}\approx \hat{\theta}' (2\hat{\theta})^{-1}$.

In general, in the adiabatic approximation we can still interpret the solution in terms of two eigenmodes with oscillation frequencies $\omega_{1}$ and $\omega_{2}$ and solve the wave packet propagation as in equation (\ref{eq:wavepacket_prop_solution}). When the initial condition is $\vec{\Phi}_0=(h_0,0)$, the amplitude of $h$ can be expressed in terms of the matrix $\hat{Q}_0$ as:
\begin{align}\label{hsolWKB}
h(\eta,k)  
&= \frac{h_0(k)}{\left[1 -\hat E_{12}(\eta_0)\hat E_{21}(\eta_0)\right]}\sqrt{\frac{|\det \hat{E}(\eta_0)|}{|\det \hat{E}(\eta)|}}
\left\{
e^{-i\int \omega_1d\eta - \int \Gamma_1 d \eta} \hat{Q}_{0,11}(\eta,\eta_0)\right. \nonumber\\
&-\left. \hat E_{12}(\eta)\hat{E}_{21}(\eta_{0})e^{-i\int \omega_2 d\eta - \int \Gamma_2d\eta}\hat{Q}_{0,22}(\eta,\eta_0)  \right\}\\
&\equiv h_0(k)\left[f_1(\eta,k)e^{-i\int \omega_1d\eta}+ f_2(\eta,k)e^{-i\int \omega_2d\eta} \right]\label{hsolWKB_2}
,
\end{align}
where recall we assume a $+x$ propagating mode, $\hat{U}=\hat{E}/\sqrt{|\det \hat{E}|}$ and $\hat{E}$ has unity numbers in the diagonals.
Here we have also implicitly assumed that the off-diagonal terms of $\hat{Q}_0$ vanish and $\theta_1\not=\theta_2$.  
We also emphasize that $\hat{Q}_0$ may not always be real, but it is a slowly varying function of the cosmological background. Therefore, the phase of $\hat{Q}_0$ does not determine the oscillation frequency of the eigenmodes, but it will contribute to additional phase off-sets.
The constant coefficient solution in Eq.~(\ref{hsolconstparam}) is recovered when $\hat{Q}_0$ is the identity matrix.

The definitions of the mixing angle and phase have the same expressions as in Eq.~(\ref{eq:mixing_angle})-(\ref{mixphase}), and now depend on time. Note that now, however, the mixing angle does not fully determine the amplitude of the second eigenmode to $h$, as it was the case for constant coefficients.
In Eq.\ (\ref{hsolWKB_2}) we have explicitly expressed the solution of $h$ in a compact form as a superposition of the propagating eigenstates, each one with its own dispersion relation $\omega_A$ 
and amplitude correction function $|f_A(\eta,k)|$. 

\subsection{GW wavepacket propagation}
\label{sec:GW-propag}
In the previous sections we have seen that the GW propagation of each $k$ mode is determined by a superposition of the propagation eigenstates, which can have both non-trivial dispersion relations as well as wavenumber or time dependent amplitudes. 
We can generically write the GW solution when only $h$ is emitted as:
\begin{equation}
    h(\eta,k)= \sum_A h_0(k)\, f_A(\eta,k)\,e^{-i\phi_A(\eta,k)}\,,
\end{equation}
where the phases are
$\phi_A=\int\omega_A d\eta$. From here we can identify distinct regimes in the propagation of the GW signal, which are summarized in Table \ref{table:summary_time}, and depend on how the propagation time of the signal $T$ compares to certain characteristic timescales.  
Here, $T_\text{mix}$ indicates when the phase interference between the two eigenmodes becomes relevant, in cases where $\phi_1\not=\phi_2$. 
GW signals associated with mergers are temporally localized in a finite wavepacket. 
Therefore,  we may also have that its associated group velocities are different, $v_{g,1}\not=v_{g,2}$, and thus the two eigenstates propagate at different speeds and may eventually decohere, leading to two individual GW signals or echoes. In this case, there is no interference between the two propagating eigenstates and thus $T_\text{mix}$ becomes irrelevant. Finally, the timescale $T_\text{broad}$ indicates the moment when each eigenstate starts suffering considerable deformations due to the fact that non-trivial dispersion relations typically lead also to dispersive group velocities $v_{g,A}(k)$. Note that each eigenstate may have its own distinct $T_\text{broad}$. In  Table \ref{table:summary_time} we show what the physical effects in the signal will be for different timescale scenarios.

\begin{table}[t!]
\centering
\begin{tabular}{ |c|c | c|} 
\hline
\rowcolor{mygray}\textbf{Timescales} & \multicolumn{2}{c|}{\textbf{Physics Effects}}\\ 
\hline
$T\ll T_\text{mix},T_\text{broad}$ &  \multirow{4}{*}{Single} & Unmodified waveform \\
\cellcolor{mygray}{$T_\text{broad}<T \ll  T_\text{mix}$ } &  & \cellcolor{mygray}{Distortion due to MDR}  \\ 
$T_\text{mix}<T<T_\text{broad},T_\text{coh}$ & & Distortion due to interference \\ 
\cellcolor{mygray}{$T_\text{mix}, T_\text{broad}<T<T_\text{coh}$} & & \cellcolor{mygray}{Distortion due to interference and MDRs}\\ 
\hline
$T_\text{coh}<T \ll  T_\text{broad}$ &   \multirow{2}{*}{Echoes} & Unmodified phase with different amplitude \\
\cellcolor{mygray}{$T_\text{coh}, T_\text{broad} < T$} & & \cellcolor{mygray}{Distorted due to MDRs} \\ 
 \hline
\end{tabular}
\caption{Summary of the propagation regimes and their effects on the signal. An unmodified waveform refers to one that propagates with the dispersion relation $\omega=ck$, and hence keeps the shape of its phase evolution during propagation. In addition, MDR refers to a modified dispersion relation. Note that we always have that $T_\text{mix}<T_\text{coh}$, thus in the top four regimes, we will detect a single GW event, whereas in the bottom two regimes we will detect GW echoes.}
\label{table:summary_time}
\end{table}

These time scales can be illustrated by 
considering the propagation of a finite wave\-packet, with a central frequency $k_0$. In this case, the phase of the Fourier components of the wavepacket of each propagating eigenstate can be expanded as:
\begin{equation}
    \phi_A = \int_{\eta_0} d\eta
    \left[ \omega_A(\eta,k_0) +  
    \frac{\partial \omega_A }{\partial k}
    (k-k_0) + \frac{1}{2} 
     \frac{\partial^2 \omega_A }{\partial k^2} (k-k_0)^2 + \ldots\right],\label{PhaseExpansion}
\end{equation}
for $A=1,2$, and $\eta_0$ the emission time. On the one hand, the zeroth-order term here determines the phase velocity $v_{\mathrm{ph},A}=\omega_A(\eta, k_0)/k_0$ of the $k_0$ mode,  
and provides a constant phase shift to the wavepacket. Since the two eigenmodes have different frequencies $\omega_A$, when they propagate coherently the total signal is expected to exhibit oscillations due to the mixing at frequency $\Delta \omega=\omega_2-\omega_1$ (see Eq.~(\ref{eq:amplitude_const})). This allows us to define the typical time scale where the phase of the two eigenmodes mix, and its associated length scale:
\begin{equation}
    \int_{\eta_0}^{\eta_0+T_{\rm mix}} d\eta \, \Delta \omega(\eta,k_0) \sim 2\pi; \quad L_\text{mix}\sim \langle v_{g,A}(k_0)\rangle T_\text{mix},
\end{equation}
where the length scale is defined as the corresponding location of the wavepacket at $T_{\rm mix}$, using the time average $\langle v_{g,A}(k_0)\rangle$ of the group velocity.  We shall see below that to define the smallest length scale at which mixing is important, $v_{g,A}$ is the group velocity of the eigenmode that propagates faster. Using the group velocity ensures that regardless of whether the initial wavepacket is a pure state in $h$ in momentum or frequency space that the mixing time and length are the same for a sufficiently broad wavepacket (see analogous discussion for neutrino mixing \cite{Cohen:2008qb} and App.~\ref{App:mono_k_omega}).
These scales will tell us whether for a given source at a fixed distance we expect to see a change in the interaction states due to mixing, as with neutrino mixing.
When mixing occurs, the amplitude of $h$ undergoes oscillatory modulation between $(|f_1| \pm |f_2|) |h_0|$ with frequency $\Delta \omega$ according to Eq.~(\ref{eq:amplitude_const}). 

The group velocity itself is related to the first order term in Eq.~(\ref{PhaseExpansion}) through the stationary phase approximation:
\begin{equation}\label{propspeed}
    v_{g,A}(\eta,k_0) = \left. \frac{\partial \omega_A(\eta,k) }{\partial k} \right\vert_{k=k_0}.
\end{equation}
If the two eigenmodes propagate at different group velocities, after a sufficiently long time, the two modes will decohere and $h$ will have two observably  distinct components that effectively propagate independently from each other. More specifically, the temporal coherence of the wavepacket is lost at the time $T_\text{coh}$ when the difference in the group velocities of each eigenstates $\Delta v_g=v_{g,2}-v_{g,1}$ 
has introduced a spatial separation larger than the width of the eigenstate wavepackets, $\sigma_A$,
\be\label{Tcoh_gaussian}
\left| \int_{\eta_0}^{\eta_0+T_\text{coh}} d\eta\, \Delta v_g(\eta,k_0) \right| \sim \sigma_A(\eta=\eta_0+T_{\rm coh}),
\ee
where, we shall see below, that due to
wavepacket distortion effects the width may evolve from its initial value. Furthermore, since each eigenstate has a different dispersion relation, they may suffer different amount of distortions. In Eq.\ (\ref{Tcoh_gaussian}) we will make the conservative choice of using the width $\sigma_A$ of the wavepacket that has elongated the most.

Thus, the coherence time $\Tcoh$ sets a relevant time scale in the GW propagation beyond GR that can be used to divide the propagation into different regimes as a function of the travel time $\Delta \eta=\eta-\eta_0$. Similar to the mixing length scale, one could define a coherence length scale as
\begin{equation}\label{eq:Lcoh}
    \Lcoh\sim \langle v_{g,A}(k_0)\rangle\,\Tcoh\,,
    \end{equation}
where $v_{g,A}$ is chosen as the group velocity of the eigenmode propagating faster.
This scale is useful when determining whether the GW signal is in the coherent or decoherent regime.
On the one hand, for $\Delta\eta \ll T_\text{coh}$, the two eigenmodes propagate coherently and hence the GW amplitude $|h|$ is given by Eq.~(\ref{eq:amplitude_const}) for nearly monochromatic wavepackets, where the third term describes the coupling between the two eigenmodes. On the other hand, for $\Delta\eta \gg T_\text{coh}$, $|h|$ has two separate wavepackets, whose amplitudes are given by $|f_1 h_0|$ (from the first eigenmode) and $|f_2h_0|$ (from the second eigenmode). 
Similarly, their time delay will be given by the difference in the group velocities of the eigenstates $\Delta v_g$ and the distance to the source.
Note that depending on the theory and the distance to the source, the propagation eigenmodes might always be in the coherent state observationally. 
Moreover, the duration of the GW signal, or its waveform width $\sigma_x$ is subject to the detector sensitivity. We will discuss more on the observational implication in Section \ref{sec:implications}.

The third term in Eq.~(\ref{PhaseExpansion}) gives the group velocity dispersion. The fact that each propagation eigenstate behaves as a dispersive wave also tell us that there could be distortions of the wavepackets in the propagation basis as well. 
If the initial wavepacket has a width
$\sigma_x$  and thus a range in $k$ of width $\sigma_k \sim \sigma_x^{-1}$, then the dispersion in group velocities across the packet will distort its shape when
\begin{equation}\label{Tbroad_gaussian}
\int_{\eta_0}^{\eta_0+T_\text{broad}}  d\eta\, \sigma_k \frac{\partial v_{g,A}}{\partial k}(\eta,k_0)\sim \sigma_x \quad \Rightarrow \quad  \int_{\eta_0}^{\eta_0+T_\text{broad}}  d\eta\, \frac{\partial^2 \omega_{A}}{\partial k^2}(\eta,k_0) \sim \sigma_x^2.
\end{equation}
For example with a Gaussian wavepacket with a velocity constant in time, the group velocity dispersion will lead to a broadening
to:
\begin{equation}\label{SigmaBroad}
    \sigma_A= \sigma_x \sqrt{ 1 + \left(\frac{\Delta\eta}{T_{{\rm broad},A}}\right)^2}\,.
\end{equation}
Note that the amplitude of the wavepacket will also diminish as $|h_{A}(\sigma_A)/h_{A}(\sigma_x)|=\sigma_x/\sigma_A$ (for $A=1,2$) due to the broadening.
In more general situations where various frequencies in the wavepacket are emitted at different times, such as in the case of  the coalescence of two compact objects, the distortions caused by group velocity dispersion can lead a shrinking of the duration of the signal as well, but we will continue to use ``broad" to denote the effect.
We recall that if $T_{\rm coh}>T_{\rm broad}$ then the broadened width of the wavepacket should be used in determining the coherence time as shown in Eq.~(\ref{Tcoh_gaussian}). For a Gaussian wavepacket, combining Eq.\ (\ref{SigmaBroad}) 
and (\ref{Tcoh_gaussian}), for constant coefficients in the EoM of $h$ and $s$, we find that 
\begin{equation}\label{TcohSimple}
    T_\text{coh}\sim \sigma_x / \sqrt{\Delta v_g^2(k_0) - \sigma_x^2/T^2_\text{broad,A}(k_0)}.
\end{equation}
For the decoherence time to remain finite one needs 
\begin{equation}\label{DecohCrit}
    \Delta v_g(k_0) T_\text{broad,A}(k_0) > \sigma_x.
\end{equation}
This is because the ratio of the wavepacket separation to the broadened width eventually becomes constant in time in a case of constant coefficients,
\begin{equation}\label{SepInifity}
    \frac{\int_{\eta_0}^{\eta_0+\Delta\eta}\Delta v_g(k_0) d\eta}{\sigma_A} \rightarrow \frac{\Delta v_g(k_0) T_\text{broad,A}(k_0)}{\sigma_x},
\end{equation}
despite the growth in both.

Next, we emphasize that for certain propagation times, higher-order terms in the expansion (\ref{PhaseExpansion}) may also be relevant. In particular, the cubic term $(1/6)\partial^3\omega_A/\partial k^3 (k-k_0)^3$  can lead to additional distortions for  propagation times longer than $T > T_{3,A}$ where $T_{3,A}\sim \sigma_x^3/(\partial^3\omega_A/\partial k^3)$. Furthermore, even when $T<T_{3,A}$, the series expansion (\ref{PhaseExpansion}) can only be approximately truncated to quadratic order if $T_{\text{broad},A}<T_{3,A}$. For toy Gaussian wavepacket, note that since $\sigma_x$ may broaden in time due to $( \partial^2 \omega_A/\partial k^2)\not=0$, then the relative difference between $T_{\text{broad},A}$ and $T_{3,A}$ can change in time. Indeed, since $T_{3,A}/T_\text{broad}\propto \sigma_x$, at short times we may have $T_\text{broad}\sim T_{3,A}$ depending on the model parameters. An analogous reasoning can be applied to higher-order terms in Eq.~(\ref{PhaseExpansion}).

Note that in this section we have Taylor expanded the effects of the dispersion relation as in Eq.~(\ref{PhaseExpansion}) as a tool to discuss the different physical effects that mixing can bring. However, in practice, in the rest of the paper we make use of the exact MDRs for both propagating eigenstates, and therefore all the effects discussed here (and higher-order corrections, if present) will be included.

From this Taylor expanded analysis, we conclude that
the phenomenological description of GW wavepacket propagation naturally involves the group velocity in determining key observational properties of the signal.
This is different from many previous works in the literature where the GW signal was interpreted as an ensemble of gravitons traveling at the particle velocity \cite{Will:1997bb,Mirshekari:2011yq}. 
Such an approach has been motivated by explicit Lorentz-violating theories of quantum gravity, while here we are dealing with spontaneous violations of Lorentz invariance due to a modified gravity propagation over cosmological backgrounds.

In addition to the changes discussed here
in the arrival time and shape of the overall wavepacket, the phase evolution of the waveform across the wavepacket provides a direct observable of GW signals. 
This can also  be understood from Eq.~(\ref{PhaseExpansion}) and the fact that the group velocity depends on the frequency, $v_{g,A}(k)$, which will change the arrival time and phase of each frequency component within the wavepacket, as well as their relative phases with respect to GR. If the phase evolution of gravitational waves in GR is known, then it is possible to test for deviations from GR.
In particular, these waveform distortions due to the phase are the analogue of amplitude distortions and occur when $T>T_{\rm broad}$.  They can be straightforwardly tested in the decoherence regime $T> T_{\rm coh}$, when there is no additional mixing between the two eigenstates. In this regime, the propagation of each eigenstate can be considered independently, resembling the case of a single tensor mode in Eq.~(\ref{eq:prop_single}) with dispersion relation (\ref{eq:dispersion_single}). 
The phase modification of each eigenstate with respect to GR can be generically expressed as:
\begin{equation}
    h_A(\eta,k)\;\propto \; h_\text{fid}(\eta,k)\,e^{-i\Delta\phi_{\text{mg},A}(\eta,k)}\,,
\end{equation}
where $h_\text{fid}(\eta,k)=h_0(k)e^{-ik(\eta-\eta_e)}$ for an initially emitted plane wave $h_0(k)$ at $\eta_e$, and \begin{equation}\label{DeltaPhik}
    \Delta\phi_{\text{mg},A}(\eta,k) = \int_{\eta_e}^{\eta} \lp\omega_A(k)-k\rp d\eta'\,.
\end{equation}
In Section \ref{sec:implications}, we will apply the formalism developed here to realistic waveforms to analyze these phase distortions. Note that throughout this section we have been assuming that both eigenstates have the same initial condition and thus fiducial waveform in momentum space. However, realistic GW initial conditions are typically given in frequency space, $h_\text{fid}(\omega)$, as they describe a temporally varying signal detected at a fixed location. 
As discussed in Appendix \ref{App:mono_k_omega}, if the difference between the modified dispersion relations of each eigenstate is small compared to $\sigma_k$, 
then initial conditions given in $\omega$ space can be straightforwardly translated into initial conditions in $k$ space, where both eigenstates have approximately the same momentum $k$, validating thus our assumptions.

It is important to note that this modification of the phase is only due to the propagation, and hence it is independent of the emission mechanism of the wave. In other words, there are no constraints on the type of waveform to be considered in $h_0(k)$.

\section{Examples of GW mixing}
\label{sec:examples_mixing}

Now that we have seen how to solve analytically the mixed propagation of $h$ and $s$, we proceed to numerically corroborate our results with a toy Gaussian wavepacket and consider particular examples of each type of mixing. In addition, this section will help gain physical intuition on the relevant time scales and parameters that lead to observational effects. We will discuss the prospects of detecting such observational signatures for realistic waveforms from compact binary coalescences in Section \ref{sec:implications}.

\subsection{Initial Conditions}

Since for the moment we are more interested in understanding the physical implications of the various GW propagation effects than quantifying the effects on a specific GW merger signal, we will work with an initial wavepacket described at emission, which we take to be $\eta_0=0$ in this section, by a carrier frequency $k_0$ modulated by a Gaussian wave packet of width $\sigma_x$, i.e.
\begin{equation}
\tilde{h}_0(x) =  2\cos(k_0 x) e^{-x^2/2\sigma_x^2}
=( e^{i k_0 x} + e^{-i k_0 x} )  e^{-x^2/2\sigma_x^2},
\end{equation}
propagating in the $+x$ direction. At the time $\eta_0$ we also choose $s(\eta_0)=0$, in order to isolate the effects due to mixing of degrees of freedom only.

Conveniently, the Fourier transform is the sum of Gaussians centered at $k_0$ and $-k_0$:
\begin{equation}
h_0(k) = \frac{1}{\sigma_k} \left[ e^{-(k-k_0)^2/2\sigma_k^2} +
 e^{-(k+k_0)^2/2\sigma_k^2} \right]
\,,
\end{equation}
where $\sigma_k=1/\sigma_x$.   
Since the initial wavepacket is real, note that 
$h_0(-k)=h_0^*(k)$ and because the propagation maintains its reality, we need to calculate only $k>0$ components.   
We will also work in
the limit that $k_0\gg \sigma_k$ so that to good approximation the first term carries 
the $k>0$ components.
For convenience, we will display the real space
wavepacket constructed from the $k>0$ components in this way, e.g
for the initial wave:
\begin{equation}
\tilde h_0(
    x) \rightarrow  \frac{1}{\sqrt{2\pi}}
    \int_0^\infty dk e^{ikx} \frac{1}{\sigma_k} e^{-(k-k_0)^2/2\sigma_k^2} \approx
    e^{ik_0 x} e^{-x^2/2\sigma_x^2},
\end{equation}
which has the benefit that the modulus 
extracts the Gaussian envelope without having to average over the carrier oscillations.   Despite the simplicity of this toy model, it captures all the relevant effects of mixing, dispersion and wavepacket decoherence.

In the rest of this section, we consider separately different types of interactions between $h$ and $s$, and analyze their effects on the modulated Gaussian wave packets. 
Even though we will consider separately the effects of velocity, mass, friction, and chiral mixing, a general cosmological model may have a combination of these types of mixing. In that case, at least at linear order in the mass, friction and chiral matrix elements compared to $k^2$, the expressions for the eigenfrequencies of the propagating eigenstates will be the superposition of the linear corrections beyond GR that we find for each separate example. 

The numerical examples in this section will have an arbitrary normalized unit for $1/k$,
$\sigma_x$, the typical timescales $T_{\rm mix}$, $T_{\rm coh}$, $T_{\rm broad}$, and the dimensional model parameters. In all the examples that will follow, we use units of wavepacket width $\sigma_x$.

\subsection{Velocity Mixing}\label{sec:velmixing}

Velocity mixing occurs when there are interactions between the tensor modes at quadratic order in $k$. The EoM that includes this type of mixing are given by
\begin{equation}\label{velocity_eom}
\left[\hat{I} \frac{d^2}{d\eta^2} +\bpm c_h^2 & c_{hs}^2 \\ c_{hs}^2 &  c_s^2\epm k^2  \right] \bpm h \\ s \epm =0,
\end{equation}
where in general each tensor mode has a different speed $c_A^2$ and the mixing is governed by $c_{hs}^2$.  The velocity matrix is assumed to be real and positive definite, in order to ensure stability of the solutions. In that case, we find  the eigenfrequencies to be real (i.e.\ $\Gamma_A=0$) and linear in $k$, with dispersion relations given by
\begin{align}
&\omega_{1}^2 = k^2\left( c_h^2+\frac{1}{2}\dc^2 +\frac{1}{2}\sqrt{4c_{hs}^4+\dc^4}\right) \,,\label{vel_mixing_vg1}\\
&\omega_{2}^2 = k^2 \left(c_h^2+\frac{1}{2}\dc^2 -\frac{1}{2}\sqrt{4c_{hs}^4+\dc^4} \right)\,,\label{vel_mixing_vg2}
\end{align}
if $\dc^2=c_s^2-c_h^2\leq 0$, so that in the no-mixing limit, that is $c_{hs}=0$, we get that $\omega_{1}$ describes the frequency of $h$ and $\omega_{2}$ that of $s$. Whereas, if $\dc^2>0$, then we will define $\omega_1^2$ with a minus sign in front of the square root, and $\omega_2^2$ with a plus sign.

The mixing matrix $\hat{E}$ is real and satisfies $\eM_{12}=-\eM_{21}$. This matrix is completely determined by the mixing angle $\Theta_g$ (fixing the mixing phase to $\phi_g=\pi$), which follows:
\be \label{eq:tg_vel_mixing}
\tan^2\Theta_g=\frac{4c_{hs}^4}{\left(|\dc^2 |+ \sqrt{\dc^4+4c_{hs}^4}\right)^2}\,.
\ee
Note that the mixing angle is frequency independent and thus it only rescales the amplitude of each eigenstate. In the no mixing limit, $c_{hs}\rightarrow 0$, we have that $\Theta_g\rightarrow 0$, regardless of the sign of $\dc^2$. For tensor modes with the same speed, $\dc^2=0$, the mixing is maximal, that is $\Theta_g=\pi/4$. 

The mixing time is simply given by $T_{\rm mix}= 2\pi/\Delta \omega$. Since the dispersion relations are linear in $k$, there will be no distortions of the signals associated to a MDR. In other words, the broadening time $T_{\rm broad} \rightarrow \infty$.   The group velocities in this case can also be directly computed as $v_{g,A}=\omega_A/k$, which is again frequency independent and equal to the phase velocity.   
Therefore, the coherence time $T_{\rm coh} = \sigma_x k/\Delta \omega$ is independent of $k$ and leading to a splitting of the original wavepacket into two signals or echoes,
with an amplitude rescaling of $|f_1|,|f_2|$ for $T>T_{\rm coh}$ but no waveform distortion. 
 The overall signal can be however distorted due to the interference of the eigenstates transitioning to the decoherence regime. The group velocity of each eigenstate could in general be different from the speed of light either because $c_{h,s}\ne 1$ or due to the mixing with the other mode.  As we will discuss in Sec.\ \ref{sec4:vel}, this makes velocity mixing testable with multi-messenger GW events.

\begin{figure}[!htb]
\centering
\includegraphics[width=\columnwidth]{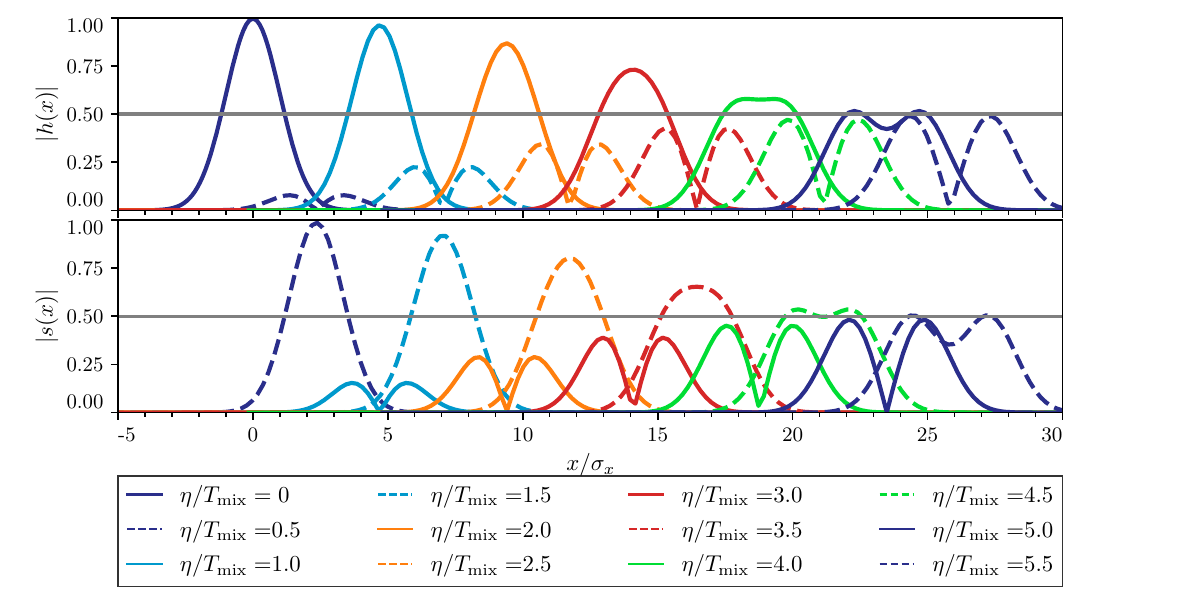}
\caption{\label{fig:velocity}
Propagation of a narrow Gaussian wavepacket for the case of velocity-mixing from the coherence to the decoherence regime. 
We have chosen $c_{hs}^2=0.1$, $\dc^2=0$, and tune $c_h^2=0.9$ to set $v_{g,1}= 1$. We also choose $k_0=12$ and $\sigma_{x}=1$.
The grey horizontal lines indicate the maximum amplitude of the two Gaussian propagating eigenmodes, which are both equal to half of the initial amplitude in this example. 
}
\end{figure}

Fig.~\ref{fig:velocity} shows a numerical example of the propagation of a toy GW signal described by a narrow Gaussian wavepacket, with each curve showing a temporal snapshot of the signal, from the coherence regime to the transition to decoherence. The top panel shows the amplitude $|h|$ of the GW signal, and the bottom shows the second tensor field $|s|$.
Here we have chosen  $k_0=12$, $\sigma_x=1$, $c_{hs}^2=0.1$ , $\dc^2=0$, and tune $c_h^2=0.9$ to set $v_{g,1}=1$. For this choice, the mixing and coherence timescales are $T_{\rm mix}=5.0$ and $T_{\rm coh}=9.5$. 
In addition, the mixing angle is given by $\tan^2\Theta_g=1$, which 
rescales the amplitude of the two propagating eigenstates to half of the initial amplitude as indicated by the horizontal grey lines.
During the coherence regime, the two eigenstates interfere and, as a result, the overall amplitude of the signal oscillates with a period of $T_{\rm mix}$. 
We show the snapshots at $n T_{\rm mix}/2$ where $n$ is an integer. We see that $h$ suffers destructive interference at odd $n$, while $s$ at even $n$. As the two eigenmodes separate, the $h$ and $s$ wavepackets become double peaked and both even and odd $n$ saturate to $1/2$ amplitude. This separation is apparent earlier in the destructive interference cases due to the sensitive cancellation required. Since this mixing behavior and approach to decoherence is similar in the cases that follow, we hereafter omit showing $s$ and also align the $h$ wavepacket to the arrival of the first peak in order to display a larger range of propagation times.

\subsection{Mass Mixing}\label{sec:massmixing}
Mass mixing commonly appears in models of modified gravity, however it may appear in combination with other types of mixings. Mass mixing only is a feature of massive bigravity \cite{deRham:2010kj,Hassan:2011zd}, which effectively propagates one massless graviton interacting with one massive graviton. In this section, we consider general scenarios with only mass mixing interactions.
The general EoM are given by:
\begin{equation}
\left[\hat{I} \frac{\d^2}{\d\eta^2} +\bpm c_h^2 & 0 \\ 0 &  c_s^2\epm k^2 +\bpm m_h^2 & m_{hs}^2 \\ m_{hs}^2 & m_s^2 \epm \right] \bpm h \\ s \epm =0,
\end{equation}
where we have assumed that the fields $h$ and $s$ are canonically normalized so that the mass matrix is symmetric, and we assume $c_{h,s}^2>0$ and the mass matrix to be real and positive definite to ensure stability of the solutions in the no-mixing limit.
The associated eigenfrequencies are therefore always real (i.e.~$\Gamma_A=0$) and given by:
\begin{align}
&\omega_{1}^2 =  \lp c_h^2+\frac{1}{2}\dc^2 \rp k^2 +\frac{1}{2}M^2 -\frac{1}{2}\sqrt{(\dc^2k^2+\Delta m^2)^2+4m_{hs}^4} \label{Mass_Mixing_w1}\,,\\
&\omega_{2}^2 = \lp c_h^2+\frac{1}{2}\dc^2 \rp k^2 +\frac{1}{2}M^2 +\frac{1}{2}\sqrt{(\dc^2k^2+\Delta m^2)^2+4m_{hs}^4} \label{Mass_Mixing_w2}\,,
\end{align}
if $\dc^2k^2+\Delta m^2 \geq 0$, so that in this case $\omega_1$ describes the propagation of $h$ in the no mixing limit of $m_{hs}=0$. In the  case of  $\dc^2k^2+\Delta m^2 \leq 0$, we define $\omega_{1}^2$ and $\omega_{2}^2$ with a plus and minus sign in front of the square root term, respectively.
Here, we have introduced the sum of the squared masses $M^2\equiv m_h^2+m_s^2$, their difference $\dm^2=m_s^2-m_h^2$, as well as the difference in the speeds $\dc^2=c_s^2-c_h^2$. Notice that $\Delta m^2$ is the mass difference between $h$ and $s$, and not of the propagating eigenstates.

In this case, the mixing matrix $\hat{E}$ is real and  $\eM_{12}=-\eM_{21}$. This matrix is fully determined by the mixing angle (the mixing phase is fixed to $\phi_g=\pi$), which is explicitly given by:
\be \label{eq:tg_mass_mixing}
\tan^2\Theta_g=\frac{4m_{hs}^4}{\left(|\dc^2k^2 +\dm^2| + \sqrt{(\dc^2k^2+\Delta m^2)^2+4m_{hs}^4}\right)^2}\,.
\ee
Note that whenever $\dc^2\neq0$, the mixing angle will be frequency dependent, possibly introducing distortions in the wavepacket even in the decoherence regime. In addition, in the no-mixing limit, we have that $\Theta_g=0$, regardless of the sign of $\dc^2k^2 +\dm^2$.

In the following, we consider particular sub-cases of mass mixing and analyze their effect on the GW propagation. Note that, in general, both eigenmodes will have different group velocities, according to Eq.\ (\ref{propspeed}). As a consequence, these modes will decohere after sufficiently long times. From Eq.\ (\ref{eq:amplitude_const}), since $\Gamma_A=0$ and $\phi_g=\pi$, we expect the amplitude of each detected wavepacket, $|f_1h_0|$ and $|f_2h_0|$, to be given by:
\begin{equation}\label{ModeAmpl_MassMixing}
    f_1=  \frac{1}{1+\tan^2\Theta_g}, \quad   f_2 = f_1\tan^2\Theta_g\,,
\end{equation}
thus being fully controlled by the mixing angle $\Theta_g$.

\subsubsection{Small speed difference}\label{massmixingequalspeed}
We consider the case in which the difference in velocities is smaller than the mixing term: $|\Delta c^2k^2|\ll |m_{hs}^2|$. A specific case is when $\Delta c^2=0$. 
Because of the similarities with massive bigravity theory, we will focus on the case in which $m_{hs}^4=m_h^2m_s^2$ so that there is one massless and one massive eigenmode, as is the case in the ghost-free theory of massive bigravity \cite{deRham:2010kj,Hassan:2011zd}. In particular, from (\ref{Mass_Mixing_w1})-(\ref{Mass_Mixing_w2}) we find in this regime that 
\begin{align}
    &\omega_{1}^2\approx c_h^2k^2 + \frac{m_h^2}{M^2}\Delta c^2k^2\,, \label{omega1_bigravity}\\
    &\omega_{2}^2\approx c_h^2k^2+M^2 + \frac{m_s^2}{M^2}\Delta c^2k^2\,\label{omega2_bigravity},
\end{align}
where we see that in the limit of $m_h\rightarrow 0$ (so that $\dc^2k^2+\Delta m^2 \geq 0$), we obtain that $\omega_1$ and $\omega_2$ describe the propagation of $h$ and $s$, respectively.
The mixing angle also simplifies to:
\begin{equation}
    \tan^2\tg \approx \left(\frac{m_{hs}}{m_s}\right)^4 \left( 1 - 2\frac{\Delta c^2k^2}{M^2}\right) \,, 
\end{equation}
which vanishes in the no-mixing limit of $m_{hs}\rightarrow 0$.
Due to the non-zero value of $M$, the group velocities of the two eigenmodes will differ and the initial wavepacket will eventually decohere. Explicitly, their group velocities are given by: 
\begin{equation}\label{v12massmixing1}
    v_{g,1}\approx  c_h \left( 1+ \frac{1}{2}\frac{\Delta c^2}{c_h^2}\frac{m_h^2}{M^2} \right) , \quad v_{g,2}\approx   \frac{c_h^2 k}{\sqrt{c_h^2k^2+M^2}}\left(1+
    \frac{1}{2}\frac{\Delta c^2}{c_h^2}\frac{m_s^2}{M^2}\frac{(c_h^2k^2+2M^2)}{(c_h^2k^2+M^2)}\right).
\end{equation}
The typical time scales that describe the mixing, coherence and broadening of the signal are:
\begin{align}
    T_{\rm mix} & \approx \frac{2\pi}{\sqrt{c_h^2k^2+M^2}-c_hk} \\
     T_{\rm coh} & \approx  T_{\rm mix} \frac{\sqrt{c_h^2k^2+M^2}}{2\pi c_h} \sigma_A\\
     T_{\text{broad}, 1} & \approx \sigma_x^2\frac{c_h M^6}{3 m_{hs}^4\dc^4k}; \quad T_{\text{broad}, 2}\approx 
     \sigma_x^2\frac{\lp c_h^2k^2+M^2 \rp^{3/2}}{c_h^2M^2}\,,
\end{align}
where we distinguish the broadening time of each of the eigenstates.

\begin{figure}[!htb]
\centering
\includegraphics[width=0.55\columnwidth]{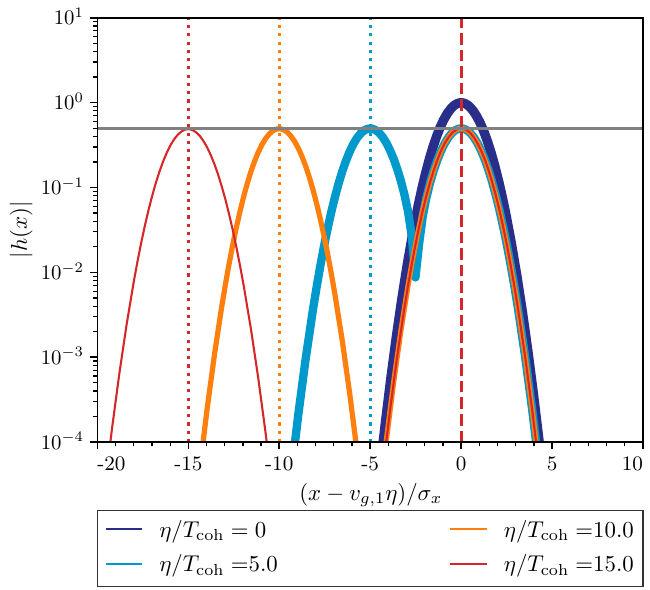}
\caption{\label{fig:mass-ch=cs}
Propagation of a narrow Gaussian wavepacket, in the case of mass-mixing and same speeds $c_h=c_s$. The grey horizontal line indicates the amplitude of the envelopes for the two eigenmodes, which coincide in this example. 
The vertical lines show the positions of the wavepackets predicted from the group velocities, with dashed and dotted lines for the first and second eigenmodes, respectively.
Parameters: $m_s^2=m_h^2=m_{hs}^2=1$, $c_h^2=c_s^2=1$, $k_0=100$, $\sigma_{x}=1$. 
}
\end{figure}

A toy model of the propagation of $h$
is shown in Fig.~\ref{fig:mass-ch=cs}, where $\Delta c^2=0$. 
Here we are plotting the evolution with respect to the group velocity of the first eigenmode $v_{g,1}$, which coincides in this case with the decoupled velocity of $h$. For the choice of $m_s^2=m_h^2=m_{hs}^2=1$, $c_h=c_s=1$, $k_0=100$ and $\sigma_{x}=1$, the mixing angle is such that $\tan^2\Theta_g=1$, so the amplitudes of the two eigenmodes contributing to $h$ are equal $f_1=f_2=1/2$  
(described by the horizontal grey line). 
The vertical dashed line indicates the position of the first eigenmode, who is at the origin, whereas the position of the second wavepacket is indicated with dotted vertical lines, and it is calculated using the group velocity difference $\Delta v_g=v_{g,2}-v_{g,1}$ given in (\ref{v12massmixing1}).
The propagation time is scaled by the coherence time $T_{\rm coh}$. 
As discussed in Sec.~\ref{sec:GW-propag}, when the propagation time is much larger than the coherence time $\eta\gg T_{\rm coh}$, the propagation enters the decoherent regime so that one eigenmode lags behind another and the wave packet splits into two separate parts. 
We confirm that the two eigenmodes decohere when $\eta\gg T_{\rm coh}$ and the wavepacket splits into two Gaussians. Note that the Gaussians associated to both eigenmodes do not exhibit any visible distortions on the timescales shown. In particular, in this example the first wavepacket is not distorted at all since $\partial^2\omega_1/\partial k^2=0$ (i.e.~$T_{\text{broad},1}\rightarrow \infty$), whereas for the second wavepacket we find $\partial^2\omega_2/\partial  k^2\approx 2m_{hs}^2/c_h k^3$. Therefore, in this example the change in width $\sigma_A^2-\sigma_x^2$ (see Eq.\ (\ref{SigmaBroad})) is highly suppressed or exactly vanishing. 

\subsubsection{Small mass mixing}

In the case in which there is a different sound speed between $h$ and $s$ the mixing is modified. Here we consider the limit when the mixing term in the mass matrix is small compared to the difference in frequencies associated with the difference in sound speeds. 
In this limit we have that  $|m_{hs}^2/(\dc^2 k^2)|\ll 1$ (for $\Delta c^2\not=0$), 
and from (\ref{Mass_Mixing_w1})-(\ref{Mass_Mixing_w2}) we find the eigenfrequencies to be approximately given by:
\begin{align}
&\omega_{1}^2 \approx  c_h^2k^2 +m_h^2 -\frac{m_{hs}^4}{\dc^2 k^2+\dm^2} +\mathcal{O}\lp\frac{m_{hs}^8}{\dc^8 k^8}\rp \,,\\
&\omega_{2}^2 \approx  c_s^2k^2 +m_s^2 +\frac{m_{hs}^4}{\dc^2 k^2+\dm^2} +\mathcal{O}\lp\frac{m_{hs}^8}{\dc^8 k^8}\rp \,,
\end{align}
which makes $\omega_1$ to be associate to $h$ in the no-mixing limit, regardless of the value of $\dc^2 k^2+\dm^2$. Here we see that the correction to the dispersion relations of $h$ and $s$ due to their couplings will scale as $k^{-2}$ at high wavenumber. The mixing angle is
\begin{equation}\label{MixAngle_MassMix2}
\tan^2\Theta_g \approx \frac{m_{hs}^4}{\lp \dc^2k^2+\dm^2\rp^2} +\mathcal{O}\lp\frac{m_{hs}^8}{\dc^8 k^8}\rp\,.
\end{equation}
In this case, the group velocities of the two modes are explicitly given by:
\begin{align}
&v_{g,1} \approx 
\frac{c_h^2k}{\sqrt{c_h^2k^2+m_h^2}} +\frac{m_{hs}^4}{\lp\dc^2k^2+\dm^2\rp^2} \frac{k\lp 2\lp c_h^2k^2+m_h^2\rp \dc^2 +c_h^2\lp \dc^2k^2+\dm^2\rp \rp}{2\lp c_h^2k^2+m_h^2\rp^{3/2}} 
+ \ldots \,,\label{eq:mass-v1}\\
&v_{g,2} \approx
\frac{c_s^2k}{\sqrt{c_s^2k^2+m_s^2}} -\frac{m_{hs}^4}{\lp\dc^2k^2+\dm^2\rp^2} \frac{k\lp 2\lp c_s^2k^2+m_s^2\rp \dc^2 +c_s^2\lp \dc^2k^2+\dm^2\rp \rp}{2\lp c_s^2k^2+m_s^2\rp^{3/2}} +  \ldots
\,,
\end{align}
with corrections of order $\mathcal{O}\lp\frac{m_{hs}^8}{\dc^8 k^8}\rp$ and higher.
Also, the typical time scales describing mixing, coherence, and broadening are:
\begin{eqnarray}
    T_{\rm mix} &&\approx \frac{2\pi}{\sqrt{\vphantom{c_h^2} c_s^2k^2+m_s^2}-\sqrt{c_h^2k^2+m_h^2}}, 
\end{eqnarray}
\begin{equation}
    T_{\rm coh} \approx \frac{\sigma_{A}}{k\lp \frac{c_s^2}{\sqrt{c_s^2k^2+m_s^2}}-\frac{c_h^2}{\sqrt{c_h^2k^2+m_h^2}} \rp},
\end{equation}
\begin{equation}
    T_{\text{broad},1} \approx \sigma_x^2\frac{\lp c_h^2k^2+m_h^2\rp^{3/2}}{c_h^2m_h^2} ;\quad  T_{\text{broad},2}\approx \sigma_x^2\frac{\lp c_s^2k^2+m_s^2\rp^{3/2}}{c_s^2m_s^2}.
\end{equation}

\begin{figure}[h!]
\centering
\includegraphics[width=0.55\columnwidth]{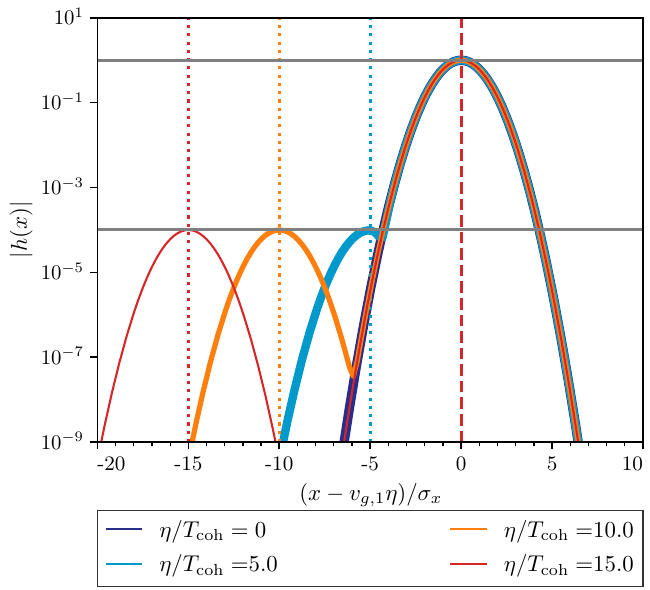}
\caption{\label{fig:mass-mixing}
Wave packet propagation of the $h$ field for small mass-mixing case, at different times. 
The two horizontal lines indicate the amplitudes of the envelopes for the two eigenmodes. 
The vertical lines show the positions of the wavepacket, with dashed and dotted lines for the first and second eigenmodes, respectively.
Parameters: $m_s^2=m_h^2=1$, $m_{hs}^2=10$, $c_h^2=1$, $c_s^2=0.9$, $k_0=100$, $\sigma_{x}=1$. 
}
\end{figure}

A toy model calculation of this example is presented in Fig.~\ref{fig:mass-mixing}, where we show $h$ at different times and follow the same plotting conventions of Fig.~\ref{fig:mass-ch=cs}. Here we have chosen $c_h^2=1$, $c_s^2=0.9$, $m_s^2=m_h^2=1$ and $m_{hs}^2=10$, $\sigma_x=1$. Contrary to the previous case, now the mixing (and thus the amplitude of the second mode) is suppressed as $\tan^2\Theta_g\sim \mathcal{O}(10^{-4})$ due to the difference in speeds of the eigenmodes. 
Since the group velocities  of the two eigenmodes are different, after $T_\text{coh}$ they decohere and propagate as two separate wavepackets. The horizontal lines show the theoretical expectation for the amplitude of each wavepacket, according to Eq.~(\ref{ModeAmpl_MassMixing}). 
Note that in this case, none of the velocities of the two wavepackets coincide with the naive velocity of $h$, that is $v_h=c_h^2k/\sqrt{c_h^2k^2+m_h^2}$, due to the presence of the mass mixing. 
In the small mixing limit, as shown in Eq.~(\ref{eq:mass-v1}) the group velocity of the first eigenstate $v_{g,1}$ deviates from $v_h$ by $\mathcal{O}\lp \frac{m_{hs}^4}{\lp\dc^2k^2+\dm^2\rp^2} \rp$. 
In this example, there is no visible broadening of the Gaussian, as the broadening timescale (timescale determining when the variance changes by an order 1 factor) is estimated to be  $T_\text{broad}\sim 10^4 T_\text{coh}$. 

\subsection{Friction Mixing}\label{sec:friction_mixing}
Friction mixing appears when there are derivative interactions between the metric and the additional modified gravity fields. This happens for instance in vector-tensor models such as multi-Proca theories with internal global symmetries \cite{Allys:2016kbq}, which may also include the presence of additional mass mixing interactions. Nevertheless, in this section we consider general models with only friction mixing for pedagogical reasons.
The EoM are given by:
\be
\lb \hat{I}\frac{\d^2}{\d\eta^2} + \bpm 0 & -2\alpha \\ 2\alpha & 4\dnu  \epm\frac{\d}{\d\eta} +\bpm c_h^2 & 0 \\ 0 &  c_s^2\epm k^2\rb \bpm h \\ s \epm =0\,, 
\ee
where we have not included a friction term for $h$ because it can always be reabsorbed through a field redefinition of both fields $\vPhi\rightarrow e^{-\int \nu_h d\eta/2}\vPhi$ without affecting the rest of the terms in the EoM. 
In general, explicit expressions for the 
eigenfrequencies are complicated and not particularly illuminating.
 In order to get some intuition let us consider different sub-classes of limiting situations, from the simplest to the more involved.

\subsubsection{$\Delta\nu=0$ and $\Delta c^2=0$}
\label{Friction_example1}
Whenever we only have the non-diagonal entries $\alpha$ and $c_s=c_h$, the eigenfrequencies are real and simplify to
\begin{align} 
\omega_{1}&=\sqrt{c_h^2k^2+\alpha^2}-\alpha\,,  \label{Fricion_Example1_w1}\\ 
\omega_{2}&=\sqrt{c_h^2k^2+\alpha^2}+\alpha\,, 
\label{Fricion_Example1_w2}
\end{align}
with $\Gamma_1=\Gamma_2=0$. 
In fact, they only differ by a constant factor, i.e.\ $\Delta\omega=2\alpha$, and therefore the group velocities  of both eigenmodes are the same
\begin{equation}
v_{g,1}=v_{g,2} = \frac{c_h^2 k}{\sqrt{c_h^2k^2+\alpha^2}}.
\end{equation}
Contrary to the mass mixing case, the two eigenmodes propagate coherently all the time, and the oscillations of the GW signal are described by the mixing timescale $T_\text{mix}=2\pi/\alpha$.

For the eigenfrequencies in Eq.~(\ref{Fricion_Example1_w1})-(\ref{Fricion_Example1_w2}), the mixing matrix $\hat{E}$ will be imaginary, such that  $\hat{E}_{12}^-=\hat{E}_{21}^-=i$. Therefore, the mixing angle fully determines the mixing matrix, and it is given by $\Theta_g=\pi/4$ (and the mixing phase is fixed to $\phi_g=\pi$). Therefore, the solution of $h$ reads
\be\label{eq:h-friction-dc0}
\vert h(\eta)\vert^2=\frac{1}{2}|h_0|^2\lp1+\cos\lb2\alpha \eta\rb\rp\,.
\ee
From here we observe that the phase and frequency evolution of the total wave will be preserved, and we will only see a time modulation of the amplitude.

\begin{figure}[!htb]
\centering
\includegraphics[width=0.55\columnwidth]{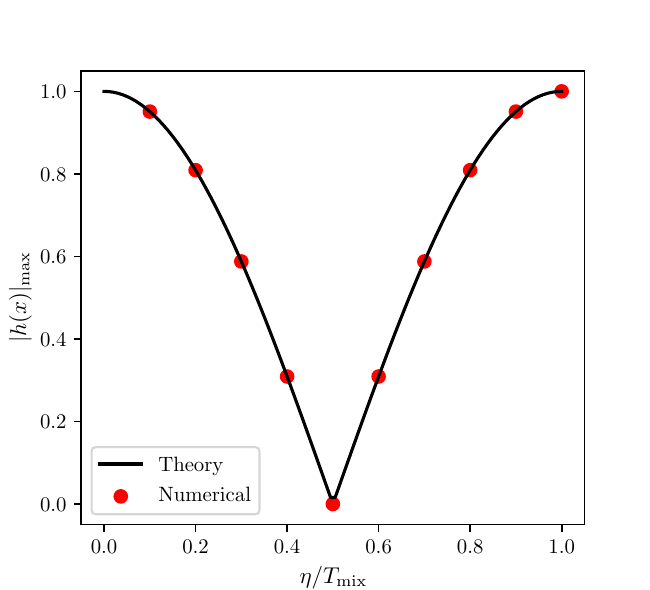}
\caption{\label{fig:friction-ch=cs}
Maximum amplitude of coherent propagation of $h$ at different times, in the case of friction-mixing. $T_\text{min}= 2\pi/\alpha$ is the phase evolution time for one period. 
The black line is the theoretical prediction of the amplitude Eq.~(\ref{eq:h-friction-dc0}) while the red dots are the numerical calculations.
Parameters: $\alpha=1$, $c_h^2=c_s^2=1$, $k_0=100$, $\sigma_{x}=1$. }
\end{figure}

A toy model calculation for a Gaussian wavepacket is shown in Fig.~\ref{fig:friction-ch=cs}. Since, the only property of $h$ that changes according to Eq.~(\ref{eq:h-friction-dc0}) is the amplitude, we only plot the maximum amplitude of the Gaussian wavepacket $h$ at different times. We fix the values of the parameters to $\alpha=0.1$, $c_h=c_s=1$, $\sigma_x=1$, and $k=100$. The black line shows the prediction from Eq.~(\ref{eq:h-friction-dc0}), and the red dots confirm the numerical results.

\subsubsection{$\Delta\nu$=0}\label{FrictionCase1}

If we now allow for a different speed between the tensor modes $c_s\neq c_h$, the eigenfrequencies read
\begin{align}
&\omega_{1}^2 = \frac{(c_h^2+c_s^2)k^2+4\alpha^2-\sqrt{16c_h^2k^2\alpha^2+(4\alpha^2+\dc^2k^2)^2}}{2}\,,\\
&\omega_{2}^2 = \frac{(c_h^2+c_s^2)k^2+4\alpha^2+\sqrt{16c_h^2k^2\alpha^2+(4\alpha^2+\dc^2k^2)^2}}{2}\,,
\end{align}
if $\dc^2 \geq 0$, so that $\omega_1$ describes the propagation of $h$ in the no-mixing limit of $\alpha\rightarrow 0$.

The mixing matrix $\eM$ is found to be always imaginary, but $|\hat{E}_{12}|\not=|\hat{E}_{21}|$ in general situations, and hence the mixing angle does not always determine the mixing matrix. Nevertheless, since $\hat{E}_{12}\hat{E}_{21}$ is real, the mixing angle does always determine the solution for $h$ when the initial conditions for $s$ vanish. Even though there is friction in this example, the eigenfrequencies are real and thus the amplitude of the eigenmodes do not exhibit an exponential decay due to the friction mixing. Explicitly, since $\Gamma_A=0$ and $\phi_g=\pi$, the amplitude of the two eigenmodes contributing to $h$ are given by Eq.~(\ref{ModeAmpl_MassMixing}).

Next, we analyze various limiting cases:
 
\paragraph{A) Small mixing.} In the limit of $|\alpha^2/(\Delta c^2\, k^2)|\ll 1$ the eigenfrequencies simplify to 
\begin{align}
\omega_{1}&= c_hk\lp1-2\frac{\alpha^2}{\Delta c^2k^2}+\mathcal{O}\lp\frac{\alpha^4}{\Delta c^4k^4}\rp\rp\,, \label{eq:t1_friction_small_alpha}\\
\omega_{2}&= c_sk\lp1+2\frac{\alpha^2}{\Delta c^2k^2}+\mathcal{O}\lp\frac{\alpha^4}{\Delta c^4k^4}\rp\rp\,, \label{eq:t2_friction_small_alpha}
\end{align}
which are defined such that $\omega_1$ describes the propagation of $h$ in the no-mixing limit, for any value of $\dc^2$.
The group velocities of each eigenmode are different and explicitly given by:
\begin{align}
&v_{g,1}=  c_h\lp1+2\frac{\alpha^2}{\Delta c^2k^2}+\mathcal{O}\lp\frac{\alpha^4}{\Delta c^4k^4}\rp\rp\,, \label{eq:vh_small_alpha}\\
&v_{g,2}=  c_s\lp1-2\frac{\alpha^2}{\Delta c^2k^2}+\mathcal{O}\lp\frac{\alpha^4}{\Delta c^4k^4}\rp\rp\,,
\end{align}
In this limit, the mixing matrix is described by:
\begin{align}
&\hat{E}_{12}\approx 2i\frac{c_s\alpha}{\Delta c^2 k}\,,
\qquad
\hat{E}_{21} \approx 2i\frac{c_h\alpha}{\Delta c^2 k}\,,
\end{align}
and hence the mixing angle approximates to:
\be \label{MixAngleFriction1}
\tan^2\Theta_g\approx 4\frac{c_hc_s\alpha^2}{\Delta c^4 k^2}\,.
\ee
Note that this case is very similar to the mass mixing case in the regime of small mixing (see Eq.~(\ref{MixAngle_MassMix2})), in the sense that the mixing angle is suppressed by both the small friction/mass mixing compared to the speed difference $\Delta c^2k^2$. However, here $\tan^2\Theta_g$ scales as $k^{-2}$ whereas in the mass mixing case it scales as $k^{-4}$, for large $k$. Therefore, we have a parametrically smaller suppression with friction mixing. 
The typical time scales of mixing, coherence, and broadening are given by:
\begin{equation}
    T_{\rm mix} \approx \frac{2\pi(c_h+c_s)}{\dc^2k}, 
\end{equation}
\begin{equation}
    T_{\rm coh} \approx \frac{(c_h+c_s)\sigma_A}{\dc^2}, 
\end{equation}
\begin{equation}
    T_{{\rm broad},1} \approx \frac{\dc^2k^3\sigma_x^2}{4c_h\alpha^2};\quad T_{{\rm broad},2} \approx \frac{\dc^2k^3\sigma_x^2}{4c_s\alpha^2}.
\end{equation}

\paragraph{B) Small mixing and smaller speed difference.} In the limit in which both $\alpha/(c_hk)$ and $(k\Delta c^2)/\alpha c_h$ are small, then we get
\begin{align}
\omega_{1}&= c_hk\lp1+\frac{\Delta c^2}{4c_h^2}-\frac{\alpha}{c_hk}+\frac{1}{2}\frac{\alpha^2}{c_h^2k^2}+\mathcal{O}\lp\frac{\alpha^{4}}{c_h^4k^4}\rp\rp\,, \label{eq:t3_friction_small_alpha}\\
\omega_{2}&= c_hk\lp1+\frac{\Delta c^2}{4c_h^2}+\frac{\alpha}{c_hk}+\frac{1}{2}\frac{\alpha^2}{c_h^2k^2}+\mathcal{O}\lp\frac{\alpha^{4}}{c_h^4k^4}\rp\rp\,, \label{eq:t4_friction_small_alpha}
\end{align}
for $\dc^2\geq 0$. In the case of $\dc^2<0$, the expressions of $\omega_1$ and $\omega_2$ are swapped.
The group velocity of the first eigenmode is:
\be \label{eq:vg_dispersion_Dc}
v_{g,1}= c_h\lp1+\frac{\Delta c^2}{4c_h^2}-\frac{1}{2}\frac{\alpha^2}{c_h^2k^2}+\cdots\rp\,,
\ee
and the second eigenmode has a speed that is suppressed by $\Delta v_g \approx -(k\Delta c^4)/(8c_h^2\alpha)$. Due to this non-vanishing difference, the two eigenmodes will decohere after sufficiently long times, in particular after $T_\text{coh}\sim \sigma_A/\Delta v_g$. 
Since $\Delta v_g$ can be very small, the decoherence time can be very long.  
Also note that there is a non-trivial group velocity  of the eigenmodes at leading order in $k$. In this case, the anomalous speed also affects the mixing matrix:
\be
\hat{E}_{12}=\hat{E}_{21} \approx i\lp1-\frac{k|\Delta c^2|}{4c_h\alpha}\rp
\ee
and thus the mixing angle is given by 
\be \label{eq:mixing_angle_dc}
\tan^2\Theta_g\approx1-\frac{k|\Delta c^2|}{2c_h\alpha}\,.
\ee
The typical time scales of mixing, coherence, and broadening are given by:
\begin{equation}
    T_{\rm mix} \approx \frac{\pi}{\alpha} 
    \lp 1 -\frac{\dc^4k^2}{32c_h^2\alpha^2}\rp,
\end{equation}
\begin{equation}
    T_{\rm coh} \approx \frac{8c_h^2\alpha\sigma_A}{\dc^4k} ,
\end{equation}
\begin{equation}
    T_{\text{broad},1} \approx T_{\text{broad},2}\approx  \frac{c_hk^3\sigma_x^2}{\alpha^2} \lp1+\frac{\dc^4k^3}{16c_h\alpha^3} \rp.
\end{equation}
Note that this case is similar to the case of mass mixing with small speed differences, where typically $\tan^2\Theta_g\sim 1$ and thus both eigenmodes contribute equally to $h$, according to Eq.\ (\ref{ModeAmpl_MassMixing}). 
However, here the distortions of the Gaussian wavepackets can be sizeable by the time the two eigenmodes decohere or even prevent them from reaching decoherence, as discussed in Eq.\ (\ref{DecohCrit}),
decoherence will be achieved if $\Delta v_g T_\text{broad,A}/\sigma_x\simeq k^4\dc^4\sigma_x/(8c_h^2\alpha^3)>1$.

\begin{figure}[!htb]
\centering
\includegraphics[width=0.46\columnwidth]{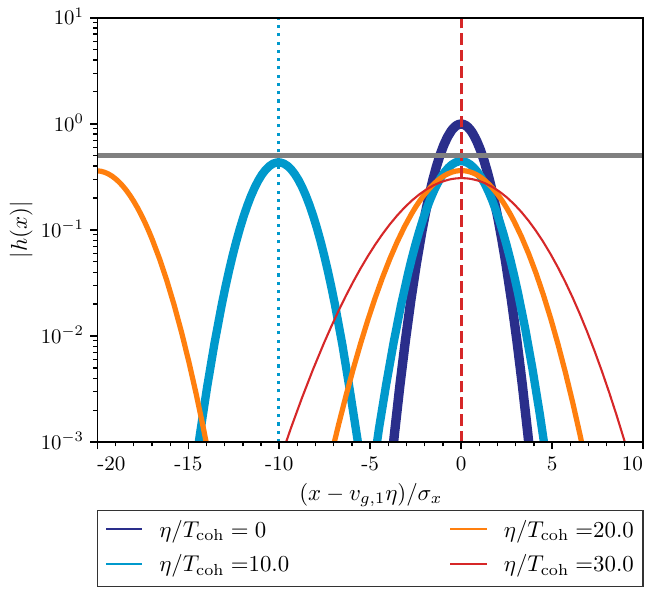}
\includegraphics[width=0.51\columnwidth]{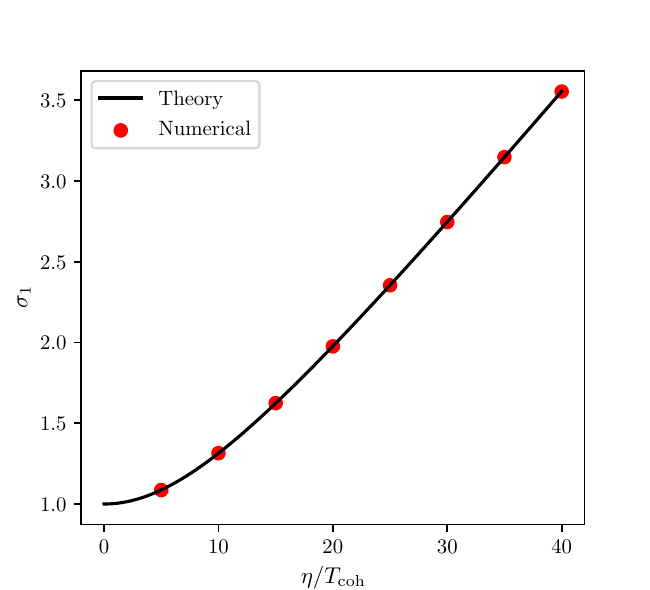}
\caption{\label{fig:friction-small-mixing-2}
On the left panel we show the wavepacket propagation for small friction-mixing case in the limit both $\alpha/(c_hk)$ and $(k\Delta c^2)/\alpha c_h$ are small. The two eigenmodes decohere after $T_\text{coh}$ and then exhibit considerable distortions that are reflected in the spread of the Gaussians since $T_\text{broad} \sim 9.6 T_\text{coh}$. On the right panel, we show the {first eigenmode} wavepacket width evolution as function of time. 
The black line is the theoretical prediction of the width Eq.~(\ref{SigmaBroad}) while the red dots are the numerical calculations {that correspond to the left panel curves centered at 0}.
The model parameters in this example are $\alpha=1$, $c_h^2=1$, $c_s^2=0.999$, $k_0=100$, $\sigma_{x}=1$.
}
\end{figure}

A toy model calculation is shown in Fig.~\ref{fig:friction-small-mixing-2}. 
On the left panel we show the evolution of the Gaussian wavepacket. The expected amplitudes of each eigenmodes is shown in the horizontal grey line. 
In this case we see that there are visible distortions to the wavepacket on the timescales observed. For the parameters considered in this example, we get that $\sigma_{1}/\sigma_x \approx 2.0$ and $\sigma_{2}/\sigma_x \approx 1.8$ at the time $20T_\text{coh}$. 
Since each eigenmode conserves energy independently, their amplitudes decay to compensate for the spread of the Gaussian, and therefore their amplitudes do not match exactly the expected grey lines. In this example, a slight skewness of the first wavepacket is also visible, which brings about $\sim 1\%$ distortions in the Gaussian. This happens because the cubic higher order term in the phase expansion (\ref{PhaseExpansion}) is such that $T_3/T_\text{broad}\sim k\sigma_x\sim 0.01$. In the right panel of Fig.~\ref{fig:friction-small-mixing-2} we explicitly show the evolution of the width of the wavepackets as a function of time. The black line shows the theoretical expectation of the width evolution, based on the quadratic expansion in Eq.~(\ref{SigmaBroad}). The red dots show the numerical results, which agree with the expected theoretical results in the black solid line.

\subsubsection{$\dc^2$=0}\label{Friction_with_Decay}
Assuming $c_h=c_s$, when $\Delta\nu\geq0$ (to ensure stability of the solution in the no-mixing limit), the set of four eigenfrequencies can be complex and given by:
\begin{align}
\theta_{1\pm}&=\pm\sqrt{c_h^2k^2- (|\onu| - \dnu)^2}+ i\dnu -i|\onu| \,,\label{eq:t3_friction} \\
\theta_{2\pm}&=\pm\sqrt{c_h^2k^2-(|\onu| + \dnu)^2}+ i\dnu+ i|\onu| \,, \label{eq:t4_friction}
\end{align}
if $\alpha\leq \Delta\nu$. Here, we have introduced the frequency parameter $\onu\equiv \sqrt{\alpha^2-\dnu^2}$ associated to the friction mixing.  In general, we use the following convention for taking the square root of a complex number: $\sqrt{A^2e^{i2\phi}}=|A|e^{i\phi}$, and therefore we have that $\omega_\nu=i|\omega_\nu|$ if $\alpha\leq \Delta\nu$. Here, we have defined the eigenfrequencies such that in the no-mixing limit of $\alpha\rightarrow 0$, $\theta_{1\pm}$ and $\theta_{2\pm}$ describe the propagation of $h$ and $s$, respectively. In addition, from Eqs.\ (\ref{eq:t3_friction})-(\ref{eq:t4_friction}) we see manifestly that both eigenfrequencies have the structure $\theta_{\pm}=\pm \omega + i\Gamma$, for some real $\omega$ and $\Gamma$ parameters determining the oscillation frequency and decay rate of the propagation modes, respectively.

For $\alpha>\Delta\nu$, we define the eigenfrequencies in a way that is disconnected from that found in (\ref{eq:t3_friction})-(\ref{eq:t4_friction}), since when $\alpha>\Delta\nu$ the set of four eigenfrequencies must be paired differently in order to have the structure $\theta_{\pm}=\pm \omega + i\Gamma$. In this case, we thus define $\theta_{1\pm}$ and $\theta_{2\pm}$ as
\begin{align}
\theta_{1\pm}&=\pm\sqrt{c_h^2k^2+(\pm\onu + i\dnu)^2} + i\dnu \pm \onu \,,\label{eq:t1_friction} \\
\theta_{2\pm}&=\pm\sqrt{c_h^2k^2+(\pm\onu - i\dnu)^2}+ i\dnu \mp \onu \,. \label{eq:t2_friction}
\end{align}
Notice that even though it is not manifest, Eq.\ (\ref{eq:t1_friction})-(\ref{eq:t2_friction}) do satisfy that only their real part change signs with the $\pm$ solutions, whereas their imaginary part does not change signs, when $\alpha>\Delta\nu$.
Since these definitions are not applicable in the no-mixing limit (as they require $\alpha>\Delta\nu$), there is no generic association between $\theta_1$ and the field $h$, thus the labels of 1 and 2 are chosen arbitrarily. In this case, we have chosen them in a convenient way so that for the $-$ propagating mode (the one propagating forward), the expressions in (\ref{eq:t3_friction})-(\ref{eq:t4_friction}) and (\ref{eq:t1_friction})-(\ref{eq:t2_friction}) are actually the same. 

Focusing in the $-$ propagating mode, we obtain the propagation speeds of the two eigenstates as $v_{gA}=\Re(\partial \theta_{A-}/\partial k)$. Regardless of the parameter values of the model, we find
\begin{equation}
    v_{g,1}=\Re\left(\frac{c_h^2k}{\sqrt{c_h^2k^2+(\omega_\nu - i\Delta\nu)^2}}\right), \quad v_{g,2}=\Re\left(\frac{c_h^2k}{\sqrt{c_h^2k^2+(\omega_\nu + i\Delta\nu)^2}}\right)\,.
\end{equation}
In the case of $\alpha\leq \Delta \nu$, $\omega_\nu=i|\omega_\nu|$ and both expressions for $(\partial \theta_{A-}/\partial k)$ are real, and generically different. Therefore, in this case, the two eigenstates will eventually reach decoherence.
On the other hand, in the case of $\alpha>\Delta \nu$, we have that $\omega_\nu$ is real, and both expressions for  $(\partial \theta_{A-}/\partial k)$ are complex. Nevertheless, both speeds $v_{g,A}$ are the equal  since the arguments in both equations are related by simple conjugation, and hence their real parts $\Re$ are the same. The two eigenmodes propagate always coherently in this case, although each eigenstate will have a different exponentially suppressed amplitude.

The corresponding mixing matrix $\eM$ will have 
off-diagonal components that may be different and complex. Therefore, the mixing of $h$ is described by a non-trivial mixing angle $\Theta_g$ and mixing phase $\phi_g$, both depending on the model parameters. We explicitly find that mixing to be given by:
\begin{equation}
    \eM_{12-}\eM_{21-}=\frac{\alpha^2}{(\Delta\nu - i\omega_\nu)^2},
\end{equation}
which vanishes in the limit of no mixing, when $\alpha\rightarrow 0$.
From here we obtain that the mixing angle and phase, for the $-$ propagating mode, are given by:
\begin{align}
    & \tan^2\Theta_g=1, \quad \tan \phi_g= +\frac{2\omega_\nu \Delta\nu}{\Delta\nu^2-\omega_\nu^2}, \quad  \text{if} \quad \alpha>\Delta\nu,\\
    &  \tan^2\Theta_{g} = \frac{\alpha^2}{(\Delta\nu + |\omega_\nu|)^2}, \quad \phi_g=0, \quad \text{if} \quad \alpha\leq \Delta\nu\,.
\end{align}
Notice that in the case of $\alpha>\Delta \nu$, one always has a fixed maximal mixing between both modes. However, in the limit of $\alpha\rightarrow 0$, the mixing angle vanishes. 

In the case of $\alpha\leq \Delta\nu$, the timescales of mixing, coherence, and broadening in the large-$k$ limit are given by:
\begin{align}
&  T_{\rm mix} \approx \frac{\Delta\nu \pi}{c_h k |\omega_\nu|} +\mathcal{O}(k^{-3}),\\
& T_{\text{coh}}\approx \sigma_A \left(\frac{c_hk^2}{2|\omega_\nu|\dnu} +\frac{3}{4}\frac{ (|\omega_\nu|^2 +\dnu^2) }{c_h \dnu |\omega_\nu|} +\mathcal{O}(k^{-2})\right),\\
& T_{\text{broad},1}\approx \sigma_x^2\left(\frac{c_hk^3}{(|\omega_\nu|-\dnu)^2} -\frac{3k}{2c_h}  +\mathcal{O}(k^{-1}) \right),\\
&T_{\text{broad},2}\approx  \sigma_x^2\left(\frac{c_hk^3}{(|\omega_\nu|+\dnu)^2} -\frac{3k}{2c_h}  +\mathcal{O}(k^{-1}) \right) . 
\end{align}
On the other hand, in the case of $\alpha>\Delta \nu$, the timescales of mixing, coherence, and broadening are given by:
\begin{align}
    T_{\rm mix} &= \frac{\pi}{\omega_\nu},\\
    T_{\text{broad},1} = T_{\text{broad},2} &=  \sigma_x^2\Re\lp\frac{\lp c_h^2k^2+(\omega_\nu-i\Delta\nu)^2\rp^{3/2}}{c_h^2(\omega_\nu-i\Delta\nu)^2} \rp,
\end{align}
and $T_{\text{coh}}$ is infinite since $v_{g,1}=v_{g,2}$.

\begin{figure}[t!]
\centering
\includegraphics[width=0.55\columnwidth]{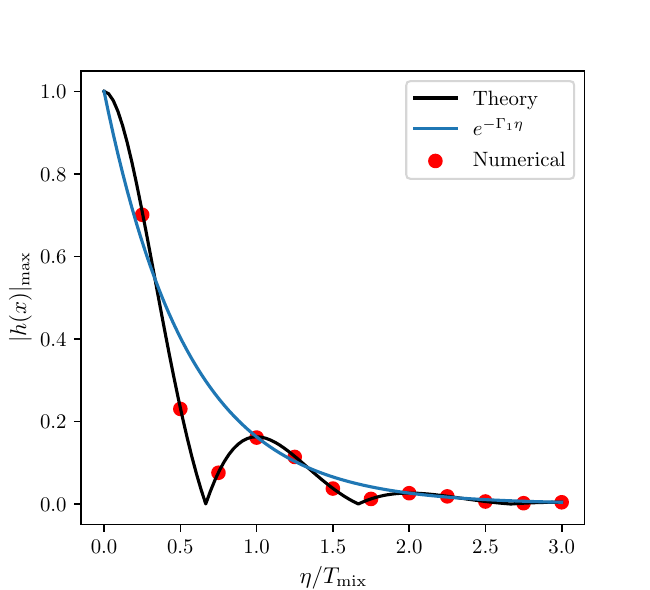}
\caption{\label{fig:friction-Delnu}
Maximum  amplitude of $h$ at different times, in the case of friction-mixing with $c_h=c_s$ and non-vanishing $\Delta\nu$. The black line is the theoretical prediction of the amplitude Eq.~(\ref{eq:amplitude_const}) while the red dots are the numerical calculations. 
The cyan line corresponds to exponential suppression of the amplitude of the wavepacket. 
Parameters: $\alpha=0.1$, $\Delta\nu=0.05$, $c_h^2=c_s^2=1$, $k_0=100$, $\sigma_{x}=1$.  
}
\end{figure}

A toy model calculation with $\alpha>\Delta\nu$ is shown in Fig.~\ref{fig:friction-Delnu}, for the values $\Delta\nu=0.05$, $\alpha=0.1$, $c_h=c_s=1$, $k_0=100$, and $\sigma_x=1$. Since the modes propagate coherently, we only show the maximum of $h$ as a function of time, using as reference the mixing time $T_\text{mix}$. The black line describes the theoretical prediction, according to Eq.~(\ref{eq:amplitude_const}), while the red dots indicate the numerical results. 
Note that since the eigenfrequencies are complex, the amplitudes of both eigenmodes are suppressed by $e^{-\Gamma_{A}\eta}$ (recall $\Gamma_A$ are the imaginary components of the eigenfrequencies (\ref{eq:t1_friction})-(\ref{eq:t2_friction})). 
We observe that the amplitude of $h$ is a combination of the exponential suppression due to friction, together with an oscillatory behavior of the two modes that have $\omega_{1}\not=\omega_2$. In this particular example, we have $\Gamma_1\approx \Gamma_2$, and both modes contribute equally as $\tan^2\Theta_g=1$, and the mixing phase is $\phi_g=+(2/3)\pi$. The exponential suppression of the envelope is shown in the cyan line.
Notice that because the mixing phase alters the overall amplitudes of $|f_1| \approx |f_2|$, $h$ not only oscillates due to mixing but also can exceed this envelope.

\subsection{Chiral Mixing}\label{sec:chiral}
Chiral mixing appears in theories that break parity symmetry, such as in the case for Yang-Mills theories \cite{BeltranJimenez:2018ymu}, and vector-tensor magnetic gaugid theories \cite{Piazza:2017bsd}. In the latter, for special choices of model parameters, chiral mixing is the only form of mixing present.
Motivated by these models, in this section we isolate the effects of  chiral mixing. 
Let us consider the following polarizations-dependent EoM: 
\begin{equation}
\left[\hat{I} \frac{\d^2}{\d\eta^2} +\bpm c_h^2 & 0 \\ 0 &  c_s^2\epm k^2 \pm\bpm \mu_h & \gamma \\ \gamma & \mu_s \epm k \right] \bpm h_{L,R} \\ s_{L,R} \epm =0,
\end{equation}
where the $\pm$ correspond to the left-handed and right-handed circular polarizations.
The associated eigenfrequencies are different for the two polarizations:
\begin{align}
&(\omega_{1;L,R})^2 = \lp c_h^2+\frac{1}{2}\dc^2 \rp k^2 \pm\frac{1}{2}\mu_{\rm tot}k -\frac{k}{2}\sqrt{4\gamma^2 +( k\dc^2 \pm \dmu)^2} \,,\label{w1LR}\\
&(\omega_{2;L,R})^2 = \lp c_h^2+\frac{1}{2}\dc^2 \rp k^2 \pm\frac{1}{2}\mu_{\rm tot}k +\frac{k}{2}\sqrt{4\gamma^2 +( k\dc^2 \pm \dmu)^2} \label{w2LR}\,,
\end{align}
if $k\dc^2 \pm \dmu>0$, so that in the no-mixing limit of $\gamma=0$, $\omega_1$ and $\omega_2$ describe the propagation of the field $h$ and $s$, respectively. Here, we have introduced $\mu_{\rm tot}\equiv \mu_h+\mu_s$ and $\dmu=\mu_s-\mu_h$. Note that these eigenfrequencies are real, and therefore the eigenmodes will not exhibit exponential decays in their amplitudes, i.e.~$\Gamma_A=0$. We also note that the chiral terms $\mu_{h,s}$ and $\gamma$ affect the group velocities of the eigenmodes and therefore, for each polarization, the two eigenmodes are generally expected to propagate at different speeds even when $c_h=c_s$. Note that even if $\gamma=0$, GWs would still be chiral if $\mu_h\not=0$, and left and right circular polarizations would propagate birefringently with different  $\omega_{1;L}$ and $\omega_{1;R}$.
However, if in addition $\gamma\not=0$, then the detected GW signal $h$ in the most general case would be the superposition of 4 propagating modes: 2 eigenstates for each of the 2 polarizations given in Eq.\ (\ref{w1LR})-(\ref{w2LR}). 

For the case with mixing, i.e.\ $\gamma\not=0$, we define the mixing angles and relevant regime timescales for the two propagating modes left and right separately. 
Since the eigenfrequencies and the matrix $\eM$ are real, the solution of $h$ is fully determined by the mixing angle only, with a mixing phase $\phi_g=\pi$. In particular, we have that:
\begin{equation}
    \tan^2\Theta_{g;L,R}= \frac{4\gamma^2}{\left(|\Delta c^2 k\pm \dmu | + \sqrt{4\gamma^2+(\Delta c^2k \pm \dmu)^2}\right)^2},
\end{equation}
which vanishes in the no-mixing limit.
Furthermore, the amplitudes of the two eigenmodes contributing to $h_L$ and $h_R$ will have the same functional form as that for the mass mixing case, given by Eq.~(\ref{ModeAmpl_MassMixing}). 

We emphasize that, contrary to the cases of friction and mass mixing, besides modifications in the dispersion relation and amplitude of the signal, chiral mixing also leads to changes in the polarization of the GW, which can be probed with multiple GW detectors.
Indeed, if we start with a given polarization at emission, then during propagation this polarization may rotate due to the chirality of the solution. More specifically, if a source emits only left-handed or only right-handed polarization, then the detected signal will have the same polarization as the emitted one since left and right polarizations do not mix with each other. However, if a source emits any other type of polarization, then it will change during propagation due to the different admixtures of $L$ and $R$, which propagate differently, e.g. 
for $+$ and $\times$: 
\begin{equation}\label{LR_pluscross_relation}
    h_+=\frac{1}{\sqrt{2}}(h_L+h_R); \quad h_\times=\frac{i}{\sqrt{2}}(h_L-h_R).
\end{equation}
Therefore, it becomes relevant to characterize the polarization of the detected signal in order to understand how it changed during its propagation. As explained in Appendix \ref{App:polarization}, for each wavepacket, one can calculate the total $L$ and $R$ polarization content of the wave (which will be a superposition of the propagating eigenstates), and characterize the total signal with four parameters $\{A, \phi, \beta,\chi\}$, where $A$ determines the total amplitude of the signal, $\phi$ its phase, and the angles $\beta$ and $\chi$ its polarization.
In particular, the parameter $\beta$ describes the degree of circular polarization through the ratio between the semi-major and semi-minor axes of a general elliptical polarization (with $\beta=0$ describing a linear polarization, and $\beta=\pm \pi/4$ a circular polarization), and $\chi$ gives the orientation of that elliptical polarization.
At emission, the wave will have a fixed $\beta$ and $\chi$ polarization parameters, but as the signal evolves in time, its polarization can change and have new $\beta$ and $\chi$ parameters that depend on how long the signal has been propagating for. The change in $\beta$ due to propagation is called amplitude birefringence, whereas the change in $\chi$ is called phase birefringence (see a review of parity-violating theories with a single tensor field that exhibits these two types of effects in \cite{Zhao:2019xmm}).
The former is associated with a change due to propagation in the relative amplitude of $L$ vs.\ $R$ and the latter in their relative phase; in particular:
\begin{equation}
    \frac{h_R}{h_L}=re^{4i\chi}; \; r=(\cos\beta+\sin\beta)/(\cos\beta-\sin\beta),
\end{equation}
where both angles range between $[-\pi/4,\pi/4]$, and thus $r\geq0$.

In general, these angles may also depend on wavenumber, but for the Gaussians considered in this section, we can calculate these angles at the central wavenumber $k=k_0$. From Eq.\ (\ref{hsolconstparam}), we know that the most general solution to left and right polarizations define these relative amplitudes and phases and hence $\Delta \chi$ and $\Delta \beta$:
\begin{equation}\label{Chiral_h_sol}
h_{p}(\eta,k)=  h_{0p}(k)\cos^2\Theta_{gp}  \sqrt{1+\tan^4\Theta_{gp}+2\tan^2\Theta_{gp}\cos(\Delta\omega_{p}\Delta\eta)}e^{-i\theta_p'}, 
\end{equation}
where $p$ is a subscript indicating each polarization $p=L,R$. Also, we have defined $\Delta\omega_{p}=\omega_{2;p}-\omega_{1;p}$, and $\tan \theta_p'=(\sin\omega_{1;p}\Delta\eta+\tan^2\Theta_{g;p}\sin\omega_{;2p}\Delta\eta) /(\cos\omega_{1;p}\Delta\eta+\tan^2\Theta_{g;p}\cos\omega_{2;p}\Delta\eta)$. 

In the coherence regime, the polarization angles $\beta$ and $\chi$ can be calculated superposing all the wavepackets and calculating the net $h_L$ and $h_R$ components of the signal, according to Eq.\ (\ref{Chiral_h_sol}). On the other hand, in the decoherence regime, one can characterize the polarization content of each echo formed. For example, when $c_h=c_s$, for small deviations from GR we will find that only two echoes form: one containing the propagating modes $\omega_{1;R}$ and $\omega_{1;L}$, and another one containing $\omega_{2;L}$ and $\omega_{2;R}$. For each echo, one can calculate its own $\beta$ and $\chi$ angles in order to describe its observed polarizations. In a more general case with $c_s\not=c_h$, four echoes will form, each one with a purely circular polarization left or right.

Next, we explore in more detail two limiting cases:

\subsubsection{$\Delta c^2=0$} \label{sec:gaussian_chiral_dc0}
For $\Delta c^2=0$, the eigenfrequencies simplify to:
\begin{align}
&(\omega_{1;L,R})^2 =  c_h^2k^2  \pm\frac{1}{2}\mu_{\rm tot}k \pm\frac{k}{2}\sqrt{4\gamma^2 +  \dmu^2} \,,\label{omega1_chiral1}\\
&(\omega_{2;L,R})^2 = c_h^2 k^2  \pm\frac{1}{2}\mu_{\rm tot}k \mp\frac{k}{2}\sqrt{4\gamma^2 +  \dmu^2} \,,\label{omega2_chiral1}
\end{align}
if $\Delta\mu\leq 0$ so that in the no-mixing limit of $\gamma=0$, $\omega_{1}$ and $\omega_2$ describe the propagation of $h$ and $s$, respectively. The associated group velocities are:
\begin{align}
    v_{g1;L,R}=\frac{ 4c_h^2k \pm \mu_{\rm tot}\pm\sqrt{4\gamma^2+\dmu^2}}{2\sqrt{4c_h^2k^2\pm 2\mu_{\rm tot} k \pm 2\sqrt{4\gamma^2+\dmu^2}k }} \\
    v_{g2;L,R}= \frac{ 4c_h^2k \pm \mu_{\rm tot}\mp\sqrt{4\gamma^2+\dmu^2}}{2\sqrt{4c_h^2k^2\pm 2\mu_{\rm tot} k \mp 2\sqrt{4\gamma^2+\dmu^2}k }}.
\end{align}
From here we see that generically the group velocities of the four propagating eigenmodes are different, and thus for a general initial wavepacket containing both left and right polarizations, one expects to see four echoes for long enough times, each one with a purely circular polarization. However, in the large-$k$ expansion, we see that there are only two distinct group velocities: 
\begin{align}
   &  v_{g1;L}\approx v_{g1;R}\approx c_h+\frac{1}{32c_h^3k^2}\left( \mu_\text{tot} +\sqrt{4\gamma^2+\Delta\mu^2}\right)^2 + \mathcal{O}(k^{-3}), \label{Chiral_v11}\\
    & v_{g2;L}\approx v_{g2;R}\approx c_h+\frac{1}{32c_h^3k^2}\left( \mu_\text{tot} -\sqrt{4\gamma^2+\Delta\mu^2}\right)^2 + \mathcal{O}(k^{-3}).\label{Chiral_v22}
\end{align}
Therefore, from here we see that right after decoherence of the left-handed wavepackets $\omega_{1;L}$ and $\omega_{2;L}$ is reached (and similarly for right-handed), only two echoes will be observed, each one containing a combination of both right and left polarizations. 

Nevertheless, for long enough times, four echoes may be observed, since the eigenmodes $\omega_{1;L}$ and $\omega_{1;R}$ may eventually decohere, and similarly for the pair $\omega_{2;R}$ and $\omega_{2;L}$. 
In order to decohere, the separation of the eigenmodes needs to be larger than their width, which could be changing due to broadening as discussed in Eq.~(\ref{DecohCrit}). 
In particular, their decoherence timescales will be given by their group velocity difference:
\begin{align}
    v_{g1;R}-v_{g1;L}\approx \frac{1}{32c_h^5k^3}\left( \mu_\text{tot} +\sqrt{4\gamma^2+\Delta\mu^2}\right)^3 + \mathcal{O}(k^{-5}),\label{Deltav_1}\\
    v_{g2;L}-v_{g2;R}\approx \frac{1}{32c_h^5k^3}\left( \mu_\text{tot} -\sqrt{4\gamma^2+\Delta\mu^2}\right)^3 + \mathcal{O}(k^{-5}).\label{Deltav_2}
\end{align}
Combining with the $T_\text{broad}$ expression given below, the criteria for the full decoherence of the four eigenmodes reads
\begin{equation}\label{DecohCritChiral}
    \frac{1}{2}(\mu_{\rm tot}\pm\sqrt{4\gamma^2+\dmu^2})\frac{\sigma_x}{c_h^2} > 1,
\end{equation}
which must be satisfied for both $\pm$ signs for the two pairs of $(\omega_{1;L}, \omega_{1;R})$ and $(\omega_{2;L}, \omega_{2;R})$ to decohere. Note that this condition does not depend on $k$ at leading order, and therefore whether decoherence is reached depends only on the choice of parameters.

On the other hand, the mixing angle simplifies to:
\begin{equation}
    \tan^2\Theta_{g;L,R}= 
    \frac{4\gamma^2}{\left(\sqrt{4\gamma^2+\dmu^2} + |\dmu| \right)^2},\label{Chiral_MixingAngle}
\end{equation}
so that the mixing angle vanishes in the no-mixing limit of $\gamma\rightarrow 0$.
Notice that, in this case with $\Delta c^2=0$, the two mixing angles are the same.
Depending on the parameter choice, these mixing angles may be small or large, 
Because of the relation between $\Theta_{g;L}$ and $\Theta_{g;R}$, the amplitude of the four independent wavepackets $|h_0f_{A;L,R}|$ 
satisfy the following relations: 
\begin{equation}
    \frac{|f_{1;L}|}{|f_{1;R}|}=\frac{|f_{2;L}|}{|f_{2;R}|}= 1,
\end{equation}
where $h_{0L}$ and $h_{0R}$ are the initial conditions for left and right polarizations. 
The typical timescales of mixing, coherence, and broadening in the large-$k$ limit are given by:
\begin{align}
&T_{\text{mix};L,R}\approx  \frac{ 4\pi c_h}{\sqrt{4\gamma^2+\Delta\mu^2}} \left( 1\pm \frac{\mu_\text{tot} }{4c_h^2k}\right),\\
&T_{\text{coh};L,R}\approx \sigma_{A;L,R}\frac{8c_h^3k^2}{\mu_\text{tot}\sqrt{4\gamma^2+\Delta\mu^2}}\left(1 \pm \frac{(\Delta\mu^2+4\gamma^2+3\mu_\text{tot}^2)}{4\mu_\text{tot}^2c_h^2k}\right), \label{Tcoh_chiral1}\\
&T_{\text{broad},1;L,R}\approx \sigma_x^2\frac{16c_h^3k^3}{(\mu_\text{tot}+\sqrt{4\gamma^2+\Delta\mu^2})^2}\left(1+ \frac{3(\pm\mu_\text{tot}\pm\sqrt{4\gamma^2+\Delta\mu^2})}{4c_h^2k}\right) , \\
&T_{\text{broad},2;L,R}\approx \sigma_x^2\frac{16c_h^3k^3}{(\mu_\text{tot}-\sqrt{4\gamma^2+\Delta\mu^2})^2}\left(1+ \frac{3(\pm\mu_\text{tot}\mp\sqrt{4\gamma^2+\Delta\mu^2})}{4c_h^2k}\right) .
\end{align}

In the coherence regime, the explicit amplitudes of each circular polarization are given by
\begin{equation}
\begin{split}
    |h_{L,R}|^2=\frac{|h_{0L,R}|^2}{4\gamma^2+\Delta\mu^2}&\Bigg[2\gamma^2+\Delta\mu^2+2\gamma^2\cos\left(\frac{\sqrt{4\gamma^2+\Delta\mu^2}}{2c_h}\Delta\eta\right) \\
    &\pm\gamma^2\sin\left(\frac{\sqrt{4\gamma^2+\Delta\mu^2}}{2c_h}\Delta\eta\right)\frac{\sqrt{4\gamma^2+\Delta\mu^2}\mu_\text{tot}\Delta\eta}{4c_h^3k}\Bigg]+\mathcal{O}\left(1/k^2\right)\,,
\end{split}
\end{equation}
where we see that at leading order in $k$, both left and right polarizations suffer the same amplitude change due to propagation. It is only at order $k^{-1}$ where differences appear and hence change the degree of circular polarization $\beta$:
\begin{equation}
    \tan \beta = \frac{r-1}{r+1}\,,
\end{equation}
which is explicitly given by
\begin{equation}
\begin{split}
    \tan\beta=& \frac{r_0-1}{r_0+1} -\frac{r_0}{\left(1+r_0\right)^2}\times\\
    &\times\frac{\gamma^2\sin\left(\frac{\sqrt{4\gamma^2+\Delta\mu^2}}{2c_h}\Delta\eta\right)}{2\gamma^2+\Delta\mu^2+2\gamma^2\cos\left(\frac{\sqrt{4\gamma^2+\Delta\mu^2}}{2c_h}\Delta\eta\right)}\frac{\sqrt{4\gamma^2+\Delta\mu^2}\mu_\text{tot}\Delta\eta}{2c_h^3k}+\mathcal{O}\left(1/k^2\right)\,,\label{beta_chiral}
\end{split}
\end{equation}
where $r_0$ describes the initially emitted polarization that, in this paper, is assumed to be given by the GR signal, $r_0=|h_{0R}/h_{0L}|$. Here we see that amplitude birefringence is suppressed by $1/k$.
In the coherence regime, the phase between the two polarizations is given by
\begin{equation}
    \tan (4\chi-4\chi_0) = \frac{\tan\theta'_L-\tan\theta'_R}{1+\tan\theta'_L\tan\theta'_R}\,.
\end{equation}
For large $k$, this expressions approximates to:
\begin{align} \label{eq:chi}
    &\tan (4\chi-4\chi_0)=\left\{2\gamma^2\sin\lp (\omega_{2;L}-\omega_{1;R})\Delta\eta\rp + 2\gamma^2\sin\lp (\omega_{1;L}-\omega_{2;R})\Delta\eta\rp \right.\nonumber\\
    &\left.+(-2\gamma^2-\Delta\mu^2+\Delta\mu\epsilon)\sin\lp (\omega_{1;R}-\omega_{1;L})\Delta\eta\rp +(2\gamma^2+\Delta\mu^2+\Delta\mu\epsilon)\sin\lp (\omega_{2;L}-\omega_{2;R})\Delta\eta\rp  \right\}\nonumber \\
    &/\left\{2\gamma^2\cos\lp (\omega_{2;L}-\omega_{1;R})\Delta\eta\rp + 2\gamma^2\cos\lp (\omega_{1;L}-\omega_{2;R})\Delta\eta\rp \right.\nonumber\\
    &\left.+(2\gamma^2+\Delta\mu^2-\Delta\mu\epsilon)\cos\lp (\omega_{1;R}-\omega_{1;L})\Delta\eta\rp +(2\gamma^2+\Delta\mu^2+\Delta\mu\epsilon)\cos\lp (\omega_{2;L}-\omega_{2;R})\Delta\eta\rp  \right\} \nonumber\\
    &+\mathcal{O}\left(1/k^2\right)\,,
\end{align}
where for brevity we have defined $\epsilon=\sqrt{4\gamma^2+\Delta\mu^2}$\,.

On the other hand, since in the high-$k$ limit we find that $T_\text{coh;R}\approx T_\text{coh;L}$ (using Eq.\ (\ref{Tcoh_chiral1}) and the fact that $T_\text{broad;L}\approx T_\text{broad;R}$) and $v_{g1;R} \approx v_{g1;L}$, then for propagation times comparable to $T_\text{coh;R}$ a signal containing generally right and left-handed polarization will split into two wavepackets: one containing the propagating modes $\omega_{1;R}$ and $\omega_{1;L}$, and another one containing $\omega_{2;L}$ and $\omega_{2;R}$. In this case, the birefringence angle $\beta$ for each of the two wavepackets is given by:
\begin{align}
    \tan\beta_1&=\frac{|h_{0R}|\cos^2\Theta_{g;R}-|h_{0L}|\cos^2\Theta_{g;L}}{|h_{0R}|\cos^2\Theta_{g;R}+|h_{0L}|\cos^2\Theta_{g;L}}=\frac{r_0-1}{r_0+1},\label{beta1_case1}\\
    \tan\beta_2&=\frac{|h_{0R}|\sin^2\Theta_{g;R}-|h_{0L}|\sin^2\Theta_{g;L}}{|h_{0R}|\sin^2\Theta_{g;R}+|h_{0L}|\sin^2\Theta_{g;L}}=\frac{r_0-1}{r_0+1},\label{beta2_case1}
\end{align}
where we used the relation $\Theta_{g;L}=\Theta_{g;R}$. We see that each wavepacket does not change its $\beta$ angle during its evolution. In addition, the angle $\Delta\chi$ for the two wavepackets that have decohered are given by
\begin{align}
\Delta\chi_1 &= \Delta\chi_0 +\frac{1}{4}\left(\omega_{1L}-\omega_{1R}\right)\Delta\eta, \\
\Delta\chi_2 &= \Delta\chi_0 +\frac{1}{4}\left(\omega_{2L}-\omega_{2R}\right)\Delta\eta,
\end{align}
where $\Delta\chi_0$ is given by the initial conditions, where $h_{0R}/h_{0L}=r_0\exp\{4i\Delta\chi_0\}$. 

\begin{figure}[!htb]
\centering
\includegraphics[width=\columnwidth]{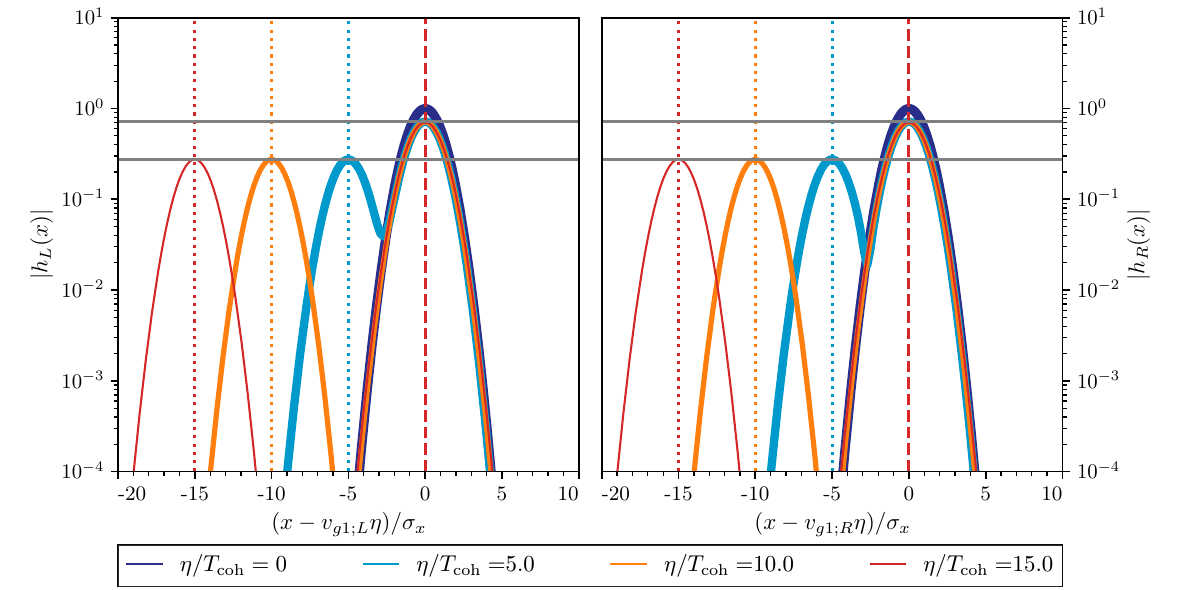}
\caption{\label{fig:chiral:dc2=0}
Wave packet propagation for the chiral mixing with $\dc^2=0$. The left and right panels show left and right polarizations respectively. 
The amplitude of the two propagation eigenmodes swap for the left and right polarizations.
Parameters: $\gamma=1$, $\mu_h=1.5$, $\mu_s=0.5$, $c_h^2=c_s^2=1$, $k_0=100$, $\sigma_{x}=1$.
}
\end{figure}

We show an example in Fig.~\ref{fig:chiral:dc2=0}, where we present separately how left and right-handed polarizations evolve in time for a Gaussian initial condition. In this example we have $\gamma=1$ and $\dmu=-1$, hence $\tan^2\Theta_{g;L,R}=(3-\sqrt{5})/2$.
We see that the amplitudes of the two left and right-handed eigenmodes are different, however the second right-handed eigenmode has the same amplitude as the second left-handed eigenmode, and similarly for the first right and left-handed eigenmodes.
For this parameter choice, we obtain that the mixing, coherent and broadening timescales are $T_{\text{mix};L,R}=\{5.64, 5.59\}$, $T_{\text{coh};L,R}=\{1.8\times 10^4, 1.7\times 10^4\}$, $T_{\text{broad},1;L,R}=\{9.2\times 10^5, 8.6\times 10^5\}$, $T_{\text{broad},2;L,R}=\{2.9\times 10^8, 2.9\times 10^8\}$. We thus confirm the previous results and see that $T_{\text{coh};L}\approx T_{\text{coh};R}$, because the propagating speed of the second right-handed mode is the same as that one of the first left-handed mode, and viceversa (see (\ref{Chiral_v11})-(\ref{Chiral_v22})). We also see that $T_{\text{broad}}\gg T_{\text{coh}}$ thus there are negligible distortions for the timescales shown in the figure.

\begin{figure}[!htb]
\centering
\includegraphics[width=0.49\columnwidth]{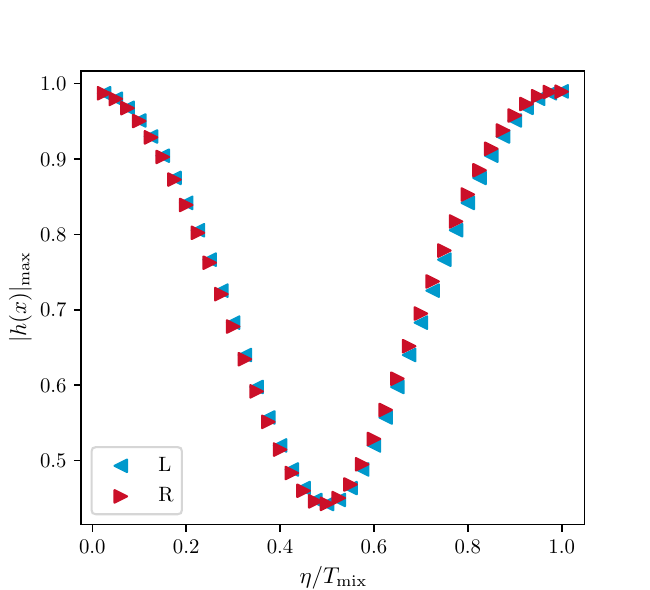}
\includegraphics[width=0.49\columnwidth]{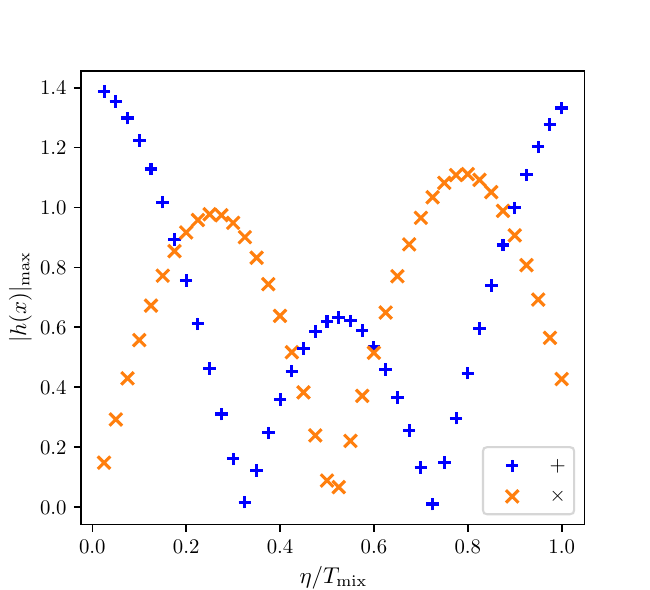}
\caption{\label{fig:chiral:dc2=0-amplitude}
Maximum amplitude of the wave packet envelope during the coherence regime for the chiral mixing with $\dc^2=0$. The initial wave packet is purely $+$ polarized.
Parameters: $\gamma=1$, $\mu_h=1.5$, $\mu_s=0.5$, $c_h^2=c_s^2=1$, $k_0=100$, $\sigma_{x}=1$.
}
\end{figure}

\begin{figure}[!htb]
\centering
\includegraphics[width=0.49\columnwidth]{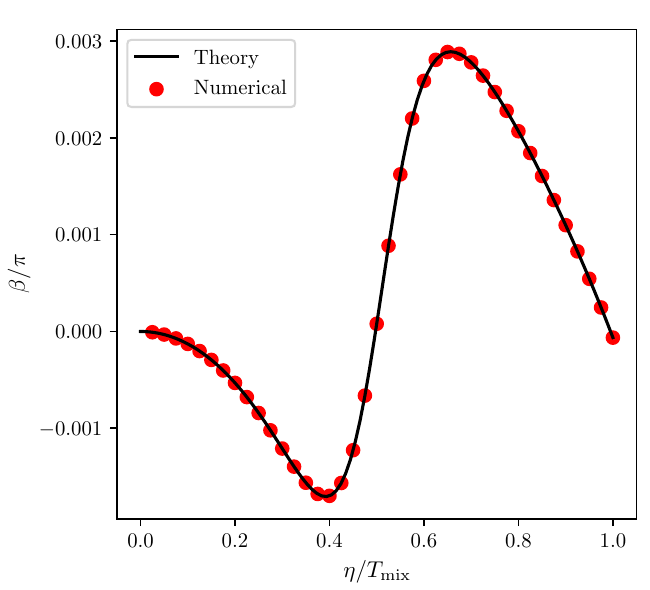}
\includegraphics[width=0.49\columnwidth]{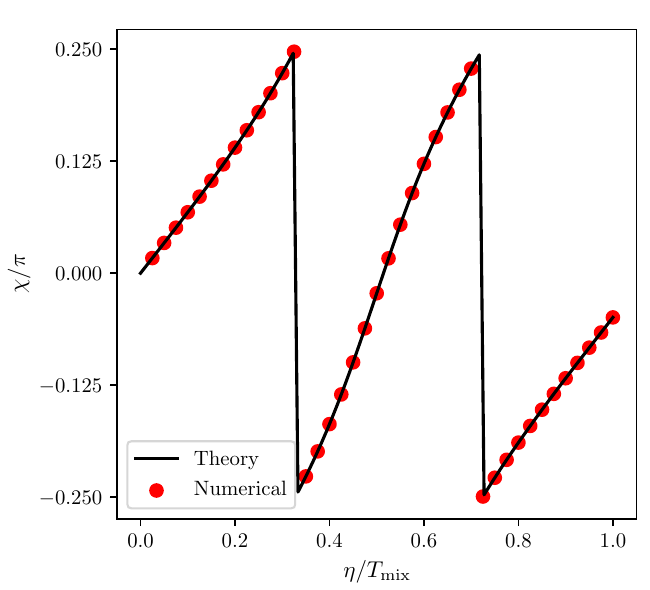}
\caption{\label{fig:chiral:dc2=0-angle}
Polarization angles $\beta$ and $\chi$ during the coherence regime for the chiral mixing with $\dc^2=0$. The initial wave packet is purely $+$ polarized.
The black lines are the theoretical predictions from Eq.~(\ref{beta_chiral}) and (\ref{eq:chi}).
Parameters: $\gamma=1$, $\mu_h=1.5$, $\mu_s=0.5$, $c_h^2=c_s^2=1$, $k_0=100$, $\sigma_{x}=1$.}
\end{figure}

Since a generic wavepacket may contain both right and left-handed polarizations, the signal will suffer amplitude and phase birefringence. We give an example of these two effects in Figs.\ \ref{fig:chiral:dc2=0-amplitude} and \ref{fig:chiral:dc2=0-angle}. We show how an initially $+$ polarized Gaussian wavepacket evolves in time, during the coherence regime (and we thus present timescales comparable to $T_\text{mix}\ll T_\text{coh}$). 
We show the amplitudes of left and right polarizations of the total signal in the left panel of Fig.~\ref{fig:chiral:dc2=0-amplitude}, where we see that they evolve slightly different in time. Indeed, this can be seen from the amplitude birefringence angle $\beta$, as shown in the left panel of Fig.~\ref{fig:chiral:dc2=0-angle} (dots are the numerical results and black line is the analytical estimate using Eq.\ (\ref{beta_chiral})). We can see that there are only small variations in $\beta$, since its changes are suppressed by $k$ according to Eq.\ (\ref{beta_chiral}).
In addition, we show the amplitudes of $+$, $\times$ polarizations of the total signal in the right panel of Fig.~\ref{fig:chiral:dc2=0-amplitude} and see how a $\times$ component is generated during the propagation due to phase birefringence. This corresponds to a rotation of the polarization during propagation, which can be explicitly seen in the  right panel of of Fig.~\ref{fig:chiral:dc2=0-angle} where we show the angle $\chi$ characterizing the phase birefringence. We can see that $\chi$ oscillates between 0 and $\pi/4$ corresponding to pure $+$ and $\times$ polarizations, respectively. We demonstrate how the numerical example (red dots) fits well the theoretical expectation (black line) of Eq.\ (\ref{eq:chi}).

\begin{figure}[t!]
\centering
\includegraphics[width=0.55\columnwidth]{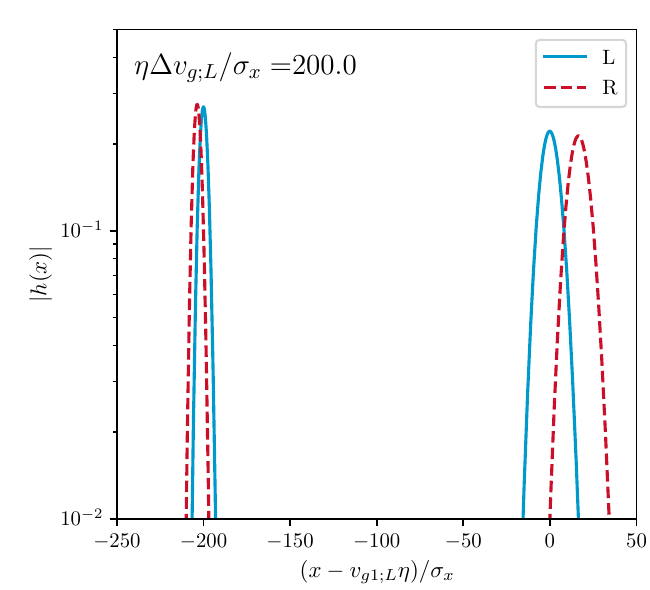}
\caption{\label{fig:chiral:4eigen}
Wavepacket propagation at time $\eta \Delta v_{g;L}/\sigma_x=200$ for chiral mixing with $\dc^2=0$ in the high-$k$ limit. 
In this example we have initially pure $+$ polarization. During propagation the peaks of the four purely left and right polarized wavepackets separate from each other, but their broadening prevents them to fully decohere.
Parameters: $\gamma=2$, $\mu_h=0.75$, $\mu_s=0.25$, $c_h^2=c_s^2=1$, $k_0=100$, $\sigma_{x}=1$.
}
\end{figure}

In Fig.~\ref{fig:chiral:4eigen} we show an example to illustrate that in the high-$k$ limit the broadening effects prevent the signal to fully decohere into four wavepackets, for this parameter choice. Here we have chosen $\gamma=2$, $\mu_h=0.75$, $\mu_s=0.25$, $c_h^2=c_s^2=1$, $k_0=100$, and $\sigma_{x}=1$, and chosen the initial wavepacket to be purely $+$ polarized. We plot the propagation of the signal in a timescale much larger than the `naive' coherence timescale between $\omega_{1,L}$ and $\omega_{2;L}$ given by $\sigma_x/\Delta v_{g;L}$, which compares the separation of the wavepackets to the initial width of the signal.
We see that the full decoherence criteria Eq.~(\ref{DecohCritChiral}) is barely satisfied. Therefore, although the peak separations of the four wavepackets keep growing and eventually become much larger than the initial width $\sigma_x$, they never become significantly larger than the actual wavepacket widths $\sigma_{1,2}$ because of their broadening effect.  
The separation of eigenmodes to width ratio Eq.~(\ref{SepInifity}) saturates to 1.5 for the slower pair and 2.4 for the faster pair for long enough times as shown in the figure.

\subsubsection{$\Delta c^2\neq0$ and small mixing}
In the limit  $|\gamma/(\dc^2k)|\ll 1$ and $|\dmu/(\dc^2k)|\ll 1$, from Eq.\ (\ref{w1LR})-(\ref{w2LR}) we find
\begin{align}
&(\omega_{1;L,R})^2 = \omega_{1,{\rm fid};L,R}^2 -\frac{\gamma^2}{\dc^2}\lp 1\mp\frac{\dmu}{\dc^2k} \rp +\mathcal{O}\lp\frac{\gamma^4}{\dc^8k^4}\rp +\mathcal{O}\lp\frac{\gamma^2\dmu^2}{\dc^6k^2}\rp \,, \label{w1LR_v2}\\
&(\omega_{2;L,R})^2 = \omega_{2,{\rm fid};L,R}^2 +\frac{\gamma^2}{\dc^2}\lp 1\mp\frac{\dmu}{\dc^2k} \rp +\mathcal{O}\lp\frac{\gamma^4}{\dc^8k^4}\rp +\mathcal{O}\lp\frac{\gamma^2\dmu^2}{\dc^6k^2}\rp \,.\label{w2LR_v2}
\end{align}
Here, we have introduced the following fiducial eigenfrequencies:
\begin{equation}
\omega_{1,{\rm fid};L,R} = \sqrt{c_h^2k^2\pm\mu_h k} \,;
\quad \omega_{2,{\rm fid};L,R} = \sqrt{c_s^2k^2\pm\mu_s k}\,,
\end{equation}
which are the frequencies when there is no mixing $\gamma=0$. Therefore, Eq.\ (\ref{w1LR_v2})-(\ref{w2LR_v2}) are such that in the no-mixing limit, $\omega_1$ and $\omega_2$ describe the propagation of $h$ and $s$, respectively, regardless of the model parameter values.
The associated group velocities in this limit are given by:
\begin{align}
    v_{g1;L,R} & =  \frac{2c_h^2k\pm \mu_{\rm h}}{2\omega_{1,\text{fid};L,R}} +\frac{( 2c_h^2k \pm \mu_\text{h})}{4\dc^2\omega_{1,\text{fid};L,R}^3} \gamma^2 +\mathcal{O}\left(\frac{\Delta\mu \gamma^2}{\dc^4k^2}\right),\\
    v_{g2;L,R} & = \frac{2c_s^2k\pm \mu_{\rm s}}{2\omega_{2,\text{fid};L,R}} -\frac{( 2c_s^2k \pm \mu_\text{s})}{4\dc^2\omega_{2,\text{fid};L,R}^3} \gamma^2 +\mathcal{O}\left(\frac{\Delta\mu \gamma^2}{\dc^4k^2}\right).\\
\end{align}
Contrary to the previous case with $\Delta c^2=0$, here we find that the group velocities of the four eigenmodes are in principle different at leading order, and hence in the decoherence regime there will be four wavepackets, with purely right and left polarization. 
Nevertheless, for some values of the parameters, it may happen that two of the four wavepackets decohere first.
In the case of $\Delta c^2>0$, the mixing angle is suppressed by
\begin{equation}
\tan\Theta_{g;L,R} = \frac{\gamma^2}{\dc^4k^2} \lp 1 \mp\frac{2\dmu}{\dc^2k}\rp 
    +\mathcal{O}\lp\frac{\gamma^4}{\dc^8k^4}\rp
    +\mathcal{O}\lp\frac{\gamma^2\dmu^2}{\dc^8k^4}\rp ,
\end{equation}
where the analogous expression for $\Delta c^2$ would be given by $(\tan\Theta_{g;L,R})^{-1}$.

Finally, we mention the typical time scales of mixing, coherence, and broadening:
\begin{align}
&  T_{\text{mix};L,R}\approx \frac{2\pi}{k(c_h-c_s)}\pm \frac{\pi (c_h\mu_s- c_s\mu_h)}{k^2c_hc_s(c_s-c_h)^2}, \\    
& T_{\text{coh};L,R}\approx \frac{\sigma_{A;L,R}}{c_h-c_s} +\frac{\sigma_{A;L,R}\lp 4\gamma^2c_h^2c_s^2-\mu_h^2c_s^3(c_h-c_s) +\mu_s^2c_h^3(c_h-c_s)\rp}{8k^2c_h^3c_s^3(c_h-c_s)^3 },\\
& T_{\text{broad},1;L,R}\approx \frac{4\sigma_x^2c_h^3\Delta c^2 k^3}{(4c_h^2\gamma^2+\mu_h^2\Delta c^2)}\lp 1\pm \frac{3(4c_h^2\Delta c^2\gamma^2\mu_h+8\gamma^2\Delta \mu c_h^4+\Delta c^4\mu_h^3)}{2kc_h^2\Delta c^2 (4c_h^2\gamma^2+\mu_h^2\Delta c^2)}\rp ,\\
&  T_{\text{broad},2;L,R}\approx \frac{4\sigma_x^2c_s^3\Delta c^2 k^3}{(4c_s^2\gamma^2-\mu_s^2\Delta c^2)}\lp 1\pm \frac{3(4c_s^2\Delta c^2\gamma^2\mu_s+8\gamma^2\Delta \mu c_s^4-\Delta c^4\mu_s^3)}{2kc_s^2\Delta c^2 (4c_s^2\gamma^2-\mu_s^2\Delta c^2)}\rp .
\end{align}

In the coherence regime, the explicit amplitudes of each circular polarization are given by
\begin{align}
    |h_{L,R}|^2=&|h_{0L,R}|^2\left[ 1 + \frac{2\gamma^2}{k^2\Delta c^4}\left(1 \mp  2\frac{\Delta\mu }{\Delta c^2k} \right)\left(\cos(\Delta \omega_{L,R} \Delta\eta)-1\right)  \right] \nonumber\\
    &+ \mathcal{O}\lp\frac{\gamma^4}{\dc^8k^4}\rp+ \mathcal{O}\lp\frac{\gamma^2\dmu^2}{\dc^8k^4}\rp,
\end{align}
where $\Delta \omega_{L,R}=\omega_{2;L,R}-\omega_{1;L,R}$. Therefore, the polarization of a given initial wave evolves such that:
\begin{align}
    & \tan\beta= \frac{r_0-1}{r_0+1} + \frac{4\gamma^2r_0}{\Delta c^4k^2(1+r_0)^2}\left[ \left( \cos(\Delta\omega_R\Delta\eta)-\cos(\Delta\omega_L\Delta\eta)\right) \nonumber \right. \\ 
    & \left. \pm  \frac{2\Delta\mu }{k\Delta c^2}\left( -2 + \cos(\Delta\omega_R\Delta\eta)+\cos(\Delta\omega_L\Delta\eta)\right) \right]+ \mathcal{O}\lp\frac{\gamma^4}{\dc^8k^4}\rp+ \mathcal{O}\lp\frac{\gamma^2\dmu^2}{\dc^8k^4}\rp.
\end{align}
Here we see that the changes in $\beta$ are more suppressed than in the case with $\dc^2=0$. In addition, the phase birefringence angle is given by:
\begin{align}
    &\tan\lp 4\chi -4\chi_0\rp=\frac{1}{2}\sec^2 \lp \Delta\omega_1\Delta\eta\rp \left\{ \sin \lp \Delta\omega_1\Delta\eta\rp + \frac{\gamma^2c^2}{\Delta c^6k^3} \left[ (k\Delta c^2 - 2\Delta\mu ) \sin (\Delta\omega_{L}\Delta\eta) \right.\right.\nonumber\\
    &\left. \left. - (k\Delta c^2+2\Delta\mu )\sin (\Delta\omega_R\Delta\eta) \right]  \right\} + \mathcal{O}\lp\frac{\gamma^4}{\dc^8k^4}\rp+ \mathcal{O}\lp\frac{\gamma^2\dmu^2}{\dc^8k^4}\rp,
\end{align}
where we have defined $\Delta\omega_1=\omega_{1;L}-\omega_{1;R}$. Similarly to the case with $\Delta c^2=0$, here we see that the angle $\chi$ varies during propagation at leading order in the large-$k$ limit.

\begin{figure}[h!]
\centering
\includegraphics[width=\columnwidth]{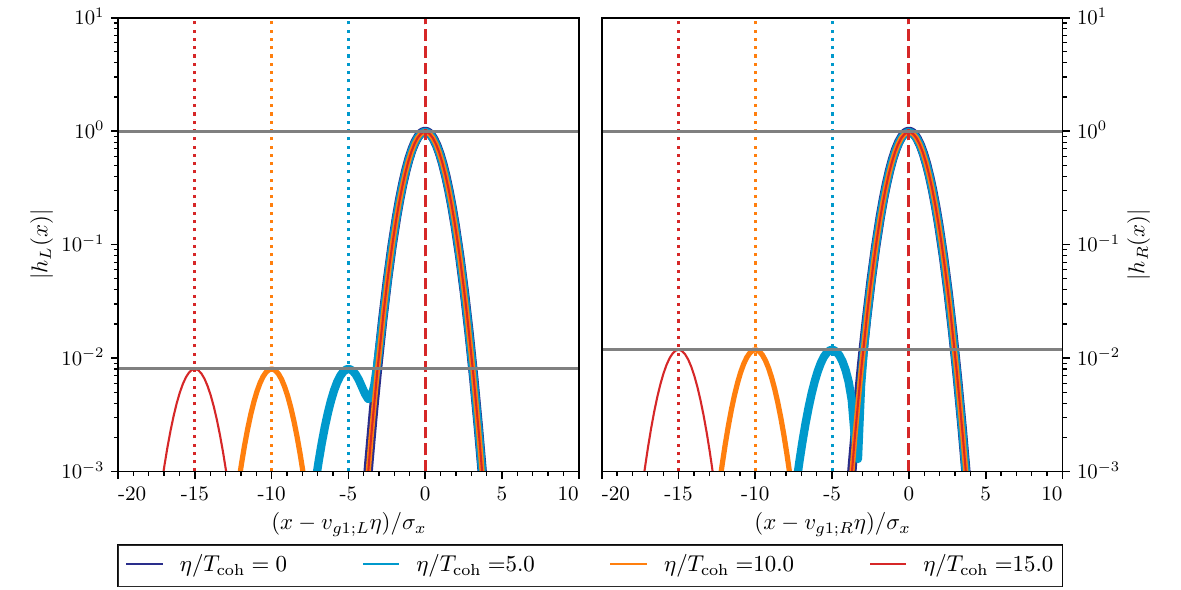}
\caption{\label{fig:chiral:dc2/=0}
Wave packet propagation for the chiral mixing in the limit $\gamma/(\dc^2k)\ll 1$ and $\dmu/(\dc^2k)\ll 1$. The left and right panels show left and right polarizations respectively. 
Parameters: $\gamma=1$, $\mu_h=1.5$, $\mu_s=0.5$, $c_h^2=1$, $c_s^2=0.9$, $k_0=100$, $\sigma_{x}=1$.
}
\end{figure}

A toy model calculation is shown in Fig.~\ref{fig:chiral:dc2/=0}. With $\dc^2=-0.1,\,k_0=100,\,\gamma=1,\,\dmu=-1$, we have $\tan\Theta_{g;L,R}\approx10^{-2}$, thus the amplitude of the second eigenstate $|f_2h_0|$ is suppressed by two orders of magnitude compared to that of the first eigenstate $|f_1h_0|$ for both left and right polarizations. The difference between the amplitudes of left and right polarizations is $|f_1|_{;L}-|f_1|_{;R} = |f_2|_{;R}-|f_2|_{;L}\approx 4\times10^{-3}$. For this parameter choice, we obtain that the mixing, coherent and broadening timescales are $T_{\text{mix};L,R}=\{1.1, 1.3\}$, $T_{\text{coh};L,R}=\{19.8, 20.0\}$, $T_{\text{broad},1;L,R}=\{1.6\times10^5, 7.3\times10^4\}$, $T_{\text{broad},2;L,R}=\{1.3\times10^5, 8.0\times10^4\}$. 
We then see that $T_{\text{broad}}\gg T_{\text{coh}}$ and hence the wave distortions are negligible for the timescales shown in the plots. 
It is important to notice that for the parameters of this example, the velocity difference between the four eigenstates  have the following hierarchy: $|v_{g1;L}-v_{g2;L}|\sim |v_{g1;R}-v_{g2;R}| \gg |v_{g1;L}-v_{g1;R}|\sim |v_{g2;L}-v_{g2;R}|$. This means that for a propagation time $\Delta\eta \sim T_{\text{coh};L,R}$ a general signal containing the four eigenstates will split into two echoes, one containing $(\omega_{1;L}, \omega_{1;R})$ and another one with $(\omega_{2;L}, \omega_{2;R})$. Each echo, will have a birefringence angles given by:
\begin{align}
    \tan\beta_1&=\frac{|h_{0R}|\cos^2\Theta_{gR}-|h_{0L}|\cos^2\Theta_{gL}}{|h_{0R}|\cos^2\Theta_{gR}+|h_{0L}|\cos^2\Theta_{gL}}\approx \frac{r_0-1}{r_0+1} - \frac{8\gamma^2\Delta\mu r_0}{(1+r_0^2)k^3\Delta c^6}+\mathcal{O}\left( \frac{\gamma^4\Delta\mu }{\Delta c^{10}k^5}\right) ,\\
    \tan\beta_2&=\frac{|h_{0R}|\sin^2\Theta_{gR}-|h_{0L}|\sin^2\Theta_{gL}}{|h_{0R}|\sin^2\Theta_{gR}+|h_{0L}|\sin^2\Theta_{gL}}\approx \frac{r_0-1}{r_0+1} + \frac{8\Delta\mu r_0}{(1+r_0^2)k\Delta c^2} +\mathcal{O}\left( \frac{\Delta\mu^2}{\Delta c^4k^2}\right).
\end{align}
Contrary to the previous case with $\Delta c^2=0$ in which the signal splits into two wavepackets containing  $(\omega_{1;L}, \omega_{1;R})$ and  $(\omega_{2;L}, \omega_{2;R})$, none of which exhibit amplitude birefringence (see Eqs.\ (\ref{beta1_case1})-(\ref{beta2_case1})), here we see that the two wavepackets exhibit amplitude birefringence, albeit small due to the $k$ suppression. In addition, the phase birefringence of the two wavepackets is given by:
\begin{align}
    \Delta\chi_1&=  \Delta\chi_0 +\frac{1}{4}\left(\omega_{1L}-\omega_{1R}\right)\Delta\eta,\\
    \Delta \chi_2&= \Delta\chi_0 +\frac{1}{4}\left(\omega_{2L}-\omega_{2R}\right)\Delta\eta .
\end{align}
For the example if Fig. \ref{fig:chiral:dc2/=0}, tt is only at $\Delta\eta \gtrsim 10^2 T_{\text{coh};L,R}$ that the signal will split into four echoes.

\section{Observational implications}\label{sec:implications}

After developing the framework for the GW propagation with additional tensor fields and applying it to toy Gaussian wavepackets, we now proceed to analyze the observational implications of a modified propagation for real compact binary coalescence signals. 
As discussed in Sec.\ \ref{sec:GW-propag}, the main difference between the coalescence scenario compared to the Gaussian model is that the coalescence signal contains observed frequencies with distinct arrival times, or equivalently, contains frequency components whose stationary phase point evolves across the wavepacket --- the famous chirping --- that current detectors measure.\footnote{It is to be noted that a compact binary coalescence can also lead to quasi-monochromatic signals if they are detected early enough in the inspiral. This is expected for example to happen for some sources detected by the future space-based detector LISA \cite{LISA:2017pwj}. Here, however, we focus on signals that have sufficient frequency evolution during the observation time.}
Thus, chirping GW signals contain information about both amplitude and phase evolution, as opposed to the Gaussian signal analyzed in the previous section where we only discussed its amplitude. 
The other main difference with respect to our analysis in Sec. \ref{sec:examples_mixing} is that we now allow parameters to vary over cosmological time scales. Therefore, in this section, we implement our WKB formalism and analyze the different propagation regimes of the signal as a function of the source redshift as opposed to its propagation time. For clarity, we will recover the speed of light $c$ in the analytical expressions of this section.

In order to compute the modifications of the waveform realistically, we must propagate a spherical wave, as opposed to the plane wave analyzed in the previous section. The main difference is the diminution of the flux due to surface area during propagation, since at the detector the wavefronts are planar to good approximation. We thus generalize the  plane wave solutions obtained in the previous sections (see  Eq.\ (\ref{hsolWKB})-(\ref{hsolWKB_2})) and schematically express the total GW signal, in terms of the source redshift $z_s$, as: 
\begin{equation}\label{EqMGh}
    h(k,z_s)= h_\text{fid}(k,z_s)
    \sum_{A}
 f_A(k,z_s) e^{-i\int_{0}^{z_s}\Delta\omega_A(k,z)/H(z)dz},
\end{equation}
where $H(z)$ is the Hubble rate, and the subscript $A$ describes all the possible propagation eigenstates that conform the total signal. 
Here we have assumed that without modified propagation effects, there is a fiducial \emph{detected} signal $h_\text{fid}(k,z_s)$ that propagates from the source to the observer according to the GR dispersion relation $\omega=ck$. Since the emitted signal is assumed to be a spherical wave, we then have that $h_\text{fid}(k,z_s) \propto e^{-ick\int dz/H(z)}/d_L(z_s)$ where the proportionality factor depends solely on the emission process, and $d_L(z_s)$ is the true luminosity distance to the source at redshift $z_s$. 
In addition, in Eq.\ (\ref{EqMGh}), we have introduced $\Delta\omega_{A}=\omega_A-ck$, the difference of the dispersion relation with respect to GR, and the functions $f_{A}(k,z)$ that are determined by the mixing matrix $\hat{E}(k,\eta)$, the damping factors $\Gamma_A(k,\eta)$ and the first-order WKB correction matrix $\hat{Q}(k,\eta)$, according to Eq.\ (\ref{hsolWKB}). 
From Eq.\ (\ref{EqMGh}), we see that waveform distortions with respect to GR may appear due to: 
\begin{itemize}
    \item Slow-varying amplitude and phase changes carried by $f_{A}\in \mathbb{C}$,
    \item Phase changes on each eigenstate due to their non-trivial dispersion relations,
    \item The way the eigenstates interfere with each other during the coherent regime to give the net GW signal.
\end{itemize}  
In all of the simple examples illustrated in this paper, only the two last mentioned effects modify the shape of the GW signal (i.e.\ its phase and amplitude frequency evolution across the signal), since frequency variations of $f_A$ (that may also affect the GW signal shape) 
are found to be highly suppressed in the large-$k$ limit and thus only produce an overall amplitude rescaling that maintains the signal's shape. 
However, we emphasize that this is not a general feature since different types of mixing could lead to non-negligible frequency-dependent variations in $f_A$.
This happens for example when there is mass, friction or chiral mixing combined with tensor modes with different speeds, $\dc^2\neq0$. This will also happen if two or more types of mixings are combined. 
During decoherence, a frequency-dependent $f_A(k)$ implies that individual echoes will be distorted even if they do not have a modified dispersion relation.

Since GW detectors observe the time evolution of GW signals, emitted and detected GW waveforms are actually provided as function of time or in temporal Fourier (frequency) space, as $h_\text{fid}(\omega,z_s)$, where $\omega$ is determined by properties of the GW source, such as the motion of a binary of compact objects. 
However, as described in more detail in 
Appendix \ref{App:mono_k_omega}, if the difference between the MDR of each eigenstate is small for the detected range of frequencies, 
then the initial conditions in $\omega$ space can be directly translated into initial conditions in $k$ space, where both eigenstates have approximately the same momentum $k$.   
In this case, one can assume that both eigenstates have the same fiducial waveform in $k$ space as well $h_\text{fid}(\omega(k))$, where the relation between $\omega$ and $k$ is obtained assuming that the deviations from GR are small. 
In practice, we thus model the frequency space signal by taking Eq.\ (\ref{EqMGh}) and approximating $\omega\approx ck$: 
\begin{equation}
   h(\omega/c,z_s) = h_\text{fid}(\omega/c,z_s)
   \sum_{A} f_A\left(\omega/c,z_s\right) e^{-i\int_0^{z_s}\Delta\omega_A(\omega/c,z)/H(z)dz}
    \equiv
   \sum_{A} h_A \,,
\end{equation}
and the real space signal can be obtained by Fourier transforming this solution:
\begin{equation}
     \tilde{h}(z_s,\eta)=\frac{1}{\sqrt{2\pi}}\int_{0}^{\infty} d\omega \Re\left\{e^{-i\omega \eta } h(\omega/c,z_s) \right\} \,.
\end{equation}
We emphasize that by making this simplifying assumption of $\omega\approx ck$, we are describing correctly the phase shifts $\Delta\omega_{A}$ and functions $f_{A}$ only at leading beyond GR. This can, however, be improved by replacing instead the exact relations $\omega_1(k)$ and $\omega_2(k)$ for the initial conditions 
 of the first and second propagating eigenstate, respectively. 
Nevertheless, in all the examples studied in this section, the leading-order deviations
from GR are enough to describe the detected waveform accurately. 
Since the phase of each eigenstate is key to assess the waveform distortions, we further define the detected eigenstates $h_A$ as the contribution of eigenstate $A$ to the \emph{detected} signal $h$.
Note that $h_A$ are different from the propagation eigenstates $H_A$ introduced in Sec.\ \ref{sec:wavepacket} since $h_A$ carries the mixing elements from $\hat{U}$ 
as well as the amplitude corrections from the WKB and damping through $f_A$. 

We emphasize that GWs carry two tensor polarizations and thus there will be a solution like Eq.\ (\ref{EqMGh}) for each polarization, which is obtained by propagating the two polarizations emitted. The polarizations are typically cast in the + and $\times $ or $L$ and $R$ basis. For coalescing binary sources, the emitted polarization depends on the inclination between the binary's angular momentum and the line of sight. The detected strain $h_s$ by a given GW detector, will be a linear combination of the two polarizations, determined by the polarization-response of the detector, which can be generally expressed as:
\begin{equation}\label{hpolarizations}
    h_s(k, \eta)=\sum_p F_p(\theta, \phi,\psi) h_p(k, \eta), 
\end{equation}
where $F_{p}$ (with $p$ indicating the two possible polarization states) are the antenna-pattern functions that depend on angular parameters characterizing the location and orientation of the source with respect to a detector; $\theta$ and $\phi$ indicate the position of the source in the sky, and the polarization/orientation angle $\psi$ indicates the orientation of the binary system. For LIGO/Virgo-type interferometers, these functions for $+$ and $\times$ polarizations can be found e.g.\ in \cite{Ezquiaga:2020gdt}. Analogous expressions for left and right-handed polarizations can be obtained by using the relations in Eq.\ (\ref{LR_pluscross_relation}). If there is no chiral mixing, then each polarization $+$ and $\times$ propagates equally, and the total strain will be described by two propagating eigenstates. However, if there is chiral mixing, then each polarization propagates differently and in principle the total strain will be described by four propagating eigenstates: two for each left and right-handed polarization. In order to measure the polarization states individually, a global detector network of at least three interferometers is necessary.

Before considering the different mixing scenarios and the details of their phenomenology, we highlight that in our formalism the distortions of the waveform follow naturally from solving the propagation equations, and the general solution is valid for any emitted signal, irrespective of its complexity (e.g.~coming from equal or unequal-mass binaries, with orbits with precession, eccentricity, etc), and it naturally selects the group velocity as the key quantity characterizing the propagation (as opposed to the particle velocity). 
Therefore, our approach is fundamentally different to the typical GW analysis that searches for phase distortions due to a MDR in an inspiraling, circular binary using the stationary phase approximation (SPA) and the particle velocity \cite{Will:1997bb,Mirshekari:2011yq}, which has been used by LIGO--Virgo \cite{LIGOScientific:2019fpa}. We have checked that the SPA with the group velocity is a limiting case of our propagation approach.

\subsection{Main effects: echoes, phase distortions, oscillations and birefringence}\label{main_effects}

\begin{figure}[h!]
\centering
\includegraphics[width=\textwidth]{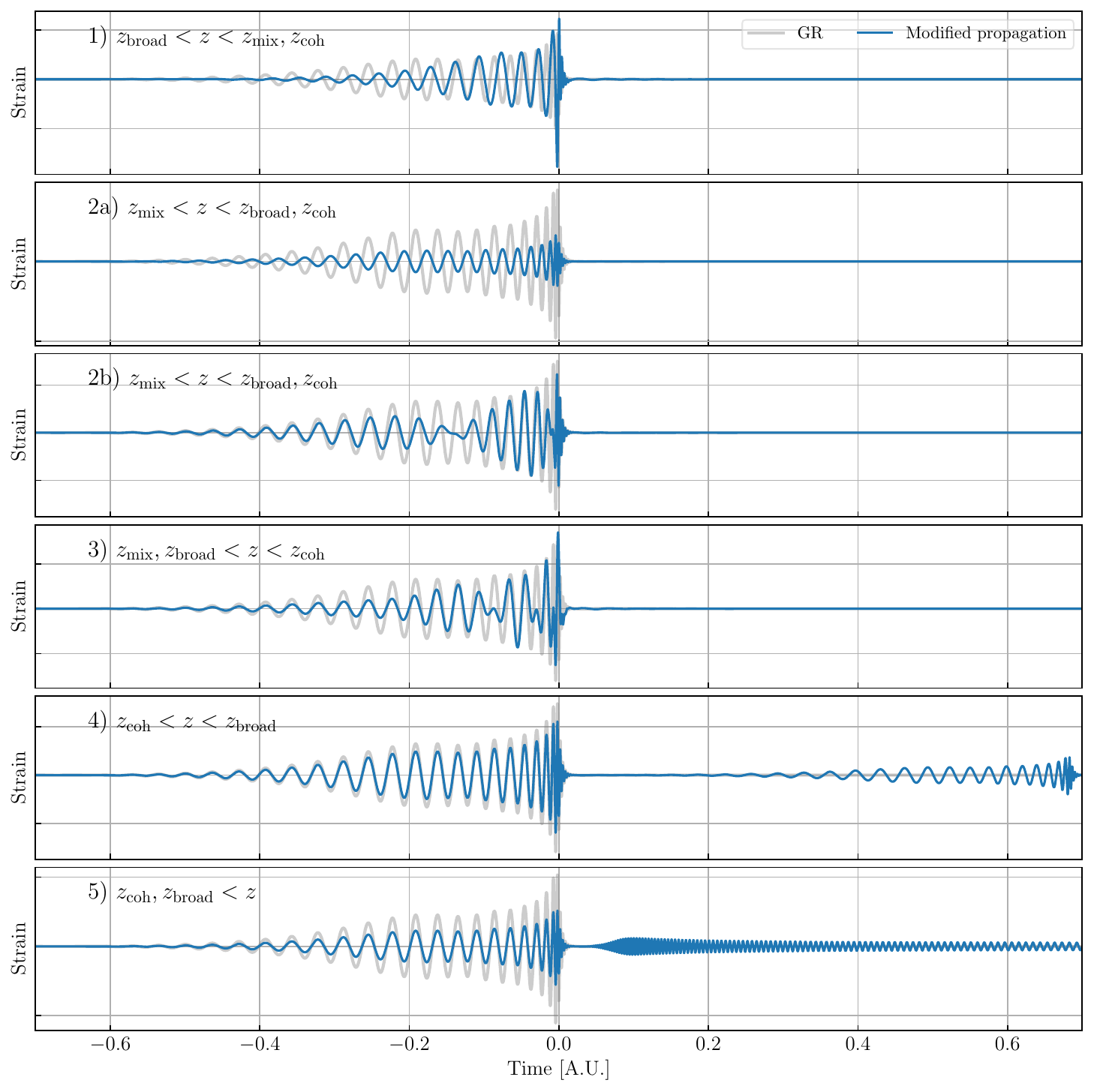}
\caption{\label{fig:summary_waveform_distortions}
Examples of the 6 different types of waveforms from a binary coalescence due to a modified GW propagation, assuming a GR emission. In grey, we show the signal that would be expected in GR, and in blue the modified signal due to a non-trivial cosmological propagation. Each panel represents a propagation epoch following Table \ref{table:summary_time2}, without showing case 0 as that is indistinguishable from GR. Specifically, each panel corresponds to: 1) single massive graviton with $m_g\sim10^{-21}\text{eV}/c^2$; 2a) friction mixing with $\nu\sim H_0$; 2b) and 3) mass mixing with $m_g\sim10^{-21}\text{eV}/c^2$; 4) velocity mixing with $\Delta v_g\sim H_0$, and 5) mass mixing with $m_g\sim10^{-21}\text{eV}/c^2$ (note that in this case $z_\text{coh}, z_{\text{broad},2}<z<z_{\text{broad},1}$). 
In all the examples we have fixed the parameters such that at least the one eigenstate propagates at the speed of light.
All these modifications could happen independently for each GW polarization leading to birefringence. Moreover, each eigenstate could be in a different propagation epoch at the same time so that, for example, only one of the echoes displays distortions.
}
\end{figure}

In full analogy with the general discussion in Sec.\ \ref{sec:GW-propag} and Table \ref{table:summary_time}, in this section 
we introduce the coherence, mixing and broadening redshifts---$z_\text{coh}$, $z_\text{mix}$, $z_\text{broad}$---which characterize different regimes in the propagation of the signal. We discuss the observational signatures that are expected in the GW signal, depending on how the source redshift $z$ compares to $z_\text{coh}$, $z_\text{mix}$, and $z_\text{broad}$. Table \ref{table:summary_time2} summarizes different observable scenarios associated to each propagation period that we also exemplify in Fig.\ \ref{fig:summary_waveform_distortions} with realistic GW waveforms from a binary coalescence that assumes the same emission of GWs as in GR, but a modified cosmological propagation. In addition, all the concrete examples presented in this section assume that the background cosmological evolution is not modified due to the presence of the additional tensor field, and hence assumed to evolve as in the $\Lambda$CDM model.

\begin{table}[h!]
\small
\centering
\begin{tabular}{ |c|c|c|}
\hline
\rowcolor{mygray} & \textbf{Regime} & \textbf{Observables}\\ 
\hline
0) & $z \ll z_\text{mix},z_\text{broad},\zcoh$ & Unmodified waveform \\ 
\rowcolor{mygray} 1) & $z_\text{broad}<z \ll z_\text{mix},\zcoh$ & Single event with modified phase evolution \\
2a) & \multirow{2}{*}{$z_\text{mix}<z<z_\text{broad},z_\text{coh}$} &  Single event with $\dLgw\not=d_L$ and constant phase shift, or\\
2b) & &  frequency-dependent amplitude modulation with phase distortions \\
\rowcolor{mygray} 3) & $z_\text{mix}, z_\text{broad}<z< z_\text{coh}$ & Single event with modified phase evolution
\\
\hline
4) & $z_\text{coh}<z<z_\text{broad}$ & Echoes with different arrival times and $\dLgw$ \\
\rowcolor{mygray} 5) & $z_\text{coh}, z_\text{broad}< z$ & Echoes with different arrival times and phase distortions  \\
 \hline
\end{tabular}
\caption{Summary of the observational effects in different propagation regimes, for a redshift source $z$. An unmodified waveform refers to one that propagates in the same way as in GR, and therefore has a dispersion relation $\omega=ck$ and its amplitude decays as $1/d_L$ for radial waves. In addition, $d_L^{GW}$ is the inferred GW luminosity distance, that may be different from $d_L$ due to the GW modified propagation. In regimes 0-3, there will be single GW event detected, whereas in 4-5 there will be multiple GW events (or echoes). The modified phase evolution and amplitude could be different for each polarization if chiral interactions are present, which leads to the additional observational effect of birefringence (not shown explicitly in this table). Note that since each eigenstate has its own $z_\text{broad}$ timescale, it could be that the various eigenstates are not in the same regime at the same time. Similarly, if there are more than two eigenstates, there may be multiple decoherence timescales, and for a given source redshift some of the eigenstates may have decohered whereas others may not. 
Moreover, whenever $z>z_\text{mix}$, there could be distortions in each waveform associated to a frequency dependent mixing angle, which may lead to additional observational signatures that, for simplicity, we do not break down in this table. 
A graphical representation of the waveform distortions associated with each of the propagation regimes is presented in Fig.\ \ref{fig:summary_waveform_distortions}.}
\label{table:summary_time2}
\end{table}

Previous works in the literature have considered some of the effects that we will discuss next in specific theories. For example, the modification of the apparent luminosity distance $\dLgw$ in gauge field dark energy theories have been studied in \cite{Caldwell:2016sut,Caldwell:2018feo}. Similarly, modification of $\dLgw$ in bigravity have been considered in the context of LIGO--Virgo \cite{Max:2017flc} and LISA \cite{Belgacem:2019pkk} observations. 
In addition, the decoherence of the wavepacket in bigravity has been explored in \cite{Max:2017kdc}. 
Lastly, a large overview of the different observational effects during the coherence regime for the different types of mixing (mass, friction, chiral) was performed in \cite{Jim_nez_2020}. 
Here, we follow a similar phenomenological approach to \cite{Jim_nez_2020} but extend their analysis discussing the decoherence regime together with the coherence period, while paying attention to the waveform distortions and phase evolution.

\subsubsection{Coherence and Echoes}
In order to understand the phenomenology of a modified propagation, the first step is to determine if the observed signal will be composed of single or multiple waveforms. Echoes of a given emitted signal can only happen when the eigenstates have different group velocities $v_{g,1}\not= v_{g,2}$. 
We define the conformal time delay between two eigenstates, for a source at $z_s$ measured by the detector today as:
\begin{align}\label{TimeDelay}
\Delta \eta (z_s)=\; & \text{min}_k[\eta_{o,\text{slowest}}(k)]-\text{max}_k[\eta_{o,\text{fastest}}(k)]\nonumber\\
\approx &\; 
 \text{min}_k\left[\eta_{e}(k)-\int_0^{z_s}\left(\frac{v_{g,\text{slowest}}(z,k)}{c}\right)\frac{dz}{H(z)} \right] \nonumber\\
 & - \text{max}_k\left[\eta_{e}(k) - \int_0^{z_s} \left(\frac{v_{g,\text{fastest}}(z,k)}{c}\right)\frac{dz}{H(z)} \right]\,,
\end{align}
where we have defined the group velocities of the fastest and slowest propagating eigenstates as $v_{g,\text{fastest}}$ and $v_{g,\text{slowest}}$. 
Here, $\eta_{o,\text{fastest}}(k)$ and $\eta_{o,\text{slowest}}(k)$ are the observed arrival times of a mode $k$ belonging to the fastest and slowest eigenstate, respectively. 
Also, the maximum and minimum functions are taken with respect to the frequency range of the GW signal that a given GW detector observes, and select the value of $k$ that maximizes and minimizes the observed arrival time. We include this because depending on the mixing model the high or low frequency modes of each eigenstate could arrive earlier or later.
The first line is such that for $\Delta \eta<0$, regardless of the propagation distortions and duration of the signal in time, the two wavepackets overlap at the detector.
In the second line of Eq.\ (\ref{TimeDelay}) we rewrite the arrival times in terms of the emitted times (which are assumed to be the same for a given $k$ value in both wavepackets) and the group velocities $v_{g,\text{fastest}}$ and $v_{g,\text{slowest}}$, and make the approximation that these two group velocities are close to the speed of light. Here, in the integral we use the
group velocities in comoving coordinates per conformal time, as we have been using throughout this paper. For a given source, such as an inspiraling binary, one can obtain the emission times between two frequency components $\eta_{e}(k_1)-\eta_e(k_2)$ from GR models of gravitational waveforms. 

Whenever $\Delta \eta(z)<0$, the wavepackets of each eigenstates are still traveling together (as they are at least partially superposing in time domain), whereas for $\Delta \eta(z)>0$ the wavepackets have a finite temporal separation among them.
Therefore, we can define the coherence redshift $\zcoh$ as:
\begin{equation}\label{zcoh}
\Delta \eta(\zcoh)=0.
\end{equation}
This quantity is the one that distinguishes between the two blocks of rows in Table \ref{table:summary_time2}. Cases $0\mbox{-}3$ correspond to a single continuous detected signal ($z<z_\text{coh}$), while $4\mbox{-}5$ display echoes ($z>z_\text{coh}$).
Note that Eq.\ (\ref{zcoh}) defines decoherence as a property of the time-domain signal. It would also be possible to distinguish the arrival of each eigenstate directly in the frequency domain. In that case, it may be possible to detect decoherence earlier than in time domain, and it will be limited by the frequency resolution rather than frequency range of the detector.

When $z>z_\text{coh}$, decoherence is achieved and multiple signals or echoes will arrive at different times, with possibly different amplitudes, polarizations, and phase evolutions. 
In order to identify these echoes with current or foreseeable future detectors, the time delay should be larger than the detector's temporal resolution, but also shorter than the span of observation projects, that is $\sim10$ years. 
Echoes can be confirmed as two events coming from the same sky location. 
The observation and characteristics of these echoes will serve to infer a number of properties about the type of mixing. For instance, the number of echoes expected depends on the number of eigenstates with different group velocities. Therefore, echo-counting could be used as a direct way to determine the number of gravitational degrees of freedom. 
For example, in theories with $N$ metric fields, the so-called multi-gravity \cite{Hinterbichler:2012cn}, there is always one massless mode and $N-1$ massive ones. Therefore, there could be up to $2N$ different echoes from the two tensorial polarizations of each metric field. 
Something similar happens in vector field theories with $N$ copies of the SU(2) group \cite{Caldwell:2016sut}. Each copy of SU(2) will have an associated effective tensor mode and thus echo.

Note that two or more similar echoes (or images) coming from the same sky location can also be produced in GR, due to strong lensing caused by inhomogeneities in the Universe. In the case of strong lensing, two images from a single source are also expected to have different arrival times, amplitudes (or magnifications), and can even have a different phase evolution as well \cite{Schneider:1992,Dai:2017huk, Ezquiaga:2020gdt, Wang:2021kzt}. However, the time delay distribution of strong lens images depends on the distribution of astrophysical objects \cite{Haris:2018vmn}, whereas for the echoes studied in this paper, it depends on the modified gravity parameters instead. 
Therefore, information about the entire population of GWs may help distinguish different time delay distributions of echoes and their origin.
In addition, in the case of modified gravity, one would expect all events further than a threshold distance to exhibit the same number of echoes, which can be searched for in the data. However, the formation of multiple images by strong lensing is an effect whose probability of occurrence grows with distance, and the number of images depends on the details of the individual lensing systems, see e.g.\ \cite{Xu:2021bfn,Mukherjee:2021qam}. Furthermore, lensing would also affect a possible EM signal from the GW source and its galaxy host, and hence one would expect multiple EM images to arrive to the observer as well \cite{Hannuksela:2020xor}, contrary to the case of modified gravity. 

Echoes may also be detected due to a modification of the GW signal during emission. This is the case for exotic compact objects (ECOs) without a horizon, where authors have shown that these objects have reflective surfaces and can therefore produce infinitely many echoes of GWs, with decreasing amplitude, and fixed time delays among them \cite{Cardoso:2016oxy,Mark:2017dnq,Micchi:2020gqy}. This prediction on number of echoes, amplitude hierarchy, and time delay distribution is fundamentally different to what is expected in the mixing models studied in this paper.

\subsubsection{Mixing} 
While in the coherence period, $z<\zcoh$, the important characteristic of the wavepacket is whether the eigenstates interfere differently after propagation or not. This can be determined computing the mixing redshift $\zmix$ as 
\begin{equation} \label{eq:z_mix}
    2\pi\sim\left\vert\int_0^{z_\text{mix}}\Delta\omega(\eta,k_*) \frac{dz}{H(z)}\right\vert\,,
\end{equation}
where, as in the previous sections, $\Delta\omega=\omega_2-\omega_1$ is the difference in the dispersion relation of the two eigenstates, and
$k_*$ is some appropriate wavenumber representative of the signal. Since generically the dispersion relation may depend on the wavenumber $k$, in practice it may be convenient to choose different values of $k_*$ in Eq.\ (\ref{eq:z_mix}) depending on the signal. In the case of inspiral binaries of compact objects, we define $z_\text{mix}$ using a value of $k$ representative for most of the duration of the signal during the inspiral, which we will choose as the minimum frequency observed by a given GW detector, $k_*=k_\text{min}$. Cases $2\mbox{-}3$ in Table \ref{table:summary_time2} display mixing.

For $z\ll \zmix,\zcoh$, case $1$, there is no effect from the additional tensor mode and the modified propagation is equivalent to that of a single tensor, which could itself still have phase deviations from GR (depending on how it compares to the broadening redshift that will be introduced next). 
When $\zmix<z<\zcoh$, cases $2\mbox{-}3$, the two propagation states interfere with each other, leading to a new net GW signal that can be scrambled compared to the original emitted signal. This can happen when both waves propagate at the same speed and hence travel together from source to observer, or when their speed difference is small enough such that they do not have time to decohere before reaching the observer. 
During this period, there could be differences with respect to GR in the phase evolution and the amplitude of the GWs, as in cases 2b-3 and 2a, respectively.  Because mixing effects depend on the source distance, observations on a population of GWs events would help constrain this coherent regime with mixing.

A special case of the regime $\zmix<z<\zcoh$ is when the mixing frequency $\Delta\omega(\eta,k)$ is nearly constant in $k$ throughout the frequency range of the detected signal. In such a case, the signal may suffer a constant phase shift and an overall rescaling in amplitude that are degenerate with simple GR waveform templates of non-precessing nearly equal-mass binaries. This corresponds to case 2a in Table \ref{table:summary_time2}, which will be discussed in more detail in subsec.\ \ref{subsec:specialcases}. 
If the GR signal contained multiple frequency components at a given time (such as from binaries with unequal masses, spin or eccentricity precession), the detected waveform in time domain will be distorted due to the constant phase shift, thus breaking the GR waveform template degeneracy. This is analogous to the constant phase shift that can appear in strong lensing in GR \cite{Schneider:1992,Dai:2017huk,Ezquiaga:2020gdt}. However, in this paper we only analyze simple waveforms from nearly-circular binaries with equal masses, in which case the aforementioned constant phase shift will be degenerate with source parameters and will lead to a GR waveform with shifted parameters. 
Note that, for a chirping GW signal, case 2a only occurs when the phase and group velocity difference between the propagating eigenstates are such that $\Delta v_{ph} \gg\Delta v_g$. Whenever the group velocity difference is comparable or larger, there will be distortions (especially near the chirp) that lead to case 2b.

\subsubsection{Broadening}
Independently of whether the signal is in the coherence or decoherence regime, waveform distortions can arise due to a dispersive (i.e.\ frequency-dependent) group velocity. Analogous to the definition of $T_\text{broad}$ in Eq.~(\ref{Tbroad_gaussian}), we define a broadening redshift $\zbroad$ for the wavepacket of each eigenmode as the redshift when the duration of the signal expected in GR $\sigma_t$ becomes comparable to the observed time delay between frequency components across the wavepacket:
\begin{equation}\label{zbroad}
    c\sigma_t \sim \left| \int^{z_{\text{broad},A}}_{0} \left(v_{g,A}(z,k_\text{max})-v_{g,A}(z,k_\text{min})\right) \frac{dz}{H(z)}\right|,
\end{equation}
where $k_\text{min}$ and $k_\text{max}$ are associated to the minimum and maximum frequencies that a detector observes from the signal\footnote{We could have again defined a slightly different condition for $z_{\text{broad}}$ specifically for temporal wavepackets, as $\int_{0}^{z_\text{broad}} (1-v_g(k_\text{max})/v_g(k_\text{min}))dz/H\sim \sigma_t$, which is expected to give a similar result as (\ref{zbroad}) in the regime we are interested in where $v_g\approx c$ and the frequency variations in $v_g$ are small.}.
Note that each eigenstate may have its own $z_{\text{broad},A}$ since they may have different group velocities $v_{g,A}$. As discussed in Sec.\ \ref{sec:GW-propag}, besides changes to the overall length/duration of a signal, a modified propagation 
also changes the phase evolution of the signal compared to GR, which will be generally given by Eq.~(\ref{DeltaPhik}). 

In contrast with the Gaussian toy-model, the duration of the observed signal $\sigma_t$ is a detector-dependent quantity since it is determined by the the overlap between the signal arriving to the observer and the sensitivity frequency range of the detectors.  
In the case of a nearly equal-mass binary coalescence of compact objects, we estimate this quantity by using the fact that the observed time interval between coalescence and the reception 
of the signal at a frequency $f_\star$ (that is sufficiently smaller than the coalescence frequency)
is approximately
\cite{Maggiore:1900zz}:
\begin{equation} \label{eq:tmerge}
    \Delta t_\text{merge} = \frac{5 t_{\mathcal{M}_z}}{(8\pi t_{\mathcal{M}_z}f_\star)^{8/3}}\,, 
\end{equation}
where $t_{\mathcal{M}_z}=G\mathcal{M}_z/c^3\simeq 5\mu\text{s}(\mathcal{M}_z/1M_\odot)$, for a binary with a redshifted chirp mass $\mathcal{M}_z$. The duration of the signal can then be approximated as $\sigma_t \simeq \Delta t_\text{merge}(f_\text{min})$, where $f_\text{min}$ is the minimum frequency a detector is sensitive to. This time-frequency relation for the evolution of binary systems may also be used to analytically obtain the emission time delay between two frequency modes, as needed for the calculation of $z_\text{coh}$ in Eq.\ (\ref{TimeDelay}). 

Cases 1, 3 and 5 of Table \ref{table:summary_time2} are all examples of waveforms with broadening. In case 1 the distortion of the waveform is purely caused by the MDR while in case 3 the total distortion is a combination of the MDR and the interference due to mixing. In case 5, each of the eigenstates could be modified due to the MDR (in analogy to case 1), as well as any frequency-dependent mixing amplitudes $f_A$, although the broadening redshifts could be different for each of them.

From the perspective of a global GW catalog/population analysis, modifications of the waveform could be seen in different ways. 
If the GW events in the catalog have all redshifts smaller than the coherence and mixing redshift of the modified gravity theory (case 1), then the population analysis could constrain a single dispersion relation valid for all the events.  
If instead the events had a $z$ such that $z_\text{broad},z_\text{mix}<z<z_\text{coh}$ (case 3), one would have to model additional time and frequency dependent amplitude corrections to the GR template.
On the other hand, if all the GW events were in the decoherence regime (case 5), echoes would lead to two sub-populations of events, each of them fitting a different dispersion relation associated to each eigenstate. In the intermediate case, when there are GW signals with $z$ above and below $z_\text{coh}$ (but still below $z_\text{mix}$), one would find that the dispersion relation of events at different redshifts changes: low-redshift would lead to a single dispersion relation and high-redshift would lead to two sub-populations with independent dispersion relations.

\subsubsection{Biased Parameter Estimation}\label{subsec:specialcases}
Here we discuss special cases of modified propagation in which individual GW signals suffer modifications that can be degenerate with GR waveform template parameters, including cases with chiral interactions that exhibit birefringence.

\paragraph{Luminosity distance and coalescence phase.}  
For signals with $z<\zbroad$, there will be no distortions of the waveform due to a dispersive group velocity, but there could still be modifications if there is mixing when $\zmix<z<\zbroad,\zcoh$ (for example case 2b). However, if the mixing frequency $\Delta\omega(\eta,k)$ is nearly constant in $k$ for the range of frequencies observed by a given detector, then the effect of interference due to $\Delta\omega$ will only introduce an overall rescaling of the amplitude and frequency-independent phase shift of an individual GW signal. For a population of GW events, since the model parameters will typically evolve on cosmological timescales, mixing will lead to a redshift-dependent phase shift and oscillatory pattern in their amplitudes. 

However, for an individual event, a frequency-independent phase shift will be degenerate with the coalescence phase parameter of GR waveform models (which indicates the location of the compact objects in the orbit at the moment of the merger) for the simple GW signals analyzed in this paper (those dominated by $\ell=|m|=2$ spherical harmonic multipoles), as shown in \cite{Ezquiaga:2020gdt}. 
In addition, the overall change in amplitude of the detected signal is degenerate with the luminosity distance of GR waveform models. Therefore, an analysis of these signals with modified propagation will lead to a biased inferred luminosity distance and coalescence phase, when using GR waveform templates. This is because, from a data analysis perspective, these parameters are constrained via parameter estimation (PE) given a template bank. 

On the contrary, if $\Delta\omega(\eta,k)$ varies in $k$ considerably across the detected signal, then interference will lead to additional phase distortions (even for simple waveforms) that cannot be generally mimicked by a change in GR waveform parameters (for example in cases 1 and 3).
In general, a degeneracy between detected individual GW signals and GR waveform templates can only happen in regimes where waveform distortions are either suppressed or mimicked by a change in GR waveform parameters (for example for case 2a with mixing, and case 4 without mixing).  

More generally, a modified GW propagation can introduce waveform distortions, which means that the inference of $\dLgw$ (and other waveform parameters) is subject to bias if the PE is not performed using beyond GR waveforms. 
Assuming that the right waveforms are being used for the amplitude and phase evolution (otherwise some unknown bias in any waveform parameter could be introduced), we can then define the modification in the $\dLgw$ from the envelope of the GW waveform, tracking the maximum amplitude at the coalescence frequency. 
The change of the luminosity distance will have a direct impact on GW observations. For a single event, if no EM counterpart is observed, a bias in $\dLgw$ will lead to a biased inferred source redshift (and source masses of the binary), or a biased Hubble rate today $H_0$. These degeneracies can however be partially broken if an EM counterpart counterpart is observed, which can constrain directly the source redshift $z$.  
GW170817 constituted the first such measurement and led to constraints in $\dLgw/d_L$ at $z\simeq0.009$, see e.g.\ \cite{Arai:2017hxj,Lagos:2019kds} for a given $H_0$. 

On the other hand, when $\zcoh<z<\zbroad$, there could be a constant rescaling in the decoherence regime, simply because we will be observing a single eigenmode that only contains partial energy of the total signal (case 4).
This again would be completely degenerate with GR waveform templates, by either shifting the Hubble rate today $H_0$ or the source redshift and masses. 
Therefore, if only sources at $z>z_\text{coh}$ are observed, the mixing could not be disentangled, unless external observations that fix the value of $H_0$ or $z$ were used in combination.

Even if individual events with modified propagation due to mixing fit well GR waveform templates, a population analysis would show a source-redshift oscillation of the GW luminosity distance that would hint to non-trivial physics. This is because if the amplitude of the GW changes with redshift, the probability of detecting a source at a given redshift and mass also changes. This, as a consequence, affects the rates of events and their observed mass distribution \cite{Ezquiaga:2021lli}, with an oscillatory pattern that could be potentially identified as modified gravity. 
However, it is important to note that there are still many uncertainties about the merger rate history of binary black holes. Therefore, this intrinsic astrophysical uncertainties would limit the capabilities to test modifications of the GW propagation from a population perspective. 

In the case of GW events that mimic GR waveforms and that are detected in the decoherence regime, all the individual events would have a constant rescaling of the amplitude. However, the two echoes will generally have different amplitudes. If only the echoes with largest amplitudes are detected  (e.g.\ if the other echo has a small amplitude under the signal-to-noise threshold of detectability, or if it has distortions that do not allow to make a detection with current match-filtering techniques), then the population of events with  $z>z_\text{coh}$ would be degenerate with a re-scaling of the local merger rate, and GW data alone could not distinguish this scenario from GR. 
However, if both echoes are detected, the GW events would appear in pairs that have the same phase evolution, and all detected pairs would have the same amplitude ratio, and one could use that information to identify the effect of modified gravity.

\paragraph{Polarization.} All the previously discussed regimes could apply separately to left and right-handed polarizations, if there is a chiral cosmological interaction. In general, since a total signal could contain both types of polarizations, the signal will be composed of four eigenstates, which may have various decoherence timescale, e.g.\ the signal may decohere into two wavepackets first, and then into four (each one of the four being only purely left or right handed). The different behavior of the left and right-handed polarizations means that the signal will change its polarization from emission to detection, leading to additional changes to source PE. The change in the polarization is characterized by the angles $\beta$ and $\chi$ that define the degree of circular polarization, and the orientation of the polarization, respectively. If these angles $\beta$ and $\chi$ are nearly independent of $k$ for the range of frequencies observed by a given detector, then a change in them during propagation will again lead to GR-like waveform with biased source parameters; the orientation $\psi$ and inclination $\iota$ of the source. This polarization content can be measured with multiple detectors. The detailed observational consequences of birefringence are model dependent (e.g.\ whether both $\beta$ and $\chi$ will vary during propagation, or only one of them), and must be analyzed case by case.

In addition to individual events, chirality could be tested in the entire population of GWs if the polarization has a special distribution. For instance, propagation could enhance the amplitude of one of the two polarizations, leading the events to exhibit preference for one specific polarization (e.g.~in dynamical Chern-Simons theories \cite{Alexander:2007kv, Okounkova:2021xjv}). Another example is if the left and right-handed eigenstates have different propagation speeds, in which case all the events with $z>z_\text{coh}$ would lead to echoes that are purely right and left handed. 
For LIGO/Virgo, there have been tests of chirality to check if sources tend to have mostly right-handed or left-handed polarization (which translates into biases of the source parameters that show preference for face-on or face-off binaries) \cite{Okounkova:2021xjv}, and this has been confirmed not to be the case. Since these propagation effects accumulate over distance, chirality may not be visible for present GW detectors (nearby sources), but may be for the next generation of GW detectors. One can also test the chirality of primordial GWs via the CMB polarization \cite{2010PhRvD..81l3529G,Gerbino:2016mqb,Thorne:2017jft,Bartolo:2018elp}.

\subsection{Velocity mixing}\label{sec4:vel}
Here we generalize the model with velocity mixing given in Eq.\ (\ref{velocity_eom}), by allowing a cosmological time dependence in the coefficients of the EoM, and applying the resulting modified propagation to a realistic GW waveform. The equations considered are:
\begin{equation}\label{velocity_eom_with_time}
\left[\hat{I} \frac{d^2}{d\eta^2} +\bpm c_h^2(\eta) & c_{hs}^2(\eta) \\ c_{hs}^2(\eta) &  c_s^2(\eta)\epm k^2  \right] \bpm a(\eta)\cdot h(\eta,k) \\ a(\eta)\cdot s(\eta,k) \epm =0,
\end{equation}
which are assumed to hold for the renormalized fields $a\cdot h$ and $a\cdot s$, as it is typically the case when the Hubble expansion of the universe is taken into consideration for specific gravity theories. In such cases, when this renormalization is performed, the presence of diagonal mass-like terms appear, which are of the order of the conformal  Hubble rate $\mathcal{H}$ squared and its time derivative ($\mathcal{H}^2=a^2 H^2$ and $d\mathcal{H}/d\eta$) and thus are neglected here when compared to the velocity matrix in the large-$k$ limit.

We start by obtaining the analytical solution of $h$, using the WKB formalism. At zeroth order in the WKB approximation, we obtain the eigenfrequencies which 
are of the form $\omega \propto k$, with the same expressions as those given in Eq.\ (\ref{vel_mixing_vg1})-(\ref{vel_mixing_vg2}). Since these dispersion relations have the same form as that in GR, velocity mixing is the simplest type of interactions because there is no distortion associated to the dispersion relation of the propagating eigenstates. 
At linear order in the WKB approximation, the calculation of $h$ simplifies greatly if all the velocity parameters $c_h$, $c_s$ and $c_{hs}$ have the same time dependence.
In this case, the propagation matrix $\hat{U}$ is constant and we find that $\hat{A}_\text{WKB}=\tM^{-1}\tM'/2$, which can be integrated exactly to obtain a simple expression for the general solution of $h$ in Eq.\ (\ref{hsolWKB}).
The final solution for $h$ in Fourier space, assuming only $h$ is initially emitted, can then be written as
\begin{align} \label{eq:amplitude_vel_mixing}
h_{+,\times}(k,\eta)&=
  h_{\text{fid}+,\times}(k,\eta)\left[ f_1(\eta)e^{-i\int_{\eta_e}^{\eta}k\left(v_{g,1}(\eta)-c\right) \d\eta'} + f_2(\eta) e^{-i\int_{\eta_e}^{\eta}k\left(v_{g,2}(\eta)-c\right) \d\eta'}\right]\,, \\
 & f_1(\eta)=\frac{\cos^2\tg}{\sqrt{v_{g,1}(\eta)/v_{g,1}(\eta_e)}}, \quad f_2(\eta)=\frac{\sin^2\tg}{{\sqrt{v_{g,2}(\eta)/v_{g,2}(\eta_e)}}},
\end{align}
where $\eta$ is the cosmological conformal time of detection, $\eta_e$ the time of emission, and $h_\text{fid}$ will be assumed to be given by GR. Here, the group velocities $v_{g,A}$ are the same as the phase velocities obtained from Eq.\ (\ref{vel_mixing_vg1})-(\ref{vel_mixing_vg2}), and the mixing angle $\Theta_g$ has the same expression as in Eq.\ (\ref{eq:tg_vel_mixing}). In this type of mixing, both polarizations $+$ and $\times$ propagate in the same way. 
From Eq.\ (\ref{eq:amplitude_vel_mixing}) we see that the propagation and phase evolution of each individual eigenstate is trivial since their phase corrections are linear in $k$ and thus correspond to a shift in the overall arrival time of the signal. In addition, the coefficients $f_{A}$ are simple rescalings of the total amplitude of the signal, which do not distort the waveform since their timescale evolution is much larger than the duration of the GW signal detected. 
Nevertheless, there can still be distortions due to interference during the coherence regime where the full signal can be re-expressed as:
\begin{equation}
    h_{+,\times}(k,\eta)=
  h_{\text{fid}+,\times}(k,\eta)\cos^2\tg \sqrt{1  +\tan^4\tg+2\tan^2\tg\cos\left(\Delta\phi_I\right)} e^{-i\theta'}e^{-i\int_{\eta_e}^{\eta}k\left(v_{g,1}(\eta)-c\right) \d\eta'} \,,
\end{equation}
where $\Delta\phi_I=\int k(v_{g,2}-v_{g,1}) d\eta$, and $\theta'$ is defined such that $\tan\theta'=\tan^2\tg\sin(\Delta\phi_I)/(1+\tan^2\tg\cos(\Delta\phi_I))$. Here, we have assumed that $v_{g,A}(\eta)/v_{g,A}(\eta_e)\approx 1$ which is valid for low-redshift sources (such as those observed by current GW detectors) or when the velocities evolve slow enough in time.
Since $\Delta\phi_I$ depends linearly on $k$, a GW signal spanning a wide range of frequencies will be expected to have frequency-dependent amplitude and phase corrections during the coherence regime.

\paragraph{Time scales.}
Since the group velocities of the two propagating eigenstates are in general different, the signal is expected to eventually reach decoherence. Furthermore, since the group velocity is frequency independent and there are no distortions of the signal in each eigenstate, the coherence redshift is simply achieved when the time delay between the eigenstates is comparable to the duration of the original emitted signal
\begin{equation}\label{zcoh_vel}
    \Big\lvert \int_0^{z_\text{coh}}\frac{\Delta v_g}{c}\frac{dz}{H(z)}\Big\rvert\sim\sigma_t\,.
\end{equation}
Then, for $z_\text{mix}<z<z_\text{coh}$, velocity mixing will correspond to case 2b of Table \ref{table:summary_time2}, with distortions in the waveform only associated to the interference between eigenstates. After decoherence, $z>z_\text{coh}$, the signal will be in case 4, with non-distorted echoes with different amplitudes.

Since in general the sound speeds of the eigenstates will be different from the speed of light, one needs to be careful to satisfy the constraints on the speed of GWs $c_g$ from GW170817 \cite{Monitor:2017mdv}, which set $|c_g/c-1|\lesssim 10^{-15}$ at $z\sim0.009$. 
On the one hand, if $c_h\neq c$, then one can fine tune the mixing so that the sound speed of the first eigenmodes is exactly equal to the speed of light today, $v_{g,1}(z=0)=c$, i.e.
\begin{equation}\label{chs_tune}
    c_{hs}^2=\sqrt{c^2-c_h^2}\sqrt{c^2-c_s^2}\,.
\end{equation}
On the other hand, if $c_h(z=0)=c$, one can have $v_{g,1}(z=0)\approx c$ by suppressing the mixing parameter $|c_{hs}^2/\dc^2|\ll 1$
(see Eq.\ (2.73) in \cite{Jim_nez_2020}). 
Note that in all the theories of tensor interactions summarized in \cite{Jim_nez_2020} (bigravity, Yang-Mills, gaugid, multi-Proca), only the Horndeski Yang-Mills interaction \cite{BeltranJimenez:2017cbn} term $L^{\mu\nu\alpha\beta}F^a_{\mu\nu}F^a_{\alpha\beta}$, where $L^{\mu\nu\alpha\beta}$ is the double dual Riemann tensor, introduces $c_{hs}^2\neq0$. All other cases lead to $c_{hs}=0$, in addition to $c_h=c$ and, sometimes, $c_s\neq c$. For that reason we will not consider together velocity mixing with other types of interactions in the following sections. 

In the case in which the mixing parameter is fine tuned according to Eq.\ (\ref{chs_tune}), one may obtain that the fractional difference in the group velocities is of $\mathcal{O}(1)$, and thus the time delay between echoes is itself cosmological, for cosmological sources. 
Therefore, interesting cases, where both echoes can be detected, are those in which $c_h^2+c_s^2\approx 2c^2$. In this limit, we find that:
\begin{equation}
    \Delta v_{g}=v_{g,2}-c\approx \frac{c_h^2+c_s^2-2c^2}{2c}\,,
\end{equation}
which can itself be tuned to give a time delay of the order of the duration of the signal when $\Delta v_g/c\sim \sigma_t H_0$. This is indeed the case exemplified in the panel 4) of Fig.\ \ref{fig:summary_waveform_distortions}, where for simplicity we assumed that the velocity matrix is time independent although the generalization is straightforward (only the arrival times of the echoes would change). 
In a more generic case, the second echo could arrive well outside any realistic observational window. 

For this mixing scenario the mixing redshift is given by the same expression of the coherence redshift (\ref{zcoh_vel}) if one substitutes $\sigma_t\to 2\pi/ck$. On the other hand, 
since there is no MDR introducing waveform distortions, $z_\text{broad}\rightarrow \infty$ for both echoes.  

\begin{figure}[h!]
\centering
\includegraphics[width=\textwidth]{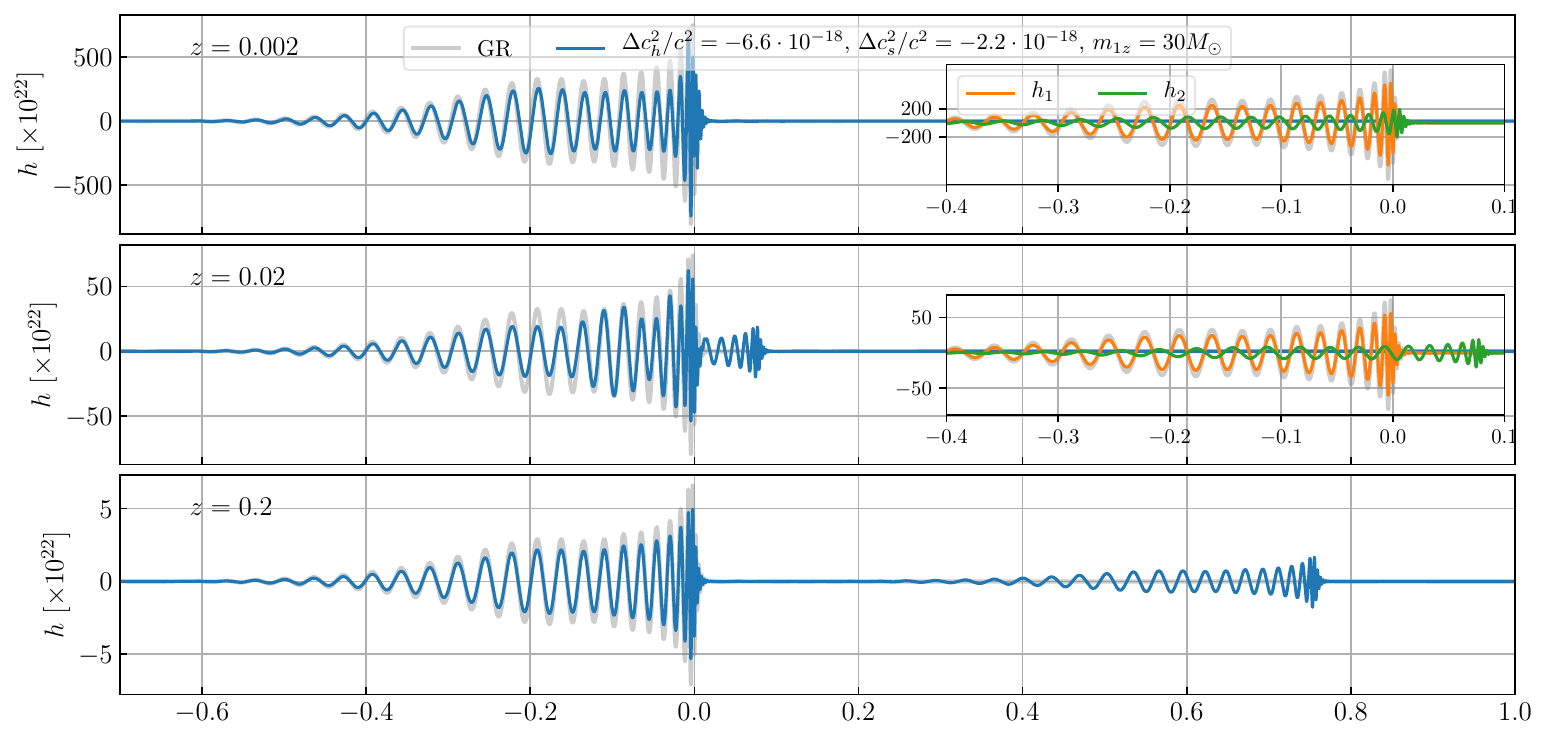}
\caption{\label{fig:distortion_velocity_mixing}
GW signal observed by ground-based GW detectors, coming from a non-spinning equal-mass black hole binary system. The three panels show the distortions of the signal suffered during propagation in a \emph{velocity mixing} model, in the case of sources with three different redshifts. At the top, we have a nearby source such that the two propagating eigenstates are in the fully coherence regime, whereas the bottom panel shows a further source such that both eigenstates are in the fully decoherence regime. In the top two panels we also show the two independent eigenstates in insets. $\Delta t =0$ is set by the arrival time of the frequency of coalescence traveling at the speed of light. The mixing is chosen so that $v_{g,1}=c$.}
\end{figure}

\paragraph{Waveform distortions.}
We now illustrate the different propagation regimes and distortions that a GW signal will exhibit due to velocity mixing.
Fig.\ \ref{fig:distortion_velocity_mixing} shows the propagation of GWs with constant velocity parameters given by $\Delta c_h^2 = c_h^2-c^2=-6.6\cdot10^{-18}c^2$, $\Delta c_s^2 = c_s^2-c^2=-2.2\cdot10^{-18}c^2$ and $c_{hs}^2 = 3.8\cdot10^{-18}c^2$, so that the eigenmodes have group velocities given by $v_{g,1}=c$ and $v_{g,2}-c=-4\cdot10^{-18}c$. We compare the modified GW signal (blue curves) and the waveform expected in GR (grey curves), when assuming that the emitted signal is the same as that in GR. This figure illustrates a waveform emitted by an equal-mass non-spinning binary black hole with redshifted component mass $m_{z}=m_{1z}=m_{2z}=30M_\odot$. 
We will use this example binary in the rest of the paper as it represents a typical detection, but our formalism is insensitive to the initial signal. 
Binaries with different masses can be more or less constraining for different types of mixing depending if the modifications are enhanced at low or high frequencies. 
In addition, we choose the binary to be  overhead ($\phi,\theta,\psi=0$) so that $F_+=1$ and $F_\times=0$, and assume it to be face-on, with inclination angle $\iota=0$, so that $|h_+|=|h_\times|$. The detected strain seen by such a single detector is then simply $h_s=h_{+}$, which is calculated using the fiducial detected signal $h_{\text{fid}+}$ obtained from the phenomenological GR model \texttt
{IMRPhenomD} \cite{PhysRevD.93.044006}. We plot the waveform that would be detected by a LIGO-type ground-based single detector, when the source is located at redshifts $z=0.002$, 0.02 and 0.2. These redshifts are chosen to represent different propagation epochs: coherence, transition to decoherence, and decoherence. 
For this example the mixing redshift is $z_\text{mix}\sim0.0009$ and the coherence redshift $z_\text{coh}\sim0.15$.
In all the panels, we choose the arrival time such that $\Delta t=0$ is the arrival time of the coalescence frequency of the GR waveform.

In the top panel of Fig.\ \ref{fig:distortion_velocity_mixing} we show the waveform during the mixing regime for a source with redshift $z=0.002$, which exhibits distortions in the amplitude and phase evolution of the signal due to interference, and thus represents an example of case 2b in Table \ref{table:summary_time2}. In the inset we show separately the detected eigenstates $h_1$ and $h_2$. For the parameters chosen here the mixing angle is $0.17\pi$ and therefore the amplitude of $h_2$ is smaller than that of $h_1$. Note that each one of these individual eigenstates has the same phase evolution as the GR one, but since they propagate at different speeds, at a given time there will be two different $k$ modes arriving (one from $h_1$ and one from $h_2$) which will lead to the distortions observed in the net signal.
At $z=0.02$, we see that the two eigenstates are transitioning to decoherence, and distortions of the waveform are clearly visible. Here we can explicitly see the time delay of the eigenmode $h_2$ due to the fact that it propagates slower than the speed of light. Finally, at $z=0.2$ we see the decoherence regime where two separate echoes are detected, each one with the same phase evolution as the GR signal, but with a different overall amplitude. This last panel is in the same regime as the one illustrated in case 4 of Fig.\ \ref{fig:summary_waveform_distortions}.

\paragraph{Biased parameter estimation.}
As previously discussed, in the coherence regime the signal is generally distorted due to frequency-dependent amplitude and phase corrections. However, in the decoherence regime both eigenstates have the same phase evolution as the originally emitted signal, with only an amplitude difference. 
In particular, each echo will have a different amplitude determined by the mixing angle $\Theta_g$ such that the apparent GW luminosity distance will be different from the true one by: 
\begin{equation}
    \left.\dLgw\right\vert_{1}\to d_L/\cos^2\tg\,,\qquad \left.\dLgw\right\vert_{2}\to d_L/\sin^2\tg\,,
\end{equation}
for the first and second echo, respectively.
On an individual event analysis, this constant rescaling will be fully degenerate with the source redshift or $H_0$. On a population analysis, if only the echoes with the highest amplitudes are detected then their change in amplitude will be degenerate with the local merger rate $\mathcal R_0$. However, if both echoes are detected, the merger rate as a function of redshift will suffer non-trivial effects that must be analyzed in detail in the future.

\subsection{Mass mixing}\label{sec4:mass}

We generalize the mass mixing example studied in subsec.\ \ref{sec:massmixing}, by including a cosmological time dependence in the coefficients of the EoM, and applying the resulting modified propagation to a realistic GW waveform.
Inspired by the cosmological behavior of massive bigravity in the late-time universe, we consider the following model:
\begin{equation}\label{MassMixing_with_time}
\left[\hat{I} \left( \frac{\d^2}{\d\eta^2}+ 
(ck)^2 \right) + a(\eta)^2c^4\bpm m_h^2 & m_{hs}^2 \\ m_{hs}^2 & m_s^2 \epm \right] \bpm a(\eta)\cdot h(\eta,k) \\ a(\eta)\cdot s(\eta,k) \epm =0,
\end{equation}
where the parameters $m_h$, $m_s$ and $m_{hs}$ are assumed to be constants, and satisfy the condition $m_{hs}^4=m_h^2m_s^2$. Similarly to the velocity mixing example, we have assumed that these equations hold for the renormalized fields $a\cdot h$ and $a\cdot s$, and neglected the mass-like terms of order of Hubble rate when compared to the mass terms $m_h$, $m_s$, and $m_{hs}$.

In this example, since all the mass mixing terms have the same time dependence, the structure of the WKB solution at lowest order remains the same as in the constant coefficient examples of Section \ref{sec:massmixing}. Indeed, following the description of Section \ref{WKBsubsection}, we find the eigenfrequencies $\omega_{A}$ and the mixing matrix $\hat{E}_{(0)}$ to have the same expressions as those found in Eq.\ (\ref{omega1_bigravity})-(\ref{omega2_bigravity}), rescaling all the masses with the scale factor. In particular, we find that $\omega_1^2=c^2k^2$ and it propagates in the same way as in GR, whereas $\omega_2^2=c^2k^2+ a^2c^4m_g^2$, with $m_g^2=m_s^2+m_h^2$. Therefore, the second eigenstate will have a non-trivial dispersive group velocity given by
\begin{equation}\label{vg_massivegravity}
    \frac{v_{g,2}}{c}\approx 1-\frac{a^2c^2m_g^2}{2k^2}\,
\end{equation}
in the large-$k$ limit.

At linear order in the WKB formalism, the calculation of $h$ simplifies since $\hat{A}_\text{WKB}=\tM^{-1}\tM'/2$ can be integrated exactly (see also \cite{Jim_nez_2020}). 
The total solution for $h$ in Fourier space, assuming only $h$ is initially emitted, can then be written as
\begin{align} \label{eq:amplitude_mass_mixing}
h_{+,\times}(k,\eta)&=
  h_{\text{fid}+,\times}(k,\eta)\left[ f_1 + f_2(k,\eta) e^{-i\int_{\eta_e}^{\eta}\Delta\omega_2(\eta',k) \d\eta'}\right]\,, \\
 & f_1=\cos^2\tg, \quad f_2(k,\eta)=\frac{\sin^2\tg}{\sqrt{\omega_2(k,\eta)/\omega_2(k,\eta_e)}},\label{f12_massmixing}
\end{align}
where the mixing angle $\Theta_g$ is constant in this case and given by $\tan^2\Theta_g=(m_{hs}/m_s)^4$, and each polarization $+$ and $\times$ propagates in the same way. Therefore, the mixing angle can be considered as a parameter independent from the mass parameter $m_g$ in the $\omega_2$ dispersion relation.
Since we will assume that deviations from GR are small, then $\omega_2(k,\eta)/\omega_2(k,\eta_e) \approx 1+c^4m_g^2/(ck)^2(a^2-a_e^2)$, and we thus find that for each eigenstate there are negligible distortions due to deviations from GR in the amplitude if $(c^2m_g)/(ck)\ll 1$. In other words, the functions $f_{1,2}(k,\eta)$ in Eq.\ (\ref{f12_massmixing}) are either exactly (for the first eigenstate) or approximately constant (for the second eigenstate) in $k$ and $\eta$, and the main distortions come from the dispersion relation in the phase and the interference of both waves. 

\begin{figure}[t!]
\centering
\includegraphics[width=\textwidth]{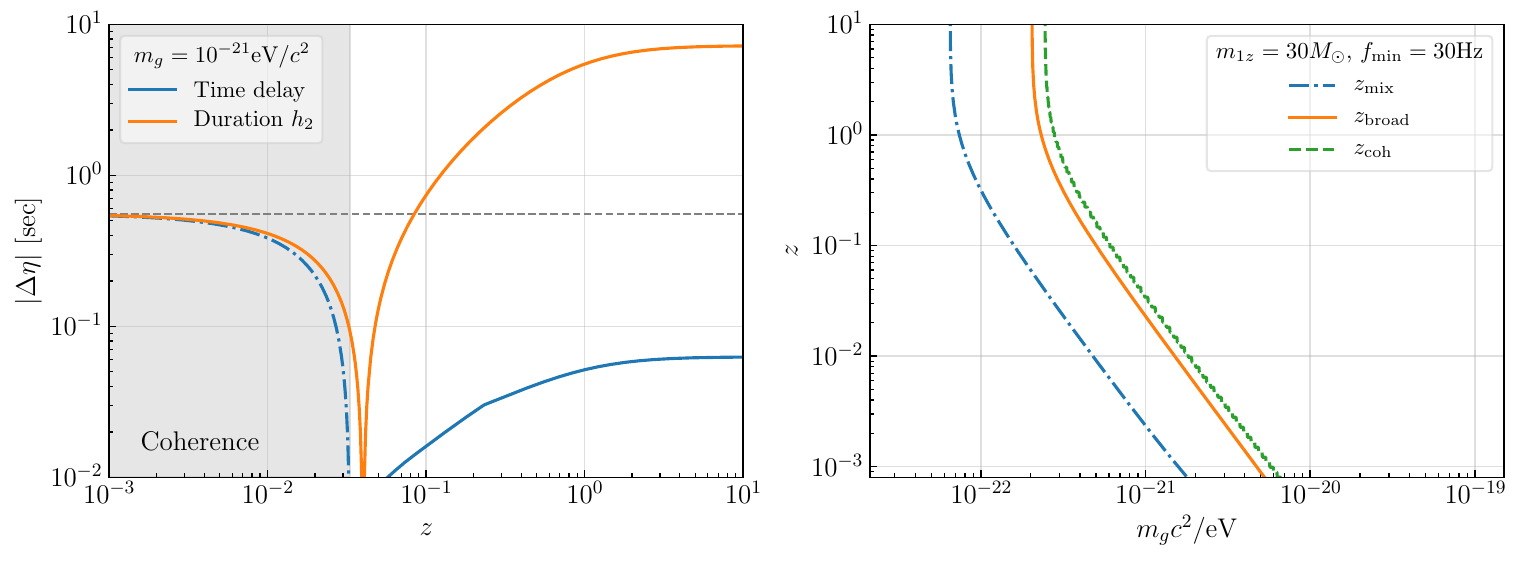}
\caption{\label{fig:dt_mass_mixing}
On the left, time delay between the propagation eigenstates in our example mass mixing theory for a non-spinning, equal mass binary with redshifted chirp mass $\mathcal{M}_z=26.1M_\odot$ as a function of the source redshift $z$. The blue line correspond to the time delay as a function of redshift. The solid line represents when the time delay is positive and the dashed-dotted line when it is negative. The dashed gray line indicates the duration of the original signal. For reference, we also display in orange the duration of the massive eigenstate. Echos are produced when $\Delta \eta>0$, in the region which is not shaded in gray. 
On the right, comparison of the mixing time scale $z_\text{mix}$, the broadening $z_\text{broad}$ and coherence $z_\text{coh}$ over a wide range of masses. We see that distortions like broadening of the second eigenmode become relevant before the two eigenstates reach the decoherence regime.
}
\end{figure}

\paragraph{Time scales.}
In this model, since the first eigenstate propagates as in GR, the time delay between the two echoes is controlled by the mass term of the second eigenstate. From Eq.\ (\ref{TimeDelay}) we can obtain the time delay at any given redshift. 
After decoherence is achieved ($\Delta \eta >0$), we obtain that
\begin{equation}\label{TimeDelay_MassiveBigravity}
    \Delta \eta (z>z_\text{coh}) \approx  \frac{c^2m_g^2}{2 k^2_\text{max}}\int_0^{z}\frac{dz}{(1+z)^2H(z)}\,,
\end{equation}
where deviations from
GR are assumed to be small. Here only the time delay from the dispersive velocity $v_{g,2}$ is present, since the $k$ values chosen for $ \text{max}_k[\eta_{o,\text{fastest}}(k)]$ and $\text{min}_k[\eta_{o,\text{slowest}}(k)]$ are both the same, given by the maximum frequency observed in the waveform, $k_\text{max}$.  
In the left panel of Fig.~\ref{fig:dt_mass_mixing} we present the result for a typical LIGO/Virgo 
source with a fixed mass $m_g=10^{-21}$eV$/c^2$.  
In this example, the time delay at LIGO/Virgo frequencies is always smaller than 1 second. 
Other parameter choices can be directly inferred using the $m_g^2$ scaling of $\Delta \eta$ in Eq.\ (\ref{TimeDelay_MassiveBigravity}). 
Therefore, the observed time delay contains physical information about the mass of the graviton. We note, however, that for sufficiently large $m_g$, the time delay can be longer than the observation time of a given detector, in which case the second signal may never be observed.\footnote{Similarly, during a given observational campaign detectors are not always online. In general, constraints using the echo time delays should take into account the duty cycle.} 
It is to be noted that, as shown in Fig.~\ref{fig:dt_mass_mixing}, the duration of the second eigenstate is parametrically larger than the time delay. This will also make a GR template-based detection of the second eigenstate for large masses difficult since the signal would be completely distorted. 
Noticeably, the time delay saturates for high redshifts of the source. In particular, for a flat $\Lambda$CDM cosmology with fractional dark matter energy $\Omega_{m,0}\approx 0.31$ we obtain that:
\begin{equation}\label{Max_td_mass_mixing}
    \Delta \eta (z\rightarrow \infty) \approx 0.26 \frac{1}{H_0} \frac{c^2m_g^2}{k^2_\text{max}}\,.
\end{equation}
This is because, in this example, the difference in group velocities between the two modes approaches zero for $z\rightarrow \infty$ due to the redshifting masses $a m_h$ and $a m_s$ in comoving coordinates. As it can be seen in the left panel of Fig.~\ref{fig:dt_mass_mixing}, the time delay may saturate at values smaller than the duration of the signal, which implies that unless the second mode is suppressed by the mixing angle, one should have already seen these echoes in the LIGO/Virgo data for this parameter value.

It is also useful to compare the broadening and coherence timescales, in order to identify if the distortions happen when the two eigenstates are propagating coherently or incoherently. We plot
$z_{\text{broad},2}$ (since that is the only one that exhibits distortions) according to Eq.\ (\ref{zbroad}) as well as $z_\text{coh}$, which corresponds to the limiting case when $\Delta \eta = 0$.\footnote{Our calculation for $z_\text{coh}$ in bigravity generalizes the $L_\text{coh}$ of \cite{Max:2017kdc} to cosmological set-ups.} 
The results for massive gravity are shown in the right panel of Fig.\ \ref{fig:dt_mass_mixing}, where we see that  $z_{\rm broad}<z_{\rm coh}$ for any value of $m_g$, confirming that in this model broadening will always happen before decoherence is reached. In other words, case 4 of Table \ref{table:summary_time2} will never happen in this model.
We therefore expect considerable phase distortions of the total GW signal during coherence, as we will confirm next. 
For completeness we also present the mixing redshift, evaluated at $k_\text{min}$, which determines when the two eigenstates begin interfering during the inspiral.

We emphasize that here we have exemplified the timescales for GWs observed by LIGO-type detectors. However, since propagation distortions depend on the frequency, different detectors may be more sensitive to these modified gravity effects. For instance, in this example of massive bigravity where the group velocity of the second eigenmode is inversely proportional to the wavenumber $k$, then low-frequency GWs detectors like LISA are expected to constrain better the parameter $m_g$. In addition, next generation GW detectors will be more sensitive, and thus make more precise observations and detect objects that are further away so that propagation effects accumulate more. For these two reasons, next generation detectors are expected to constrain modified gravity better by a few orders of magnitude. This has been found to be the case for the propagation distortions for a single massive graviton, where it has been found that the constraints on $m_g$ are expected to improve at least by 2 orders of magnitude for a single event detected with third-generation detectors \cite{Chamberlain:2017fjl,Perkins:2018tir}, compared to current LIGO constraints \cite{Abbott:2020jks}. Combining population observations may increase these constraints further by other couple of orders of magnitude, since third-generation ground-based detectors like Einstein Telescope and Cosmic Explorer are expected to detect more than $10^4$ binary black holes per year \cite{Kalogera:2019sui}.

\begin{figure}[h!]
\centering
\includegraphics[width=\textwidth]{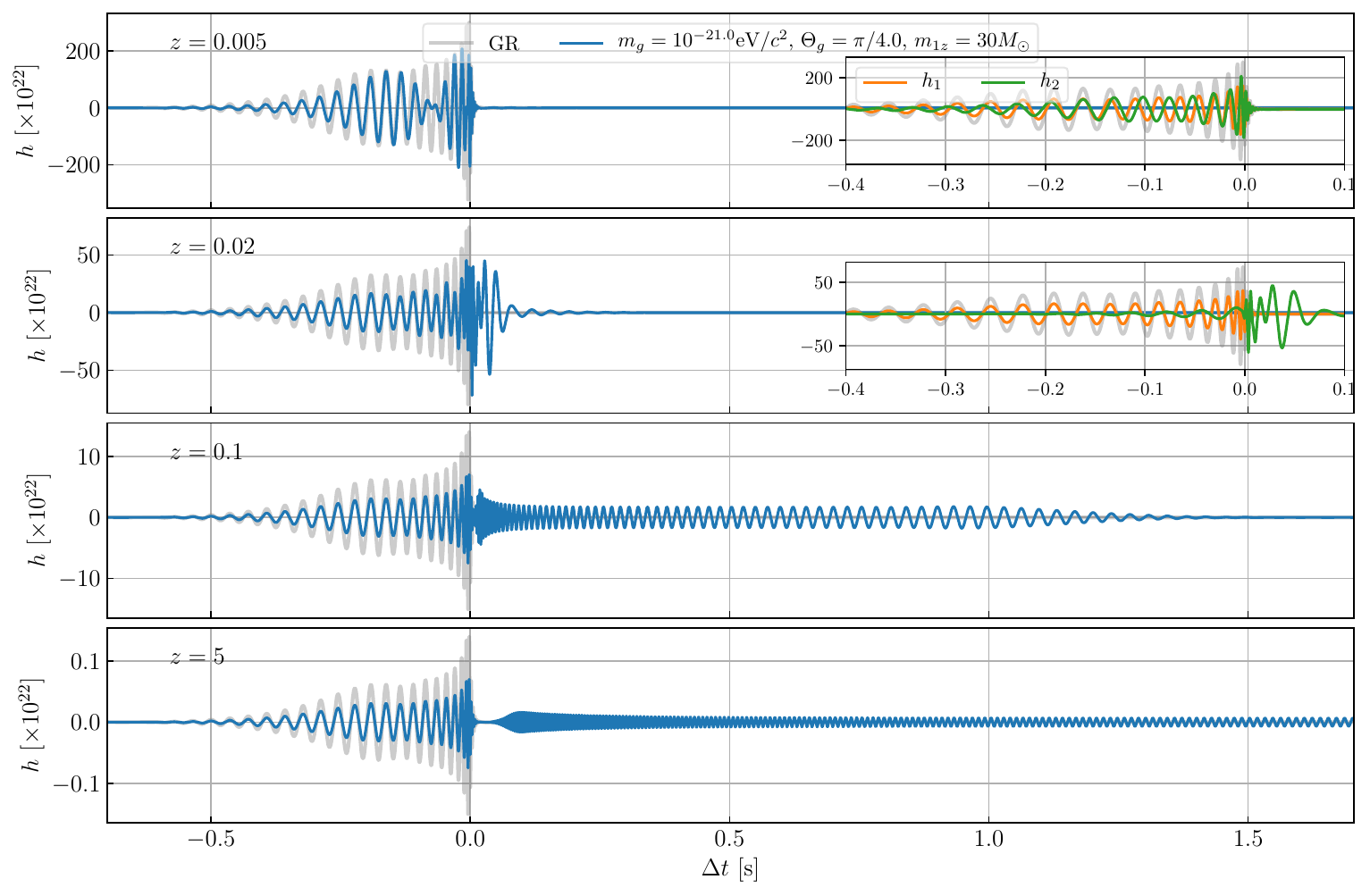}
\caption{\label{fig:distortion_mass_mixing}
GW signal observed by ground-based GW detectors, coming from a non-spinning equal-mass black hole binary system. The four panels show the distortions of the signal suffered during propagation in a \emph{mass mixing} model, in the case of sources with four different redshifts. At the top, we have a nearby source such that the two propagating eigenstates are in the fully coherence regime, whereas the bottom panel shows a far away source such that both eigenstates are in the fully decoherence regime. In the top two panels we also show the two independent eigenstates in insets. $\Delta t =0$ is set by the arrival time of the frequency of coalescence traveling at the speed of light.}
\end{figure}

\paragraph{Waveform distortions.} Next, we illustrate the distortions that a GW signal will exhibit due to the modified propagation of Eq.\ (\ref{eq:amplitude_mass_mixing}).
Fig.\ \ref{fig:distortion_mass_mixing} shows the propagation of GWs in the mass mixing model in the case of maximal mixing angle $\Theta_g=\pi/4$, and graviton mass $m_g=10^{-21}$eV$/c^2$ (blue curves), and a comparison with the waveform expected in GR (grey curves). Here we make the same choice of binary system, detector orientation, and plot conventions as in Fig.\ \ref{fig:distortion_velocity_mixing}. 
We plot the waveform that would be detected by a LIGO-type ground-based single detector, when the source is located at redshifts $z=0.005$, 0.02, 0.1, and 5. 

The top panel of Fig.\ \ref{fig:distortion_mass_mixing} shows the regime $z_\text{mix}<z<z_\text{coh},z_\text{broad}$ for the parameter values chosen, where a low-redshift source would have a single detected signal with a short mixing scale that induces modulations in the overall amplitude of the signal as well as phase shifts, that would otherwise not be present in GR for this source. Thus, this corresponds to case 2b in Table \ref{table:summary_time2}. 
We can see explicitly the mixing between the two eigenstates in the inset of the panel at $z=0.005$. Here we  see that the phase difference between the two eigenstates varies across the signal due to its $k$ dependence. Due to this variation, sometimes the two eigenstates are in phase and sometimes out of phase, leading to the amplitude modulations and shifts of peaks and troughs observed in the total blue signal. 
For low enough masses $m_g$, the mixing timescale would grow to be longer than the duration of the signal and there will not be visible modulations to the waveform, and the signal will become indistinguishable from GR for this source. For small enough mixing angles, the second eigenstate would get suppressed and the first one enhanced, so that in this case the signal would again become indistinguishable from GR, given that the first eigenstate propagates in the same way as in GR. 

The second panel, at $z=0.02$, shows a highly distorted GW signal in the regime of case 3 where $z_\text{broad,2}\lesssim z \lesssim z_\text{coh}$ (note that the first eigenstate propagates as in GR, and hence does not exhibit broadening distortions). In this case, the two eigenstates are in the transition from coherence to decoherence regime (yet still overlapping), and the waveform $h_2$ contributed by the second eigenstate is getting squeezed as it propagates, as it can be explicitly seen in the inset. This happens because typical coalescence GW waveforms have a frequency evolution that grows with time, and therefore the mass term $m_g$ makes low frequency modes to propagate slower so that in the time domain the waveform squeezes. This can be seen from the group velocity in Eq.\ (\ref{vg_massivegravity}). However, if the wave propagates for long enough, the whole waveform of the second eigenstate can be inverted in such a way that the high frequency merger reaches the observer before than the low frequency inspiral signal. This is what can be seen in the third panel from top to bottom. In this panel $z \gtrsim z_\text{coh}$ and we see that the two eigenstates have reached the limit of decoherence where they have a small ($\sim 10^{-2}$sec) temporal separation. In this model, one then has that decoherence is reached \emph{after} distortions in the second eigenstate become relevant (case 5 of Table \ref{table:summary_time2}). Note that for large enough masses $m_g$, the inversion of the second eigenstate's waveform would happen earlier, even for much lower redshift sources. 
After this waveform inversion happens, the waveform tends to stretch instead of squeeze with time, as seen in the bottom panel. In this panel we have that $z_\text{broad,2},z_\text{coh}\ll z$ and thus the two propagating states have fully decohered and reached the maximum time delay shown in Eq.\ (\ref{Max_td_mass_mixing}). In this model, this maximum time delay is shorter than the detected signal itself, and thus we expect both of the eigenmodes to be always detected in this case (provided the correct waveform templates are used to extract the signal from the noise). In addition, since the first eigenstate propagates in the same way as in GR, its waveform looks exactly like in GR in this regime, except for an overall change on its amplitude that can be easily mimicked by changes in the intrinsic parameters of the source in GR. On the contrary, the waveform of the second eigenstate is completely distorted in this decoherence regime, and cannot be mimicked by any GR waveform since the merger seems to arrive even before the inspiral.

Although a detection analysis is beyond the scope of this paper, we note that in Fig.\ \ref{fig:distortion_mass_mixing} we have compared the massive bigravity waveform to that in expected in GR from the same source, and striking differences are observed. A detailed detection analysis would compare the received signal to any GR waveform template, potentially including spin, eccentricity, etc. This may in practice make more difficult to distinguish modified gravity effects from GR, as a different choice of source parameters in GR may mimic some of the modified gravity effects. For instance, oscillatory modulations of the amplitude such as the ones seen in the first panel of Fig.\ \ref{fig:distortion_mass_mixing} are also expected in GR waveforms of sources with precession. These source parameter changes in the waveform will have to be taken into account when quantifying how well a given modified gravity waveform can be realistically distinguished from GR. 

Finally, our results for mass mixing generalize previous studies of the waveform distortions during the coherence phase in bigravity \cite{Max:2017flc,Belgacem:2019pkk,Jim_nez_2020} by including both the amplitude modulation and phase distortions. 
These effects are particularly relevant in the coherence regime and the transition to decoherence as shown in Fig.\ \ref{fig:distortion_mass_mixing}. Due to these distortions, it is possible to impose constraints on this mass mixing model with current and future GW observations. Previous analyses in massive bigravity in the coherence regime have ignored the change in the phase shifts across the frequency range of the signal, and therefore the constraints and forecasts previously obtained must be revisited in light of this mixing effect. 
GW detectors will however not be able to constrain graviton masses at the level of $m_g=10^{-33}$eV$/c^2$ that have been suggested to explain the late-time cosmic acceleration \cite{deRham:2014zqa} since in order for the exponent of Eq.\ (\ref{eq:amplitude_mass_mixing}) to be order 1 one would need to detect frequencies of order $ck\sim H_0\sim 2\cdot10^{18}$Hz. This is far beyond any planned facility.

\paragraph{Biased parameter estimation.}
Here we discuss whether there is a regime, in this mass mixing model, in which the waveform distortions are either suppressed or mimicked by a change in GR waveform parameters, but there still can be measurable amplitude changes in the signal. If this were to happen, a data analysis would bias the luminosity distance inferred from a detected event, when using GR waveform templates. 

We find that this will not happen in the coherence regime in massive bigravity, since $\Delta\omega$ scales similarly with $z$ and $k$, and therefore if time variations of the overall amplitude are visible, then the waveform distortions due to the $k$ dependence in $\Delta\omega$ will also become visible. We can see this explicitly by noting that in the mass mixing GW signal (\ref{eq:amplitude_mass_mixing}) both GW oscillations and waveform distortions enter through the exponent
\begin{equation}
    \frac{1}{2}\left(\frac{m_g}{H_0}\right)^2\left(\frac{H_0}{ck}\right)\int_0^z\frac{H_0dz}{H(z)(1+z)^2}\,,
\end{equation}
which we have factored here into three dimensionless terms. Since the integral is an order 1 quantity, one needs a large (relative to $H_0$) mass in order to compensate for the $H_0/ck$ term and have sizable amplitude modulations. At the same time, however, any order 1 change in the frequency of the signal will introduce waveform distortions. This sets a \emph{no-go} for mass mixing in the coherence regime: 
\begin{quote}
    A redshift modulation of the GW luminosity distance without waveform distortions can only occur if the relative frequency range of the signal is small, $\Delta f/f\ll1$.
\end{quote} 
This last frequency condition is not the case of coalescence binaries detected by present ground-based detectors. 
This implies that during coherence, mass mixing modifications will be seen through waveform distortions rather than in the luminosity distance, contrary to what was expected in \cite{Max:2017kdc,Belgacem:2019pkk,Jim_nez_2020}, and therefore previous constraints and forecasts obtained in the coherence regime must be revisited. Including the waveform distortions will likely improve the constraining power since GW detectors are more sensitive to the phase evolution than the amplitude changes. 
Note that this \emph{no-go} also extends to mass mixings scenarios where the tensor modes have different speeds, $\dc^2\neq0$, and the mixing is proportional to $m_{hs}^2\sim\dc^2k_*^2$ for a given frequency $k_*$ of the order of the frequencies of the signal. This is because although in this scenario it is possible to have $z_\text{coh}<z_\text{broad}$ (differently to the model of Fig.\ \ref{fig:dt_mass_mixing}), the mixing angle becomes frequency dependent introducing distortions in both echoes.

In mass mixing models with $\Delta c^2=0$, such as the one in Fig.\ \ref{fig:dt_mass_mixing}, a GR waveform template will only fit well the detected signal of the \emph{first} eigenmode after decoherence\footnote{The signal could technically also fit GR well if all the GR deviations in amplitude and frequency are highly suppressed, when $z\ll z_\text{mix}$, but this is an uninteresting case that we ignore.}. From Eq.\ (\ref{eq:amplitude_mass_mixing}) we see that for this first eigenmode the luminosity distance will be miscalculated according to:
\begin{equation}
    \left.\dLgw\right\vert_{1}\to d_L/\cos^2\tg\,,
\end{equation}
where we see that $\dLgw$ will always be biased towards larger values, since the detected amplitude of the signal will be smaller because the first eigenstate carries only a fraction of the total energy of the GW signal. This is in agreement with the low-redshift calculation of \cite{Max:2017flc}. 
A bias in $d_L$ will in turn lead to biases towards higher source redshifts and, consequently, lower source masses. 
Such constant re-scaling of the amplitude can be constrained with individual multi-messenger events (as it has been illustrated in \cite{Lagos:2019kds, Mastrogiovanni:2020gua} and others) or analyzing the population of compact binaries \cite{Ezquiaga:2021lli}.

\subsection{Friction mixing}\label{sec4:friction}

Next, we analyze a friction model, analogous to the one in Sec.\ \ref{Friction_example1}, in which both eigenstates propagate coherently all the time. Only cases 0-3 of Table \ref{table:summary_time2} are thus possible. We consider the following EoM:
\be \label{eq:cosmo_friction}
\lb \hat{I}\left(\frac{\d^2}{\d\eta^2}+
(ck)^2\right) + \bpm 0 & -2\bar{\alpha} \mathcal{H}(\eta) \\ 2\bar{\alpha} \mathcal{H}(\eta) & 0  \epm\frac{\d}{\d\eta} 
\rb \bpm a(\eta)\cdot h(\eta,k) \\ a(\eta)\cdot s(\eta,k) \epm =0\,, 
\ee
 where $\bar{\alpha}$ is assumed to be constant and dimensionless in this case. Note that actual models with friction mixing may not necessarily have the time dependence assumed here, but for concreteness we assume an $\mathcal{H}$ dependence in order to introduce an explicit hierarchy between the mixing interactions and the frequency of the signal.
Similarly to the previous mixing examples, this equation is satisfied for the renormalized fields $a\cdot h$ and $a\cdot s$, and the diagonal mass-like terms of order the Hubble rate that would appear due to this renormalization are neglected when compared to the friction matrix. Note that they will have a negligible effect even when $\bar{\alpha}\sim \mathcal{O}(1)$ since these mass terms will lead to a correction in the phase that is suppressed by $k^{-1}$ compared to the leading-order term coming from the friction matrix.
Therefore, we can safely neglect Hubble-order mass terms whenever $\mathcal{H}\ll ck$, which will hold for current and foreseen GWs detected by ground and space-based detectors, since $\mathcal{H}\sim 10^{-18}$Hz, and  LIGO-type GWs have frequencies $ck\sim 10^2$Hz, whereas LISA-type GWs have $ck\sim 10^{-2}$Hz.

In this example, the eigenfrequencies are analogous to those given by Eq.\ (\ref{Fricion_Example1_w1})-(\ref{Fricion_Example1_w2}) with an extra time dependence. Explicitly, we find that the eigenfrequencies are real and given by $\omega_1(\eta)=\sqrt{c^2k^2+(\mathcal{H}\bar{\alpha})^2}-\mathcal{H}\bar{\alpha}$, and $\omega_2(\eta)=\sqrt{c^2k^2+(\mathcal{H}\bar{\alpha})^2}+\mathcal{H}\bar{\alpha}$.
In this case, both eigenstates have the same group velocity and hence both eigenstates propagate coherently all the time and have the same broadening timescale. They propagate slower than the speed of light, at a group velocity $v_g=c^2k/\sqrt{c^2k^2+(\mathcal{H}\bar{\alpha})^2}$. 

We proceed to calculate the first-order WKB solution, according to the procedure described in Section \ref{WKBsubsection}. We find that the mixing matrix $\hat{U}$ is constant, and that $\hat{A}_\text{WKB}$ is diagonal. Explicitly, for an initial condition where only $h$ is non-vanishing, we obtain Eq.\ (\ref{EqMGh}) for $h_+$ and $h_\times$, with polarization-independent functions $f_{1,2}$ given by:
\begin{align}\label{WKBsol_friction}
  & f_1(k, \eta)= \frac{1}{2}e^{-\int_{\eta_e}^{\eta} \frac{\omega_1'}{2(\omega_1-\mathcal{H}\bar{\alpha})}\d\eta'},\quad f_2(k,\eta)=\frac{1}{2}e^{-\int_{\eta_e}^{\eta} \frac{\omega_2'}{2(\omega_2+\mathcal{H}\bar{\alpha})}\d\eta'}, 
\end{align}
where $\omega'_{1,2}$ denote derivatives with respect to conformal time of $\omega_{1,2}$, and both polarizations $(+,\times)$ propagate in the same way. In this example, the mixing angle is maximal (i.e.\ $\tan^2\Theta_g=1$), so that the two propagating eigenstates have the same amplitude initially. However, their relative amplitudes evolve in time, due to the real exponential terms in Eq.\ (\ref{WKBsol_friction}). At linear order in $\alpha$, we find that  $\int d\eta \omega_1'/(\omega_1-\mathcal{H}\bar{\alpha})\approx - \mathcal{H}\bar{\alpha}/(2c|k|)$, and  $\int d\eta \omega_2'/(\omega_2+\mathcal{H}\bar{\alpha})\approx  \mathcal{H}\bar{\alpha}/(2c|k|)$. Therefore, as long as $\bar{\alpha} \mathcal{H}_0/(ck)\ll1$, their relative time and frequency evolutions will be negligible, and we can approximate $f_{1,2}\approx 1/2$. 
Under this assumption, the solution $h_{+,\times}$ can be rewritten in a simpler way as:
\begin{equation}\label{WKBsol2_friction}
\begin{split}
  h_{+,\times}(k,\eta)&\approx 
  \frac{h_{\text{fid}+,\times}(\eta,k)}{2}e^{-i\int_{\eta_e}^{\eta}\Delta\omega_1(\eta',k) \d\eta'}  \left[1  + e^{-i\int_{\eta_e}^{\eta}\Delta \omega(\eta') \d\eta'}\right]   \\
  &\approx 
  \frac{h_{\text{fid}+,\times}(\eta, k)}{2}e^{-i\int_{\eta_e}^{\eta}\Delta\omega_1(\eta',k) \d\eta'}  \left[1  + e^{-2i\bar{\alpha}\log[(1+z_e)/(1+z)]}\right]\,,
\end{split}
\end{equation}
where in the second equality we have used that $\Delta \omega(\eta)=\omega_2-\omega_1=2\mathcal{H}\bar{\alpha}$, and $z_e$ is the redshift at emission. We emphasize that $\Delta \omega$ does not depend on $k$, and therefore the terms in the square brackets will induce a temporal modulation on the overall amplitude and phase of the signal, but will not introduce distortions to the GW phase evolution in Fourier space, for the simple waveforms analyzed here coming from nearly-equal mass binaries in orbits without precession. Frequency-dependent distortions will only come from the phase correction $\Delta\omega_1=\omega_1-ck$.

\paragraph{Time scales.}

\begin{figure}[t!]
\centering
\includegraphics[width=0.48\textwidth]{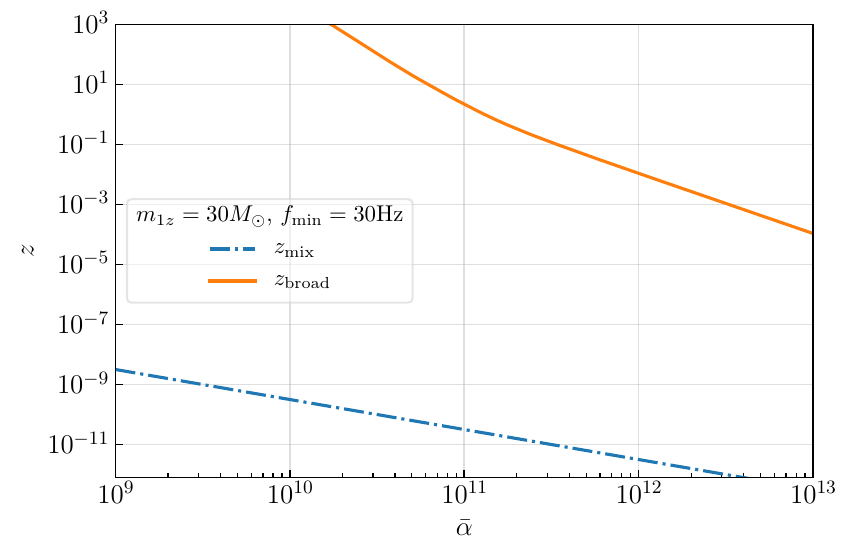}
\caption{\label{fig:zbroad_friction}
Comparison of the mixing time scale $z_\text{mix}$ and the broadening $z_\text{broad}$ for friction mixing scenarios over a wide range of parameters $\bar\alpha$ and source redshift $z$. In this scenario both eigenstates always propagate coherently so that $z_\text{coh}\to\infty$.}
\end{figure}

Since in this scenario there is no decoherence, there will be no echoes. The relevant time scales are thus given by the mixing and the broadening redshifts. The mixing scale is governed by the difference in the dispersion relations, $\Delta\omega=2\bar{\alpha} \mathcal{H}$. We can solve Eq.\ (\ref{eq:z_mix}) analytically to obtain
\begin{equation}
    z_\text{mix}\sim e^{\pi/\bar{\alpha}}-1\,.
\end{equation}
In the limit of $\bar{\alpha}\gg\pi$, one simply has $z_\text{mix}\approx \pi/\bar{\alpha}$ and the mixing could start at very low-redshifts if $\bar{\alpha}\gg 1$.
On the other hand, the broadening redshift of this model can be obtained solving
\begin{equation}
    \sigma_t\sim\lp\frac{\bar{\alpha}^2}{2c^2k_\text{max}^2}-\frac{\bar{\alpha}^2}{2c^2k_\text{min}^2}\rp\int_0^{z_\text{broad}}\frac{H(z)dz}{(1+z)^2}\,,
\end{equation}
which shares the same frequency dependence that our previous mass mixing example but a different time evolution. We plot the behavior of these two time scales in Fig.\ \ref{fig:zbroad_friction}. For $\bar{\alpha}<10^{10}$ there will be no broadening of the waveform for redshifts that will be reached with current and next-generation GW detectors. By the fact that $z_\text{broad}>z_\text{mix}$, we see that case 1 of Table \ref{table:summary_time2} will not be achieved in this example.

\begin{figure}[t!]
\centering
\includegraphics[width=\textwidth]{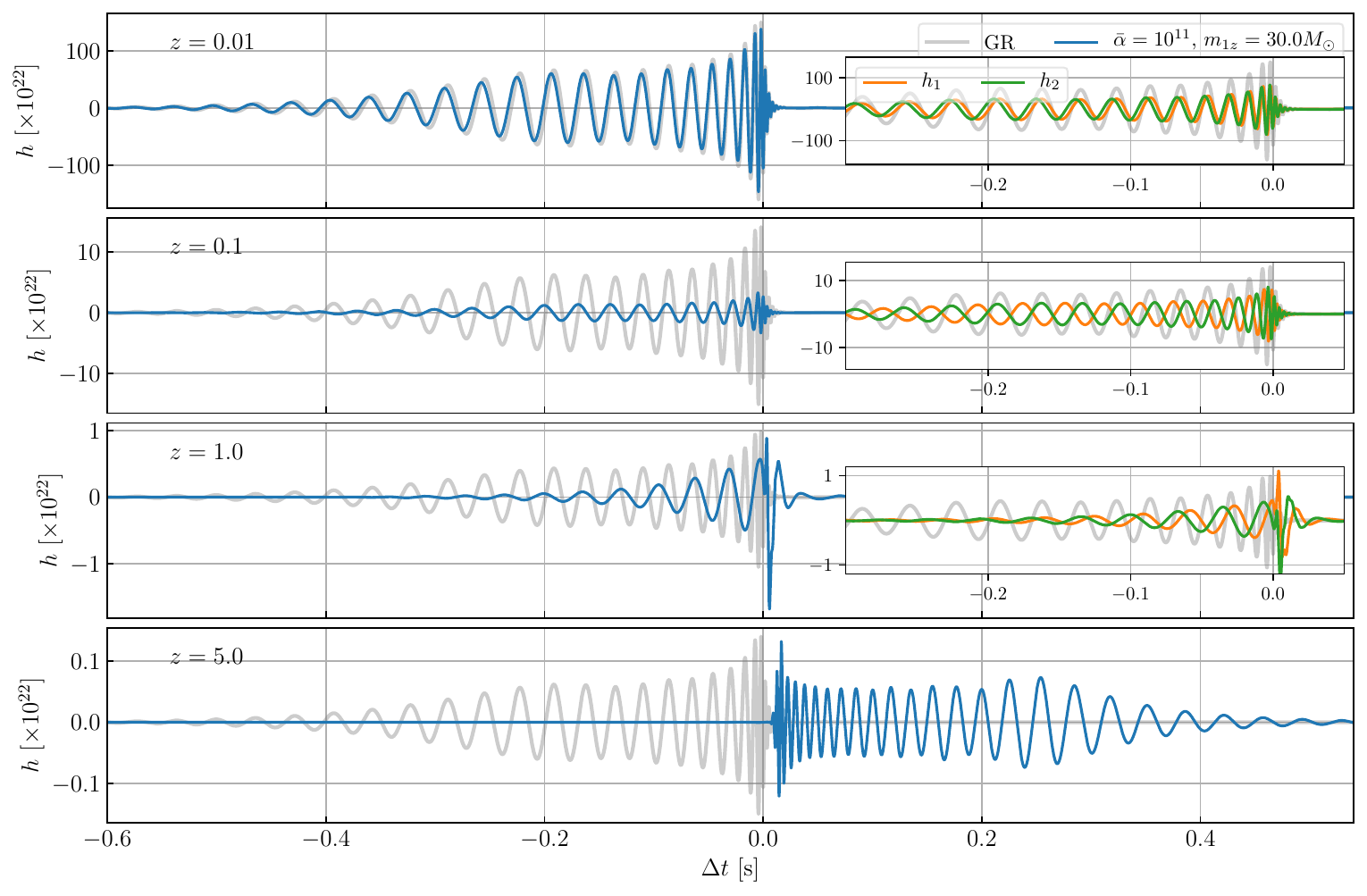}
\caption{\label{fig:distortion_friction_mixing}
GW signal observed by ground-based GW detectors, coming from a non-spinning equal-mass black hole binary system. The four panels show the distortions of the signal suffered during propagation in a \emph{friction mixing} model, in the case of sources with four different redshifts. The two eigenstates propagate coherently all the time at a group velocity slower than the speed of light. In the top three panels we also present the two independent eigenstates. In all panels $\Delta t=0$ is set by the arrival time of the frequency of coalescence traveling at the speed of light. The accumulated time delay at $z=5$ is $7$ms.}
\end{figure}

\paragraph{Waveform distortions.}
We take the same initial waveform and source location/orientation as in the previous mixing example and apply the modified propagation with friction mixing in (\ref{WKBsol2_friction}), choosing $\bar{\alpha}=10^{11}$. Such large coupling compared to the Hubble rate ($\bar\alpha=\alpha/\mathcal{H}$) is necessary in order to have detectable waveform distortions for LIGO-type detectors. Later we will show that when $\bar\alpha\sim 1-10$ there will be no distortions but instead oscillations in the GW luminosity distance. The results are plotted in Fig.\ \ref{fig:distortion_friction_mixing}. To consider a representative sample of the different propagation periods we choose the source redshifts to be 0.01, 0.1, 1 and 5 from top to bottom. 
Since the eigenstates propagate coherently, we always see one signal in the four panels. 

In the top two panels of Fig.\ \ref{fig:distortion_friction_mixing}, we represent a period where there is mixing but not broadening ($z_\text{mix} \ll z < z_\text{broad}$, case 2a), and we can observe that the amplitude oscillates with redshifts and that there is a global phase shift of the signal. The amplitude is always suppressed (or equal) to that of the GR signal, which happens because the two eigenstates have a phase difference that changes in time and therefore sometimes they are in phase (leading to a signal with the same amplitude as in GR) and sometimes out of phase (leading to a suppressed amplitude). This is explicitly seen in the inset of the panel at $z=0.1$, where there is nearly destructive interference between the two propagating eigenstates, with mild distortions caused by the MDR since in this case $z\lesssim z_\text{broad}$.

Similarly to the waveform distortions of the second mode in mass mixing, in this friction scenario the group velocity is such that low-frequency modes propagate slower squeezing the waveform initially and then inverting it. In the limit of small $\bar{\alpha}$, one finds $v_g\approx c-\bar{\alpha}^2\mathcal{H}^2/(2c^2k^2)$. At $z=1.0$, we see in the inset how both propagating eigenstates are getting squeezed and distorted due to their MDR since $z>z_\text{broad}$. Note that even though the only difference between the two eigenstates is a $k$-independent phase shift in Fourier space, this phase shift does create distortions of the waveform in real space, analogous to what has been shown to happen with the constant phase shifts induced by lensing in \cite{Ezquiaga:2020gdt}. 
At this redshift, the waveform is highly squeezed having multiple frequency components at a given time, and the same phase shift to all of these frequency components leads to distortions. This is why in the inset at 
$z=1$ we see that the detected eigenstates $h_1$ and $h_2$ have different waveforms. This panel corresponds to case 3 of Table \ref{table:summary_time2}.

At redshift $z=5$ we see that the phase distortions have accumulated for long enough that the waveform has been inverted, with the signal coming from the inspiral arriving after that coming from the merger. On top of the distortions due to the dispersion relations, in the bottom two panels we also have the same modulations of the amplitude in time due to $\Delta \omega$ because of the two modes interfering. However, since $\Delta\omega$ also introduces a frequency-independent phase shift to the signal in Fourier space, more complicated waveforms (coming from unequal masses, or precessing binary orbits) would have exhibited additional phase distortions in time domain. 
In this last panel we also explicitly see that this signal propagates slower than the speed of light, as it arrives after the signal expected in GR.
In particular, the time delay of the coalescence frequency in the blue curve is $7$ms compared to the grey curve (that by construction has an arrival time $\Delta t=0$ in this plot).

\paragraph{Biased parameter estimation.}

\begin{figure}[h!]
\centering
\includegraphics[width=0.48\textwidth]{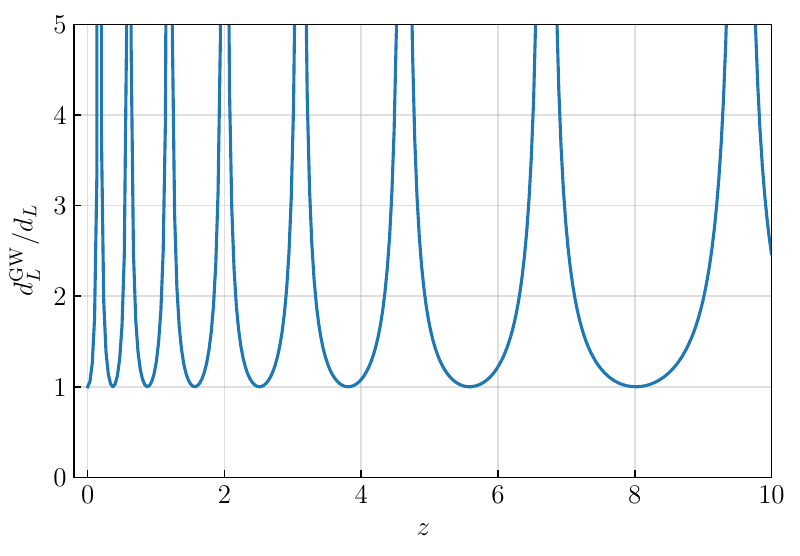}
\caption{\label{fig:dL_friction_mixing}
Modified luminosity distance during the coherence period for a friction mixing scenario. We plot the ratio of the GW and EM luminosity distance as a function of the source redshift $z$. In GR $\dLgw/d_L=1$. The ratio of the luminosity distances diverges at certain redshift when the GW signal is completely converted into the additional tensor field $s$. This example corresponds to $\bar\alpha=10$.}
\end{figure}

The case of friction mixing is fundamentally different to the mass mixing model, because here $\Delta \omega$ does not vary in $k$ across the detected signal. Indeed, in the coherence regime one can rewrite the solution (\ref{WKBsol2_friction}) in the following way:
\begin{equation}\label{Friction_dL}
    h_p(\eta,k)\approx \frac{h_{\text{fid}\,p}(\eta,k)}{\sqrt{2}}\sqrt{1+\cos\Delta\phi_I }\,e^{i\theta'(\eta)}e^{-i\int^\eta_{\eta_e} \Delta\omega_1(\eta',k) d\eta'},
\end{equation}
where $\Delta\phi_I=-2\bar{\alpha} \log[(1+z_e)/(1+z)]$, and $\tan\theta'=\sin\Delta\phi_I/(1+\cos\Delta\phi_I)$. Here we see that only $\omega_1$ may introduce frequency-dependent phase shifts to the waveform, depending on how it evolves with frequency. 
However, in this model $\omega_1$ can be approximated to:
\begin{equation}\label{Friction_omega1_approx}
    \omega_1(\eta,k) -ck \approx - \mathcal{H}\bar{\alpha} +\frac{(\mathcal{H}\bar{\alpha})^2}{2ck}+\mathcal{O}(k^{-3}),
\end{equation}
where we see that the leading-order deviation from GR does not depend on $k$, and is effectively constant during the detection of the signal since $\mathcal{H}$ variations during detection are negligible. Such a constant phase shift is degenerate with the source parameters for the simple waveforms analyzed here, where the binary is assumed to be of equal mass and move in a nearly circular orbit. The same happens with the phase shift $\theta'$ in (\ref{Friction_dL}), since it only varies on cosmological timescales and it is $k$ independent. In this scenario, then one can choose $\bar{\alpha}$ small enough such that the $k$-dependent terms in (\ref{Friction_omega1_approx}) are suppressed but $\mathcal{H}\bar{\alpha}$ and $\Delta\phi_I$ are non-negligible. In such a case, the detected signal will look like a GR waveform with an overall phase shift  and a different amplitude,  and it will correspond to an example of case 2a in Table \ref{table:summary_time2}. 
When using GR waveform templates to describe this detected signal, the amplitude will get biased in such a way that:
\begin{equation}
    \dLgw \to d_L /\sqrt{(1+\cos\Delta\phi_I)/2 }\,,
\end{equation}
which will be subject to the propagation time, in an oscillatory way. Fig.\ \ref{fig:dL_friction_mixing} shows the time variations of $ \dLgw$ in a model with $\bar\alpha=10$. Here we see that the oscillations are determined by $\Delta\phi_I$, which evolves more quickly for low redshift values. 
Importantly, in this mixing scenario the tensor $h$ can be completely converted into the tensor $s$ at certain redshifts, and therefore the luminosity distance displays poles \cite{Jim_nez_2020}. These can be easily computed from the above formula:
\begin{equation}
    z_\text{poles}=e^{\frac{\pi}{2\bar\alpha}(1+2n)}-1\,,
\end{equation}
where $n$ is a positive integer number. 

Finally, we emphasize that since  the $d_L^{GW}$ oscillations only depends on $\bar{\alpha}$ and not on $k$, the results found in this section are valid for any ground or space-based detector as long as the source is nearly circular equal-mass binary (recall that here we also find a constant phase shift that would introduce phase distortions for unequal-mass binaries or precessing orbits). This oscillatory behavior of the amplitude shows that friction interactions of order Hubble scale leave imprints that could be observed with a population of GW events in the future.

\subsection{Chiral mixing}\label{sec4:chiral}

Lastly, we study a chiral mixing scenario, analogous to that in subsec.\ \ref{sec:gaussian_chiral_dc0}, with time-dependent coefficients in the equation of motion, and apply the WKB formalism to realistic waveforms. We consider the following EoM:
\begin{equation}\label{chiral_with_time}
\left[ \hat{I}\left( \frac{\d^2}{\d\eta^2} +
(ck)^2 \right) \pm\bpm \bar{\mu}_h & \bar{\gamma} \\ \bar{\gamma} & \bar{\mu}_s \epm \mathcal{H}(\eta)ck \right] \bpm a(\eta) \cdot h_{L,R}(\eta,k) \\ a(\eta)\cdot s_{L,R}(\eta,k) \epm =0,
\end{equation}
where, for concreteness, we again assume an $\mathcal{H}$ time dependence of the mixing interactions, and $\bar{\mu}_s$, $\bar{\mu}_h$ and $\bar{\gamma}$ are assumed to be dimensionless constants. Here, the $\pm$ signs indicate the interaction sign of left and right polarizations, respectively. 
Similarly to the previous mixing examples, here we assume that the equations hold for the renormalized fields $a\cdot h_{L,R}$ and $a\cdot s_{L,R}$, and neglect the Hubble rate mass terms that appear as a consequence. Notice that, similarly to the previous friction example, these Hubble-order mass terms can be safely neglected even if the chiral coefficients are of order $\bar{\gamma}, \bar{\mu}_{h,s}\sim \mathcal{O}(1)$. 
This is due to the fact that the corrections these mass terms bring to the eigenfrequencies are suppressed by $k^{-1}$ compared to the leading-order corrections coming from chiral interactions.

The eigenfrequencies of the two propagating eigenstates are given by the WKB solution at lowest order, and have analogous expressions to those found in Eq.\ (\ref{omega1_chiral1})-(\ref{omega2_chiral1}), rescaling the chiral mixing coefficients by the Hubble rate.  Explicitly, the four eigenstates are characterized by the following eigenfrequencies in the large-$k$ limit: 
\begin{align}
    &\omega_{1;L,R}\approx ck \pm \frac{1}{4}\mathcal{H}\bar{\mu}_\text{tot} \pm \frac{1}{4}\mathcal{H}\sqrt{4\bar{\gamma}^2+\Delta\bar{\mu}^2} - \frac{\mathcal{H}^2}{32ck}\left( \bar{\mu}_\text{tot} + \sqrt{4\bar{\gamma}^2+\Delta\bar{\mu}^2}\right)^2 + \mathcal{O}(k^{-2}),\label{Chiral_w1_approx}\\
    & \omega_{2;L,R}\approx  ck \pm \frac{1}{4}\mathcal{H}\bar{\mu}_\text{tot}  \mp \frac{1}{4}\mathcal{H}\sqrt{4\bar{\gamma}^2+\Delta\bar{\mu}^2} - \frac{\mathcal{H}^2}{32ck}\left( \bar{\mu}_\text{tot} - \sqrt{4\bar{\gamma}^2+\Delta\bar{\mu}^2}\right)^2+ \mathcal{O}(k^{-2})\,,\label{Chiral_w2_approx}
\end{align}
where $\bar{\mu}_\text{tot}=\bar{\mu}_h+\bar{\mu}_s$, and $\Delta \bar{\mu}=\bar{\mu}_s-\bar{\mu}_h$. 
Here we have assumed that $\Delta\bar{\mu}\leq 0$ so that in the no-mixing limit of $\bar{\gamma}\rightarrow 0$, $\omega_{1}$ and $\omega_2$ describe the propagation of $h$ and $s$, respectively. 
Contrary to the mass mixing model, here all eigenmodes have non-trivial dispersion relations.
For the group velocities, we obtain at leading and sub-leading order that:
\begin{align}
   &  v_{g1;L}\approx v_{g1;R}\approx c+\frac{\mathcal{H}^2}{32ck^2}\left( \bar{\mu}_\text{tot} +\sqrt{4\bar{\gamma}^2+\Delta\bar{\mu}^2}\right)^2 + \mathcal{O}(k^{-3}), \label{Chiral_v1}\\
    & v_{g2;L}\approx v_{g2;R}\approx c+\frac{\mathcal{H}^2}{32ck^2}\left( \bar{\mu}_\text{tot} -\sqrt{4\bar{\gamma}^2+\Delta\bar{\mu}^2}\right)^2 + \mathcal{O}(k^{-3}).\label{Chiral_v2}
\end{align}
Note that regardless of the parameter values, the eigenstates propagate faster than the speed of light. Furthermore, low-frequency modes propagate faster than high-frequency modes, and therefore one expects to observe a waveform that elongates in time. In addition, as discussed in subsec.\ \ref{sec:chiral}, a GW signal may be generically composed by the four different eigenstates, and after coherence the signal reaches a regime with two separate wavepackets containing the modes $\{\omega_{1;L}, \omega_{1,R} \}$ and $\{\omega_{2;L},\omega_{2;R}\}$ due to the coincident group velocities in (\ref{Chiral_v1})-(\ref{Chiral_v2}). After this 2-echo decoherence is reached, the signal may split into four echoes, each one with purely right and left-handed polarization.

Next, we use the WKB approach to obtain the solution for $h$, which gives us separately the behavior of the signal for left and right-handed polarizations. Explicitly, we calculate the first-order WKB solution, and obtain that the mixing matrix $\hat{U}$ is constant, and thus the WKB correction matrix simplifies to $\hat{A}_{WKB}=\hat{\theta}^{-1}\hat{\theta}'/2$. Therefore, the solution for the GW signal can be generically expressed as Eq.\ (\ref{EqMGh}) for each component $h_L$ and $h_R$, with polarization-dependent functions $f_{A;L,R}$ given by: 
\begin{equation}
 f_{1;p}(k,\eta)=\frac{\cos^2\Theta_{gp}}{\sqrt{\omega_{1p}(\eta)/\omega_{1p}(\eta_e)}},\quad f_{2;p}(k,\eta)=\frac{\sin^2\Theta_{gp}}{\sqrt{\omega_{2p}(\eta)/\omega_{2p}(\eta_e)}},
\end{equation}
where $p=L,R$, and the mixing angles $\Theta_{g;L,R}$ are constants given by Eq.\ (\ref{Chiral_MixingAngle}). In order to obtain the total GW strain we must take into account the polarization-response of the detector as in Eq.\ (\ref{hpolarizations}), which can be rewritten in terms of left- and right-handed polarizations using Eq.\ (\ref{LR_pluscross_relation}). From Eq.\ (\ref{Chiral_v1})-(\ref{Chiral_v2}) we see that effectively the left-handed first eigenmode $h_{1L}$ propagates coherently with the right-handed first eigenmode $h_{1R}$, and similarly for the left-handed second eigenmode $h_{2L}$ and the right-handed second eigenmode $h_{2R}$. It is then useful to group the solution of the GW signal in terms of two net propagating states, given by $h_{1L}+h_{1R}$ and $h_{2L}+h_{2R}$, as 
\begin{align}\label{Chiral_WKBsol}
    h_s(\eta,k)= & e^{-i\int_{\eta_e} \Delta\omega_{1R}d\eta}\left(F_Rh_{\text{fid};R}(\eta,k)f_{1;R}(k,\eta)+ F_Lh_{\text{fid};L}(\eta, k)f_{1;L}(k,\eta)e^{-i\int_{\eta_e}\Delta\omega_{1} d\eta} \right) \nonumber \\
&+    e^{-i\int_{\eta_e} \Delta\omega_{2L}d\eta}\left( F_L h_{\text{fid};L}(\eta,k)f_{2;L}(k,\eta) + F_R h_{\text{fid};R}(\eta,k)f_{2;R}(k,\eta)e^{-i\int_{\eta_e}\Delta\omega_{2} d\eta} \right)  ,
\end{align}
where $\Delta\omega_{A;L,R}= \omega_{A;L,R}-ck$, $\Delta \omega_{1}(\eta)=\omega_{1;L}-\omega_{1;R}\approx \mathcal{H}(\bar{\mu}_\text{tot}+\sqrt{4\bar{\gamma}^2+\Delta\bar{\mu}^2})/2$,  $\Delta \omega_{2}(\eta)=\omega_{2;R}-\omega_{2;L}\approx  \mathcal{H}(-\bar{\mu}_\text{tot}+\sqrt{4\bar{\gamma}^2+\Delta\bar{\mu}^2})/2$, and both are approximately independent of $k$, with corrections of order $\mathcal{O}(k^{-2})$. 
In this chiral mixing example, it is important to identify the fiducial GR waveforms $h_{\text{fid};L}$ for left and $h_{\text{fid};R}$ for right-handed polarizations separately, as they will propagate differently. Similarly to the previous mixing examples, in practice, we will neglect the WKB first-order corrections to the amplitude functions $f_{A;L,R}$ since $|\omega(\eta)/\omega(\eta_e)-1|\ll 1$ whenever $\mathcal{H}_0|\bar{\mu}_\text{tot}\pm \sqrt{4\bar{\gamma}^2+\Delta\bar{\mu}^2}|/(ck)\ll 1$.

From Eq.\ (\ref{Chiral_WKBsol}) we see that distortions in the detected signal may be caused by how the propagating modes get individually distorted due to their non-trivial dispersion relations, how the modes interfere with each other, and how the two polarization modes interfere with each other for a given detector orientation.  

\paragraph{Time scales.}
Since left and right-handed polarizations propagate independently, we define two mixing scales associated to $\Delta\omega_L=\omega_{2;L}-\omega_{1;L}$ and $\Delta\omega_R=\omega_{2;R}-\omega_{1;R}$. At leading order for large $k$, both of these quantities have the same absolute value, $-\Delta\omega_L=\Delta\omega_R\approx \mathcal{H}\sqrt{4\bar{\gamma}^2+\Delta\bar{\mu}^2}/2$. 
Therefore, there is only one relevant scale when mixing between propagating eigenmodes becomes relevant, which is given by:
\begin{equation}\label{zmix_chiral}
    z_\text{mix}\sim e^{4\pi/\sqrt{4\bar{\gamma}^2+\Delta\bar{\mu}^2}}-1.
\end{equation}
In addition, we define the relevant time  delay as the one between the two net propagating eigenstates $h_{1L}+h_{1R}$ and $h_{2L}+h_{2R}$, which propagate with group velocities given by Eq.\ (\ref{Chiral_v1}) and (\ref{Chiral_v2}), respectively.  
Their time delay defines a two-echo decoherence regime and from Eq.\ (\ref{TimeDelay}) we find that it is explicitly given by:
\begin{equation}
\begin{split}
    \Delta \eta (z) \approx -\sigma_t &+\left[  \frac{\left(\sqrt{4\bar{\gamma}^2+\Delta\bar{\mu}^2}+\bar{\mu}_\text{tot}\right)^2}{32c^2k_\text{max}^2}
    -\frac{\left(\sqrt{4\bar{\gamma}^2+\Delta\bar{\mu}^2}-\bar{\mu}_\text{tot}\right)^2}{32c^2k_\text{min}^2}
    \right]
    \int_0^z\frac{H(z)dz}{(1+z)^2} \, .
\end{split}
\end{equation}
in the case of $\bar{\mu}_\text{tot}>0$. 
In the high $k$ limit, four-echo decoherence is achieved at timescales that are orders of magnitude longer than the two-echo decoherence timescale (recall example in subsec.\ \ref{sec:gaussian_chiral_dc0}). For this reason, we do not analyze this four-echo decoherence regime here.

\begin{figure}[h!]
\centering
\includegraphics[width=\textwidth]{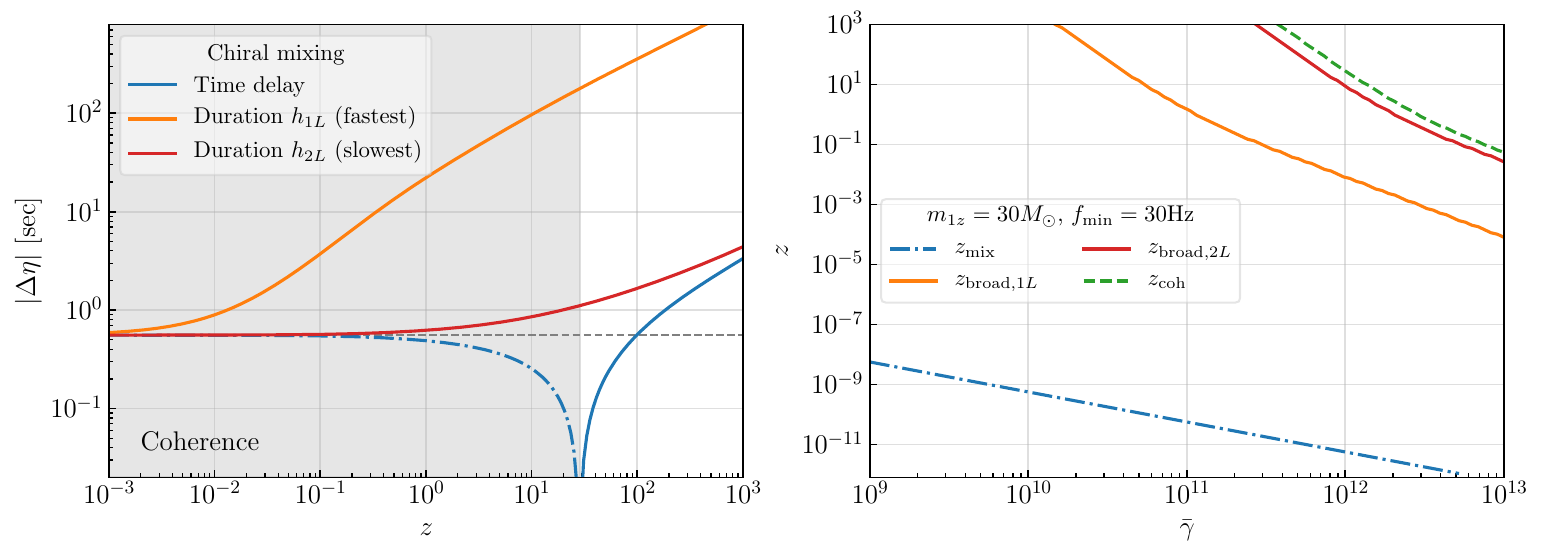}
\caption{\label{fig:dt_chiral}
On the left, time delay between the propagation eigenstates in our chiral mixing example as a function of the source redshift $z$. We use the same notation as in the mass mixing in Fig.\ \ref{fig:dt_mass_mixing}.
In this case, although the slowest mode duration increases, decoherence is eventually reached at high redshift. Note that at such redshifts the fastest mode is orders of magnitude longer than the initial signal (gray dashed line). 
Moreover, the duration of $h_{1L}$ is the same as $h_{1R}$ and similarly for $h_{2L}$ and $h_{2R}$. 
This left panel corresponds to $\bar \mu_h = 1.5\cdot 10^{12}$, $\bar \mu_s = 0.5\cdot 10^{12}$ and $\bar \gamma = 10^{12}$. 
On the right, relevant redshifts for mixing, broadening and coherence. The mixing scales of left and right modes are approximately the same, while the broadening time of the fastest ($h_{1L}$ and $h_{1R}$) and slowest ($h_{2L}$ and $h_{2R}$) net eigenstates are different. We have chosen $\bar \mu_h = 1.5\cdot \bar\gamma$ and $\bar \mu_s = 0.5\cdot \bar\gamma$.}
\end{figure}

In the left panel of Fig.\ \ref{fig:dt_chiral} we show the time delay $|\Delta \eta|$ between the two net eigenstates, and compare them to the duration of the detected eigenstates $h_{1L}$ and $h_{2L}$ (since due to Eq.\ (\ref{Chiral_v1})-(\ref{Chiral_v2}) that duration will be the same as the duration of the net eigenstates themselves), for the same binary system detected by LIGO/Virgo as in the previous examples. 
For the parameter values chosen in this example, we find that the time delay is short compared to the duration of both eigenstates (i.e.\ in most cases the signal will be in the coherence regime 3 of Table \ref{table:summary_time2}), and thus only very high redshift sources may be able to achieve this two-echo decoherence regime.
It is to be noted that when two-echo decoherence is achieved, the fastest net eigenstate has suffered very large distortions due to its MDR (case 5), which can be seen by the large duration of the signal $h_{1L}$. 
Similarly, the time delay between the fastest net eigenstate and a possible counterpart propagating at the speed of light is always parametrically smaller than its duration for merging binaries, scaling as $(f_\mathrm{max}/f_\mathrm{min})^2$, implying that waveform distortions would be the dominant effect. For example, the multi-messenger time delay for GW170817 would be $\sim 1$ sec for $\bar\gamma\sim 10^{14}$. However, for such couplings the duration of $h_{1L}$ would be already $\sim10^4$ times the original one.

We also compare the different time scales as a function of the parameters of the mixing in the right panel of Fig.\ \ref{fig:dt_chiral}, when assuming that $\bar{\mu}_h=1.5\bar{\gamma}$ and $\bar{\mu}_s=0.5\bar{\gamma}$. 
As shown in Eq.\ (\ref{zmix_chiral}), there is a single mixing redshift since left and right eigenstates start to mix at approximately the same time. For the parameters of this figure, mixing happens at very low redshifts. In addition, we see that the fastest net eigenstate broadens parametrically earlier than the slowest one due to their difference in the group velocity. Finally, in the plot we can see that, for $\sim30M_\odot$ binaries, only for large enough values of $\bar{\gamma}$, decoherence can be reached for the redshifts that will be probed by the next-generation of GW detectors.

\begin{figure}[t!]
\centering
\includegraphics[width=\textwidth]{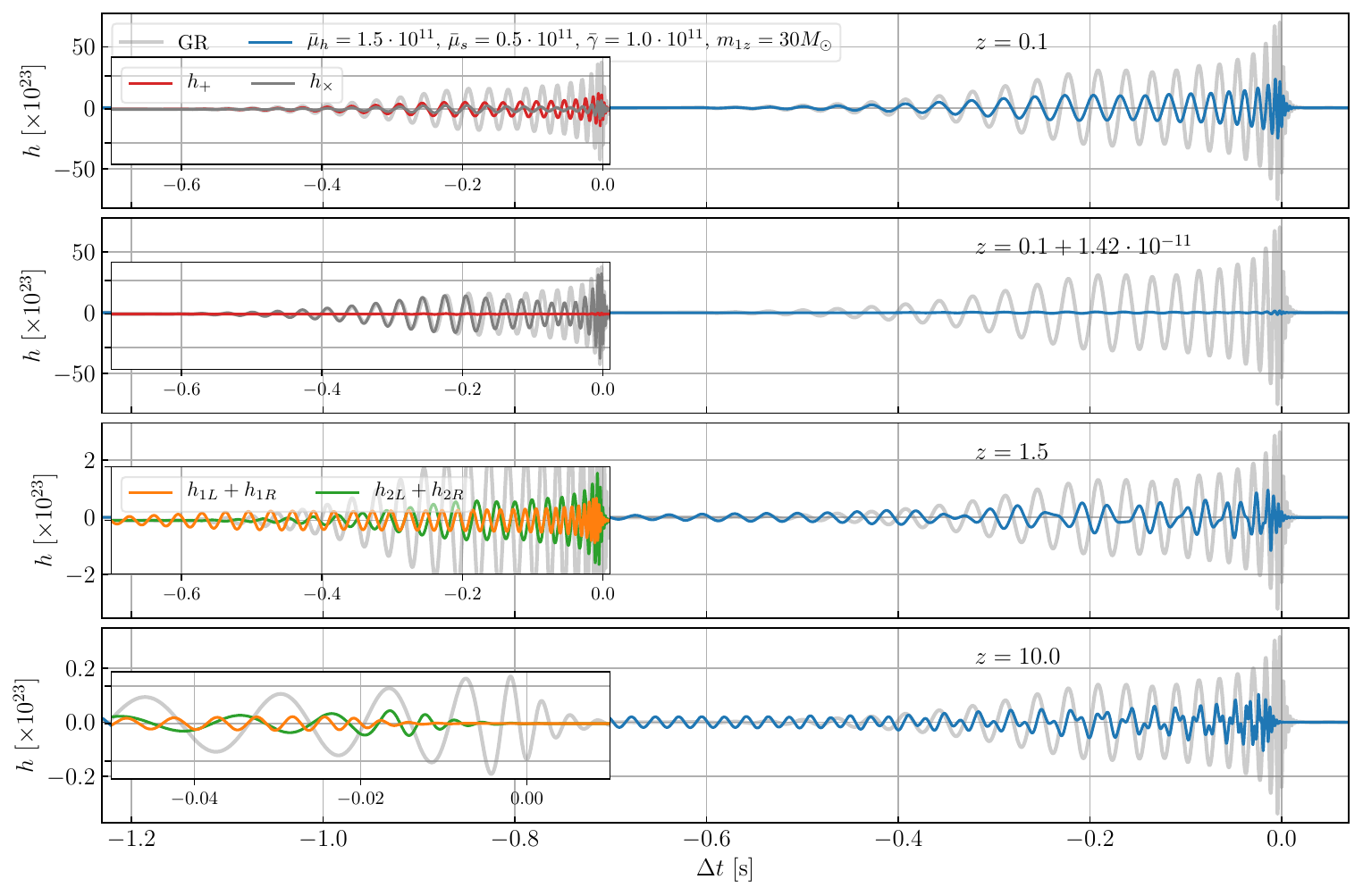}
\caption{\label{fig:distortion_chiral_mixing} GW signal observed by ground-based GW detectors, coming from a non-spinning equal-mass black hole binary system when there is \emph{chiral mixing}. The binary system is edge-on with $h_+$ only being emitted, and the detector is oriented to detect only this polarization ($F_+=1$, $F_\times=0$).  
The four panels show the distortions of the signal suffered during propagation at four different redshifts. The two eigenstates propagate coherently all the time as the waveform gets elongated. The two upper inset plots display the polarization composition ($h_+$ and $h_\times$) of the signal, while the lower two inset plots show the individual eigenstates that are composed respectively of $h_{1L}$ and $h_{1R}$, and $h_{2L}$ and $h_{2R}$. The combination of $h_{1L}$ and $h_{1R}$ is the fastest one and elongates more.
The time delay between this net eigenstate and the GR signal at the frequency of coalescence reaches 12ms at $z=10$. In all the panels $\Delta t =0$ corresponds to the arrival time of the frequency of coalescence traveling at the speed of light.}
\end{figure}

\paragraph{Waveform distortions.}
We show the observed signal due to chiral interactions in Fig.\ \ref{fig:distortion_chiral_mixing} for the same binary system as in the previous examples, but with a different orientation. In particular, we choose an edge-on binary so that the source only emits $h_+$ polarization. We keep the detector orientation to be only sensitive to this polarization fixing $F_+=1$ and $F_\times=0$. 
In this case, the GW strain (\ref{Chiral_WKBsol}) simplifies to:
\begin{align}\label{Chiral_WKBsol_plus}
    h_s(\eta,k)\approx  &\;  \frac{h_{\text{fid}+}(\eta,k)}{2} e^{-i\int_{\eta_e} \Delta\omega_{1R}d\eta}\left(\cos^2\Theta_{gR} + \cos^2\Theta_{gL} e^{-i\int_{\eta_e}\Delta\omega_{1} d\eta} \right)\nonumber \\
& +    \frac{h_{\text{fid}+}(\eta,k)}{2}e^{-i\int_{\eta_e} \Delta\omega_{2L}d\eta}\left(\sin^2\Theta_{gL} + \sin^2\Theta_{gR} e^{-i\int_{\eta_e}\Delta\omega_{2} d\eta} \right).
\end{align}
We also choose the parameters such that $\tan^2\Theta_{g;L,R}=(3-\sqrt{5})/2$ and hence we do not have maximal mixing as in the previous friction and mass mixing examples. In the top panel, at $z=0.1$, we see that the total signal is in the mixing regime where $z_\text{mix}\ll z < z_\text{coh}, z_\text{broad}$. 
Here we see that, due to the mixing, there is a frequency-dependent modulation of the amplitude of the signal in addition to phase differences (case 2b). 
As the GW propagates, its polarization composition ($h_{+,\times}$) will vary due to the different propagation equations for $h_{L,R}$. In the inset of the top panel we see that the signal has developed a non-vanishing $\times$ polarization during propagation.
In general, the frequency of this polarization rotation is determined by the order of magnitude of the couplings. In this case, with $\bar\gamma\sim10^{11}$, the frequency is very high and this is why in the second panel we need to fine-tune the redshift to capture this specific rotation of the polarization.
At this redshift, the signal is mostly $h_\times$ as shown in the inset plot. This can be contrasted with the polarization composition of the first panel, which was still mostly $+$ polarized. 

In the third panel, at $z=1.5$, we observe the mixing effects in the total signal in addition to distortions due to the dispersive group velocities. This is confirmed in the inset, where we show the two net propagating eigenstates separately: $h_1$ contains the modes $\{\omega_{1;L}, \omega_{1;R}\}$ and $h_2$ contains $\{\omega_{2;L}, \omega_{2;R}\}$. For this redshift, we have that $z_\text{mix},z_{\text{broad},2} <z\ll  z_{\text{broad},1}, z_\text{coh}$, where $z_{\text{broad},A}$ is the broadening timescale of the net propagating mode $h_A$ (both left and right handed have the same timescale, so we do not make a polarization distinction here). Therefore, the mode $h_2$ exhibits considerable elongation in the duration of its signal.
Since $h_2$ is broader than $h_1$, both net wavepackets only mix on a partial section of the total signal. Note that each net wavepacket $h_1$ and $h_2$ exhibit also distortions due to interference since they are both composed by the superposition of two propagating eigenmodes with different phases. If analyzed independently, $h_2$ will correspond to case 3, while $h_1$ to case 2b.

At $z=10$, we have that $z_\text{mix},z_{\text{broad},2} \ll z \ll  z_{\text{broad},1}, z_\text{coh}$, and thus the dispersion relation of $h_2$ has made the total signal elongate more than twice the duration expected in GR.   
Note that for this example, the time delay between the fastest mode and the GR signal at the frequency of coalescence is smaller than 12ms for sources below $z=10$. This time delay can be seen clearly in the inset of the bottom panel, where the modified gravity signal is seen to arrive before the GR one.

Overall, from this example we observe that although the group velocity of the two net eigenstates are different, there is never decoherence because the time delay between the eigenstates is smaller than the duration of the slowest mode, $h_2$. 
However, other parameter choices may achieve decoherence at $z<10$. 

Next, we provide another example of chiral interaction where we discuss in more detail how the polarization rotates from emission to detection, and how this rotation can be degenerate with source parameters of GR waveforms templates. 

\paragraph{Biased parameter estimation.}

Finally, we analyze scenarios in which individual events with chiral mixing could still mimic GR waveforms. In the coherence regime, this can indeed happen for appropriate parameter values that suppress the $k$ dependence of the phase shifts of the waveform.
More explicitly, we see in Eqs.\ (\ref{Chiral_w1_approx})-(\ref{Chiral_w2_approx}) that in the high-$k$ limit the leading-order deviations from GR of the eigenfrequencies are constant in $k$, and distortions are suppressed by a factor of order $k^{-1}$. Furthermore, these leading-order deviations will typically vary on cosmological timescales, which are much longer than the duration of the signal, so these deviations are effectively constants in time as well for individual events. 
In addition, the mixing scale determined by $\Delta\omega$ is also constant at leading order, and thus one can write a similar solution to the one in (\ref{Friction_dL}) for friction interactions. 

\begin{figure}[t!]
\centering
\includegraphics[width=\textwidth]{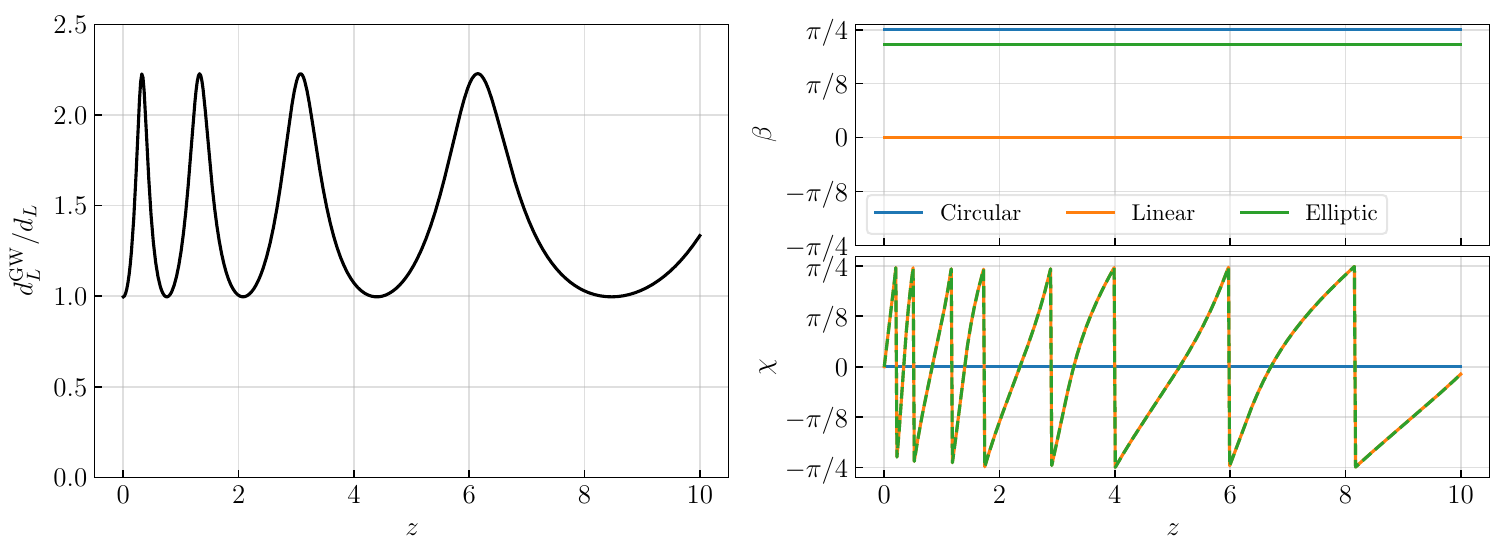}
\caption{\label{fig:dL_chiral_mixing}
Modified GW luminosity distance (left) and polarization angles $\beta$ and $\chi$ (right) during the coherence period for a chiral mixing scenario with different initial polarizations, as a function of the source redshift $z$. We consider circular ($h_+=h_\times$, $h_L=0$ for face-off), linear ($h_\times=0$, $h_L=h_R$) and elliptic ($h_{+,\times}\neq0$, $h_{L,R}\neq0$) initial polarization. The oscillations of $\dLgw$ are the same for all initial states. The polarization state described by $\beta$ remains the same, regardless of the initial polarization of the signal, since variations are suppressed by $1/ck$. The polarization orientation angle $\chi$ however rotates for an initially linearly or elliptically polarized signal. This example corresponds to $\bar{\mu}_h=15$, $\bar{\mu}_s=5$ and $\bar{\gamma}=10$.}
\end{figure}

In a case with initially only left-handed polarization, as it is the case of a circularly polarized face-on ($\iota=0$) source where $|h_+|=|h_\times|$, the polarization state will be preserved through the propagation. In this scenario, the strain is given by:
\begin{equation}
h_s(\eta,k)\approx  h_{\text{fid};L}(\eta,k)F_L\cos^2\Theta_{gL}  \sqrt{1+\tan^4\Theta_{gL}+2\tan^2\Theta_{gL}\cos(\Delta\phi_{IL})}e^{-i\theta_L'} e^{-i\int \Delta\omega_{1L}d\eta} ,\label{h_chiral_left}
\end{equation}
where $F_L$ describes the antenna response to left-handed polarization. Here, we also have that $\Delta\phi_{IL}=\int\Delta\omega_{L} d\eta$, such that $\Delta\omega_{L}=\omega_{2;L}-\omega_{1;L}\approx \mathcal{H}\sqrt{4\bar{\gamma}^2+\Delta\bar{\mu}^2}/2+\mathcal{O}(k^{-1})$. Also, $\tan \theta_L'=(\tan^2\Theta_{g;L}\sin\Delta\phi_{IL}) /(1+\tan^2\Theta_{g;L}\cos\Delta\phi_{IL})$. Here we have again neglected the amplitude corrections caused by the terms $\sqrt{\omega(\eta)/\omega(\eta_e)}\approx 1+\mathcal{O}(k^{-1})$. When all the $k^{-1}$ distortions are negligible, the leading term in $\Delta\phi_{IL}$ can still become relevant and induce visible distortions to the GW amplitude such that a single event would have an apparent luminosity distance given by:
\begin{equation}\label{dL_chiral_left}
    \dLgw \to d_L /\left[ \cos^2\Theta_{gL}  \sqrt{1+\tan^4\Theta_{gL}+2\tan^2\Theta_{gL}\cos(\Delta\phi_{IL})}\right]\,.
\end{equation}
For a population of events at different redshifts, the modulation of the GW luminosity distance for this example is presented in the left panel of Fig.\ \ref{fig:dL_chiral_mixing}, in a model where $\bar{\gamma}\sim \bar{\mu}_s\sim\bar{\mu}_h\sim \mathcal{O}(1-10)$. 
In the chiral model considered in this section, $\Theta_{g;L}$ is exactly constant, but since $\Delta\phi_{IL}$ varies in time, there will be an oscillatory behavior of the biased luminosity distance that we see in Fig.\ \ref{fig:dL_chiral_mixing}. This oscillatory behavior of the amplitude shows that chiral interactions of order Hubble scale leave clear observable imprints that could be tested with population of GW events in the future. We emphasize that the same result will be obtained for any initial emitted polarization, if we define the
luminosity distance as $(\dLgw)^{-2}\propto |h_+|^2 + |h_\times|^2= |h_L|^2 + |h_R|^2$. From Eq.\ (\ref{dL_chiral_left}) (and its analogous solution for $h_R$) we explicitly obtain that
\begin{align}
\left( \frac{d_L}{\dLgw}\right)^2 &=\frac{|h_L|^2 + |h_R|^2}{|h_{\text{fid};L}|^2+|h_{\text{fid};R}|^2} =\frac{|h_L|^2\left(1+r^2\right)}{|h_{\text{fid};L}|^2\left(1+r_0^2\right)}\nonumber\\
    &\approx \cos\Theta_{gL}^4\left[1+\tan\Theta_{gL}^4+2\tan\Theta_{gL}^2\cos(\Delta\phi_{IL})\right]\left(1+\mathcal{O}(k^{-1})\right),
\end{align}
where $r_0^2=|h_{\text{fid};R}/h_{\text{fid};L}|^2$ and $r^2=|h_{L}/h_R|^2$, and in the last line we have made a large-$k$ approximation. From here we see that, at leading order, the evolution of $\dLgw$ is the same for any initial polarization as that for an initially circularly polarized wave, according to Eq.\ (\ref{dL_chiral_left}). For chiral interactions comparable to the Hubble scale, this is always the case.  
In comparison with our friction mixing example in Fig.\ \ref{fig:dL_friction_mixing}, the chiral mixing does not display a maximal mixing and, therefore, the luminosity distance does not have poles. 
From Eq.\ (\ref{h_chiral_left}) we also see that the phases $\theta'_L$ and $\int \Delta\omega_{1;L}d\eta$ will induce a $k$-independent phase shift to the total signal, which will be degenerate with source parameters of GR waveforms, for simple signals coming from nearly-equal mass non-precessing binary systems. 

In addition to a change in the luminosity distance, chiral mixing induces a change in the polarization of the signal, which can be observed if multiple GW detectors are present. As illustrated for a toy Gaussian wavepacket in Section \ref{sec:gaussian_chiral_dc0}, we can use two angles $\beta$ and $\chi$ to describe the state of the detected polarization (see also Appendix \ref{App:polarization}), and analyze how much it changed from emission to detection.
The evolution of these angles as a function of redshift is displayed in the right panels of Fig.\ \ref{fig:dL_chiral_mixing}, in three scenarios where the initial polarization was circular with $(\beta,\chi)=(\pi/4,0)$, linear with $(\beta,\chi)=(0,0)$ and elliptical with $(\beta,\chi)=(\pi/4.64,0)$. In general, a change in the angle $\chi$ will bias the detected orientation angle of the LIGO-type antenna pattern function in Eq.\ (\ref{hpolarizations}), such that $\psi\rightarrow \psi+\Delta\chi$. In addition, the angle $\beta$ will be degenerate with the inclination of the source $\iota$, for GW binary signals dominated by a (2,2) spherical harmonic mode (analogous to what is found in dynamical Chern-Simons theories \cite{Okounkova:2021xjv}). 
In these simple cases, the observed angle $\beta$ (that describes the initial emitted polarization in addition to changes suffered during propagation) will be related to the observed inclination $\iota$ by:
\begin{equation}
 \left \lvert \frac{h_R(\eta,k)}{h_L(\eta,k)}\right\rvert=\frac{1+\tan\beta}{1-\tan\beta}=\frac{(1-\cos\iota)^2}{(1+\cos\iota)^2}.
\end{equation}
However, for GW binary sources that emit higher spherical harmonics, either because they have unequal masses or because they exhibit precession due to misaligned spins or orbital eccentricity, the angle $\beta$ will not be degenerate anymore with $\iota$ if multiple detectors can measure the polarization of each individual spherical harmonic. This is because GR predicts fixed relationships between the ratios $|h_R/h_L|$ for each harmonic given a value of $\iota$, but changing the angle $\beta$ will break these fixed relationships.

One can see in Fig.\ \ref{fig:dL_chiral_mixing} that, independently to the initial polarization of the wave, the signal does not exhibit amplitude birefringence, with $\beta$ remaining constant. However, the polarization orientation $\chi$ of the linearly and elliptically polarized wave rotates during the propagation. Because of this change in the polarization orientation $\chi$, a given detector might not detect signals that would be detectable in GR at certain redshifts, and vice-versa. 
For example, if the source-detector geometry is such that $F_{+}=1$ and $F_{\times}=0$, an initially linearly polarized GW $h_+$ will be missed by the detector at the redshifts in which the polarization has been transformed to linear $h_\times$, that is, whenever $\chi$ changed from 0 at emission to $\chi=\pi/4$ at detection. 
Moreover, the evolution of $\chi$ is determined by Eq.\ (\ref{eq:chi}) if one substitutes $\Delta\eta$ by $\log(1+z)/\mathcal{H}$. 
We emphasize that the general behavior found here for $\beta$ and $\chi$ is model dependent, since other theories with chiral interactions may for instance predict large variations of $\beta$ \cite{Okounkova:2021xjv}.

\section{Discussion}\label{sec:discussion}

Gravitational waves (GWs) from a compact binary coalescence are excellent targets to test cosmology, since they have well-modelled signals whose propagation is sensitive to the background expansion and field content of the Universe. In this paper we have developed a general parametrized approach (agnostic to specific modified gravity theories) to analyze the propagation of GWs over cosmological backgrounds in scenarios where GWs interact with an additional tensor mode. Due to these interactions, even if the emission is not modified w.r.t.\ general relativity (GR) (as assumed in this work), the extra tensor mode will be generated during propagation. As a result, GWs will have a non-trivial propagation that can lead to changes in the amplitude, phase, and polarization of the emitted GW signal.
We summarize our main results as follows:
\begin{itemize}
    \item Due to the interactions between the spacetime metric and the additional tensor mode, GWs are given by a superposition of signals---the \emph{propagating eigenstates}---, whose wavepackets can propagate coherently and interfere with each other at the detector, or propagate decoherently and lead to echoes in the detector. We parametrize the possible tensor interactions, and apply the WKB formalism to obtain the general detected GWs as the superposition of eigenstates (see Sec.\ \ref{sec:wavepacket}). We discuss how these eigenstates may have modified dispersion relations that cause the effect of decoherence and induce amplitude and phase distortions of GWs.
    
    \item We introduce three main timescales characterizing the general propagation of these models with interacting GWs---\emph{mixing}, \emph{decoherence}, and \emph{broadening}---, determining respectively when the  interference between the eigenstates evolves between emission vs.\ detection,
    when they split and can be detected as separate GW signals, and when frequency-dependent phase distortions induce considerable changes in the duration of the signal from inspiral to merger. A summary of the different timescales and associated observable effects is presented in Table \ref{table:summary_time2} and Fig. \ref{fig:summary_waveform_distortions}.
    
    \item We clarify that the \emph{group velocity} of each eigenstate is what determines the detected phase at a given frequency and hence the phase distortions of the waveform.  This is in contrast with treating GWs as an ensemble of particles traveling at the particle velocity \cite{Will:1997bb,Mirshekari:2011yq}. Moreover, we show how our WKB propagation approach can be applied to any initial GW signal, independently of the complexity of the source (whether the binary has unequal mass components, or the orbit exhibits precession or eccentricity). 
    
    \item We confirm our analytical predictions with a numerical analysis of a toy gravitational Gaussian wavepacket, for models with velocity, mass, friction and chiral tensor interactions (see Section \ref{sec:examples_mixing}), and illustrate the varied phenomenology that can happen in these models. We discuss the behavior of the amplitude of the Gaussian, its distortion and polarization content.

    \item We consider realistic GWs from a back hole binary coalescence and explore the observational effects of the modified GW propagation in \emph{single events} and discuss \emph{population analyses} with present and future GW detectors (see Section \ref{sec:implications}). We conclude that:
    \begin{itemize}
        \item Waveform distortions can appear due to 1) modifications in each eigenstate's dispersion relations, 2) interference among the eigenstates, or 3) frequency and time dependent amplitude modulations (for example if there is a frequency dependent mixing angle), although in many cases only the first two effects will be detectable.
        
        \item In a number of scenarios, a detected single GW may look like a GR waveform, but with different source parameters. This could induce a bias in the luminosity distance, coalescence phase, inclination and orientation of the source. However, a population analysis may help disentangle some of these propagation effects. For instance, the biased GW luminosity distance may oscillate with redshift, which would induce an astrophysically unexplained oscillation on the redshift distribution of GWs sources.
        
        \item If decoherence is achieved at a redshift $z_\text{coh}$ within the sensitivity of the GW detector, a population analysis would find that 
        high-redshift events ($z>z_\text{coh}$) split into sub-populations (echoes) that fit waveforms with different dispersion relations. 
        Counting echoes constraints the number of additional tensor fields that may be interacting with the spacetime metric.
        
        \item In the case of mass interactions (e.g.\ massive bigravity theories), we find the following \emph{No-go}: redshift or frequency-dependent amplitude changes of a chirping GW signal are always accompanied by phase distortions. As a result, phase distortions cannot be neglected in these models. 
        
        \item In the case of chiral mixing, we characterize the effects of amplitude and phase birefrigence. In the specific model analyzed here, amplitude birefringence was found to be suppressed for high-frequency GWs, while phase birefringence was relevant, leading to a rotation of the polarization during propagation.  
        
        \item While the propagation modifications analyzed here are generic, the case when the timescale of the tensor mixings is of order Hubble time is of particular interest. In such scenarios, we find that the distortions introduced by mass mixings will be undetectable, whereas friction and chiral mixings leave observable effects in the amplitude and phase of GW waveforms.
    \end{itemize}
\end{itemize}

Due to the rich phenomenology of deviations from GR that can be typically expected in models with tensor interactions, a numerical analysis using the current GW events detected by LIGO/Virgo may already place stringent constraints on specific modified gravity theories, although this is left for future work. In the case of massive bigravity, some current GW events were used to constrain the theory parameters \cite{Max:2017flc} and forecasts were performed for LISA \cite{Belgacem:2019pkk}, however these are conservative constraints since they neglected the phase distortions of the waveform discussed in this paper. Similarly, constraints and forecasts from populations of GW events are left for future work. 
Furthermore, predictions on the expected energy spectrum of a stochastic GW background that include these tensorial cosmological interactions could be performed in the future, using the results found in this paper on how the amplitude (and thus intensity) of GW events changes due to modified gravity.

In this paper we have assumed that the emitted signal was the same as that in GR, but this could be generalized to include also modifications at emission. The approach we used straightforwardly allows for the implementation of arbitrary initial conditions. However, we emphasize that modifications at emission are fundamentally different to those during propagation, since propagation modifications accumulate over time as the GW signal propagates, leading for instance to completely different observed signals from the same source at low and high redshift. The same is not the case for distortions during emission only.

In addition, here we have also assumed that matter only couples to the metric $h$ and thus that was the only variable determining the GW strain. However, this can be generalized, as it has been done for massive bigravity in the so-called doubly-coupled massive gravity models \cite{Comelli:2015pua, Brax:2017hxh, Akrami:2018yjz}. In such scenarios, we expect each propagating eigenstate to contribute differently to the total strain, depending on how the second tensor mode couples with matter.

Beyond the homogeneous and isotropic cosmological backgrounds analyzed here, GWs can mix with other types of fields, not just tensor modes. 
An example of this occurs when studying GW lensing around massive objects \cite{Dalang:2020eaj, Ezquiaga:2020dao}. 
Our analysis could be extended to study the propagation of GWs across inhomogeneous backgrounds, and applying our techniques to environments exhibiting a screening mechanism.
Note that screening can also have an effect, albeit small, even if there is no lensing, if the interactions between GWs and the extra field are suppressed near the source and observer \cite{Perkins:2018tir}. 

Since in this paper we have studied a few limited examples to illustrate some of the effects of tensor interactions, we may have not exhausted all possible observable effects.
However, the approach used here is general and can be straightforwardly extended to include other types of interactions (such as those involving higher spatial derivatives), multiple interactions at the same time, any cosmological time evolution of the free parameters in the tensor equations of motion, and even more than two interacting tensor fields. 
Therefore, the work done here develops all the basic ingredients to analyze more generally the propagation of GWs in any modified gravity theory in the future.

\section*{Acknowledgments}

We are grateful to Miguel Zumalac\'arregui for useful discussions at the beginning of this project and Yiming Zhong for pointing us to the wavepacket analysis of neutrino oscillations. We also acknowledge useful correspondence with Maximiliano Isi. 
JME is supported by NASA through the NASA Hubble Fellowship grant HST-HF2-51435.001-A awarded by the Space Telescope Science Institute, which is operated by the Association of Universities for Research in Astronomy, Inc., for NASA, under contract NAS5-26555. He is also supported by the Kavli Institute for Cosmological Physics through an endowment from the Kavli Foundation and its founder Fred Kavli. WH and MXL were supported by the U.S.~Dept.~of Energy contract DE-FG02-13ER41958 and the Simons Foundation. ML was supported by the Innovative Theory Cosmology fellowship at the University of Columbia.

\appendix

\section{Properties of the eigenfrequencies}\label{App:WKB}

In this appendix we provide further details about the eigenfrequencies of the propagation eigenstates. We consider first the simpler case without friction and then move to the general result. 

\paragraph{Without friction.}
\label{app:NoFriction}

Regarding the eigenfrequencies, if there is no friction then the eigenvalues satisfy the following equation:
\be
\theta^4-\theta^2 {\rm Tr}(\hat{W}) +\det (\hat{W})=0,
\ee
whose roots are given by
\be
\theta^2=\frac{1}{2}\left[ {\rm Tr}(\hat{W})\pm \sqrt{{\rm Tr}(\hat{W})^2-4\det (\hat{W})}   \right].
\ee
For a symmetric $\hat{W}$ matrix (which can always be achieved by field renormalizations), $\theta^2$ is always real. In addition, if $\det (\hat{W})>0$ and ${\rm Tr}(\hat{W})>0$ (i.e.~$\hat{W}$ has two positive eigenvalues) then $\theta^2$ will also be positive definite. Note that in order to have stable tensor perturbations we expect $\hat{W}$ to have positive eigenvalues. We thus conclude that in realistic scenarios, if friction is not present, then there will always be four real eigenfrequencies of the form $\pm \theta_1$ and $\pm \theta_2$, with $\theta_{1,2}$ real. 

\paragraph{With friction.}
Regarding the eigenfrequencies, if $\hat{\nu}$ is antisymmetric and $\hat{W}$ is symmetric, then $\theta$ will satisfy:
\be
\theta^4-\theta^2 \left[{\rm Tr}(\hat{W}) + \det (\hat{\nu}) \right] +\det (\hat{W})=0
\ee
where $\det (\hat{\nu})>0$. Similarly to the non-friction case, if $\det (\hat{W})>0$ and ${\rm Tr} (\hat{W})>0$, then the $\theta^2$ roots will be real and positive definite, leading to four real eigenfrequencies related by $\pm$ signs. Explicitly, we obtain:
\begin{equation}
    \theta^2=\frac{1}{2}\left[ {\rm Tr}(\hat{W})+ \det (\hat{\nu}) \pm \sqrt{\left[{\rm Tr}(\hat{W})+ \det (\hat{\nu}) \right]^2-4\det (\hat{W})}   \right].
    \label{Eq_thetareal}
\end{equation}

Next, we study the general case when $\hat{\nu}$ can have on-diagonal terms. In that case, $\theta$ must satisfy the following equation:
\be
\det (\hat{W})+i\theta (W_{11}\nu_{22}+W_{22}\nu_{11})  -\theta^2 [{\rm Tr}(\hat{W})+\det(\hat{\nu})]-i\theta^3 {\rm Tr}(\hat{\nu})+\theta^4=0\,.\label{Eq_eigenvalues}
\ee
As usual, there are four different solutions, but in this case the frequencies may be complex and generically expressed as $\theta=\omega+i\Gamma$, with  $\omega,\Gamma$ real. In the high-$k$ limit, separating this equation into real and imaginary components, we obtain that, if $\omega\ne 0$ (as is the case at high $k$ for all models), it will always appear with even powers, whereas $\Gamma$ will appear with both odd and even powers. Therefore $\omega\ne 0$ solutions always come in $\pm \omega$ pairs of the same $\Gamma$.
In particular, when solving for the real component of Eq.\ (\ref{Eq_eigenvalues}), we find that the possible values of $\omega^2$, given a value of $\Gamma$, are:
\begin{align}
  \omega^2&=\frac{1}{2}\left[ {\rm Tr}(\hat{W})+\det(\hat{\nu}) -3\Gamma  {\rm Tr}(\hat{\nu}) +6\Gamma^2\pm \left\{\left({\rm Tr}(\hat{W})+\det(\hat{\nu})-3\Gamma  {\rm Tr}(\hat{\nu}) +6\Gamma^2\right)^2 \right. \right.\nonumber\\
  &\left.\left. -4\det (\hat{W}) +4\Gamma (W_{11}\nu_{22}+W_{22}\nu_{11}) -4({\rm Tr}(\hat{W})+\det(\hat{\nu}))\Gamma^2 +4{\rm Tr}(\hat{\nu})\Gamma^3- 4\Gamma^4 \right\}^{1/2}   \right],\label{wsq_eigenvalue}
\end{align}
where a solution must satisfy this equation with either $+$ or $-$ but not necessarily both.
Therefore, any real value of $\Gamma$ satisfying the equations of motion will lead to a pair complex eigenvalues of the form $\theta_{\pm}=\pm \omega +i\Gamma$ satisfying Eq.\ (\ref{Eq_eigenvalues}). Since this eigenvalue equation is quartic, we then expect to have four eigenvalues that can be generically expressed as:
\begin{equation}
    \theta_{1\pm}= \pm  \omega_{1} +i \Gamma_{1}; \quad \theta_{2\pm} = \pm  \omega_{2} + i \Gamma_{2}. 
\end{equation}
In the large-$k$ limit (when $\hat{\nu}$ is small compared to $\hat{W}$), one will find that one $\omega^2$ eigenvalue is obtained from Eq.\ (\ref{wsq_eigenvalue}) with a $+$ sign in front of the square root, and the other with a $-$ sign, with Eq.~(\ref{Eq_thetareal}) as the limiting solution when $\Gamma/\omega \rightarrow 0$.
Since the real component of the eigenfrequencies determine the phase evolution of the waves, we conclude that the waves will have only two independent oscillatory solutions (and others related by $\pm$ signs), similar to the non-friction case. Each one of these two independent solutions will have an additional exponential suppression determined by $\Gamma_{1,2}$. Note that the values of $\Gamma$ may be negative, and conditions on the determinant and trace of the mixing matrices will need to be imposed to ensure stability. 

\section{Frequency vs Momentum space initial wavepacket}\label{App:mono_k_omega}

In this appendix we discuss what the appropriate initial conditions for the two eigenstates are for realistic GW emissions, focusing on comparing them in momentum and frequency space. 
If one considers waves from a compact binary coalescing, the non-trivial emission temporal profiles of GWs and possible additional fields is expected to be determined by the binary's dynamical motion, since the binary is typically assumed to be the only source present in the system (see for instance examples of scalar-tensor gravity theories with scalar radiation \cite{1989ApJ...346..366W, Will:1994fb, Lang:2014osa}). Regardless of the kind of interactions between the metric and new fields, we then expect the binary's angular velocity to determine the frequency of both the GWs and the additional fields\footnote{Mathematically, in the radiation zone, one assumes that GWs and the additional field are described by linear perturbations around a flat spacetime. If the additional fields do not couple directly to matter, and only indirectly through the metric, then the EoM are expected to have a single source term from the binary, whose structure is that one of GR since the metric is minimally coupled to matter. The time evolution of this single source term is determined by the binary's angular velocity $\Omega$, which in turn will then determine the frequency of all the fields. For instance, for a source term given by the quadrupole mass of the binary, the GW and additional field emission frequency will be given by $2\Omega$.}. 
In addition, the emission of energy in the form of additional fields can change the initial condition for GWs compared to that of GR since it will affect the binary's dynamical motion. However, in this paper we assume that the interactions between the metric and the additional field are suppressed in the strong field regime, and therefore the GW waveform emitted is approximately given by that in GR. Taking all these aspects into consideration, we therefore conclude that it is appropriate to assume that both eigenstates have approximately the same profiles for their initial conditions in frequency space, which is that of GR. 

We emphasize that even if it is appropriate to assume the same initial frequency profiles for both eigenstates, we have worked throughout this paper in momentum space. 
We find, however, that the same initial conditions can also be given in momentum space as long as both dispersion relations are such that $h_{0}(\omega)\approx h_0(k)$ and $h_0(\omega_1(k))\approx h_0(\omega_2(k))$. In order to see this explicitly, consider the toy model of constant coefficients, which is a system diagonalizable for left and right propagating waves separately, and the solutions are thus simple plane waves $e^{i(kx\pm \omega \eta)}$ in the absence of friction interactions.  
If we impose an initial condition in $k$-space $h_0(k,\eta_0)$ then, as shown in Section \ref{sec:wavepacket}, both eigenmodes will have the same profile in momentum space. The particular solution for $+x$ propagating waves is given by Eq.~(\ref{hsolconstparam}). If we then transform this solution to $\omega$-space, we find:
\begin{equation}
h(\omega,x)  
= \frac{1}{{1 -\hat E_{12}\hat E_{21}}}
\left(
e^{ + ik_1(\omega) x }h_0(k_1(\omega)) - \hat E_{12}\hat E_{21} 
e^{ + ik_2(\omega) x} h_0(k_2(\omega))\right)
.
\end{equation}
From here we see that the frequency profiles of both eigenmodes now differ from each other. This result would have been equivalent to imposing a single initial condition $h_0(\omega,x_0)$ in $\omega$ space, as long as:
\begin{equation}\label{omegakequivalence}
    |h_0(k_1(\omega))-h_0(k_2(\omega))|/|h_0(k_1(\omega))|\ll 1.
\end{equation}
for a range of relevant values of $\omega$. This is because, if there is no friction, the solutions are simple plane waves and therefore one could have found solutions directly in temporal Fourier space by solving Eq.~(\ref{eq:determinant1}) for $k$ as a function of $\omega$, instead of $\omega$ as function of $k$.
Note that the condition in Eq.~(\ref{omegakequivalence}) only requires $k_1\approx k_2$ for the initial amplitude.  
The phase difference between the propagating states of the same momentum $k$
accumulates during propagation and for
the fixed momentum space profile it is given by
\begin{equation}
\Delta \phi|_k = -[\omega_2(k)-\omega_1(k)] \Delta\eta.
\end{equation}
The phase difference between states of the same frequency $\omega$ differs across space as
\begin{equation}
\Delta \phi|_\omega = +[k_2(\omega)-k_1(\omega)] \Delta x.
\end{equation}
Since the difference between the phases is responsible for mixing phenomena, one might worry that mixing differs between the two
cases.  
Similar to the neutrino mixing case \cite{Cohen:2008qb}, the two cases give the same mixing phenomena so long as the splitting is small and the group velocities are close.    
To see this, we can approximate the momentum splitting at fixed frequency via Taylor expansion
\begin{equation}
\omega = \omega_1(k_1)=\omega_2(k_2) \approx \omega_2(k_1) + 
\frac{\partial \omega_2}{\partial k}(k_2-k_1),
\end{equation}
so that
\begin{equation}
\Delta \phi|_\omega = \Delta \phi|_k \frac{\Delta x}{\Delta \eta} \frac{1}{\partial\omega_2/\partial k}.
\end{equation}
Since the group velocity $v_g = \partial \omega/\partial k$ determines the 
propagation distance to the detector as
$\Delta x = v_g \Delta \eta$
the phase difference and hence mixing phenomena in the two cases
are nearly the same so long as the group velocities of the two propagation states are nearly the same.    This also explains why in the main text we employed the average group velocity to convert between mixing time and mixing length.

Finally, if the deviations from general relativity are small, we can also approximate $h_0(k(\omega))$ by $h_0(\omega)$ and assume the profile expected in GR.
For the toy models of Section \ref{sec:examples_mixing}, if the initial amplitudes were Gaussian wavepackets, Eq.~(\ref{omegakequivalence}) yields
\begin{equation}
[-(k_1-k_0)^2+(k_2-k_0)^2]/(2\sigma_k^2) \ll 1 .  
\end{equation}
If one has that $k_2=k_1+\Delta k$, then we need
\begin{equation}
\frac{\Delta k}{\sigma_k} \frac{(\Delta k +2(k_1-k_0))}{\sigma_k}\ll 1.
\end{equation}
This condition can easily be satisfied in the limit $k\rightarrow \infty$ as long as the phase and group velocities of the two eigenstates are similar.

More realistically, the emission of GWs in GR has frequency profiles during the binary's inspiral given by $h_0\propto \omega^{-7/6}$. In this power-law case the condition (\ref{omegakequivalence}) translates simply into:
\begin{equation}
    |\Delta k_1 (\omega)- \Delta k_2(\omega)|/|\omega|\ll 1, 
\end{equation}
where the two eigenstates have dispersion relations $k_A=\omega +\Delta k_A(\omega)$. For velocity mixing this condition would be satisfied if $c(v_{g,2}-v_{g,1})/(v_{g,1}v_{g,2})\ll 1$, whereas other mixing models would have additional requirements of high $k$.

The previous results can be generalized to the case with friction interactions and constant coefficients, where the solutions of the eigenstates were found to have exact solutions of the form $H_A\propto e^{\pm i\omega_A(k) \eta} e^{-\Gamma_A(k)\eta}e^{ikx}$, where both eigenstates were assumed to have the same $k$ because we were working in momentum space. One could have also solved exactly the equations assuming the same $\omega$ for both eigenmodes. The general relationship between $\omega$ and $k$ is discussed in  Appendix \ref{App:WKB}, where we showed that the EoM lead to 3 set of equations of the form $\Gamma_1(k)$, $\Gamma_2(k)$ and $\omega^2(k, \Gamma)$. From here we can then obtain two solutions of $k$ as function of $\omega$ by solving $\omega^2(k_1, \Gamma_1(k_1))$ and $\omega^2(k_2, \Gamma_2(k_2))$. For the two results $k_{1,2}(\omega^2)$ we then obtain $\Gamma_1(k_1(\omega^2))$ and $\Gamma_2(k_2(\omega^2))$. In this case, we have two eigenmodes that have the same $\omega$ but different $\Gamma$ and $k$: $H_A\propto e^{i\omega \eta} e^{-\Gamma_A(\omega)\eta}e^{ik_A(\omega)x}$. The matrix of eigenvectors now takes an analogous form to the one obtained in Eq.~(\ref{eq:eM}):
\be 
\hat{E}_\pm(\omega)=\bpm 1 & -\frac{\hat{W}_{12}(k_2)+i\nM_{12}(\pm \omega +i\Gamma_2)}{\hat{W}_{11}(k_2)-(\pm \omega +i\Gamma_2)^2+i\nM_{11}(\pm \omega +i\Gamma_2)} \\ -\frac{\hat{W}_{21}(k_1)+i\nM_{21}(\pm \omega +i\Gamma_1)}{\hat{W}_{22}(k_1)-(\pm \omega +i\Gamma_1)^2+i\nM_{22}(\pm \omega +i\Gamma_1)} & 1\epm\,.
\ee
On the other hand, the exact solutions $H_A\propto e^{i\omega \eta} e^{-\Gamma_A(\omega)\eta}e^{ik_A(\omega)x}$ can be transformed to frequency-domain by using the SPA assuming that $\Gamma \ll \omega$, and hence assuming that the amplitude $e^{-\Gamma \eta}$ is effectively constant for the relevant timescales. Therefore, in order to have an equivalence of the initial conditions in $k$ and $\omega$ space we require the same condition Eq.~(\ref{omegakequivalence}). The same will be valid for systems with slowly varying coefficients, compared to the scales associated to $k$ and $\omega$.

\section{GW polarization parameters}\label{App:polarization}
The most general polarization state for a monochromatic plane GW propagating along the $\hat{z}$ axis can be expressed as:
\begin{equation}
h_{ij} (\eta) = \Re\{ A_+ e^{-i(\omega \eta - \phi_+)} e_{+,ij} + A_\times e^{-i(\omega \eta - \phi_{-})} e_{\times,ij}\},
\end{equation}
where $\omega$ is the frequency of the wave, $e_{+,ij}=((1,0),(0,-1))$ and $e_{\times,ij}=((0,1),(1,0))$ are the polarization tensor basis, and the four real constant parameters $A_{+,\times}$, $\phi_{+,\times}$ fully determine the amplitude and phase of each polarization component. However, a more physically intuitive set of four real parameters that describe that signal can be used instead.  These parameters are $\{A, \phi, \beta, \chi\}$, where $A$ and $\phi$ describe the global amplitude and phase of the GW signal, whereas the angles $\beta$ and $\chi$ characterize the relation between the two polarization components. In particular, $\beta$ informs us about the degree of circular polarization that is present, and $\chi$ the orientation of the linear polarization, or more generally the semi-major axis of elliptical polarization. 
In terms of these new four parameters, the GW signal is given by:
\begin{align}
h_{ij} (\eta)  ={}& A\, \Re\Big\{  e^{-i(\omega \eta - \phi)}\left(\cos\beta \cos2\chi +i\sin\beta\sin2\chi\right) e_{+,ij} + 
\nonumber\\
& e^{-i(\omega \eta - \phi)}\left(\cos\beta\sin 2\chi -i\sin\beta\cos2\chi\right) e_{\times,ij}\Big\}.
\end{align}
The explicit relation between $\{A, \phi, \beta, \chi\}$ and $\{A_+, A_\times, \phi_+, \phi_\times\}$ is then determined by:
\begin{eqnarray}
A_+ \cos(\phi_+ -\phi) &=& A \cos\beta \cos(2\chi), \\
A_\times \cos(\phi_\times -\phi) &=& A \cos\beta\sin(2\chi), \\
A_+ \sin(\phi_+ -\phi) &=& A \sin\beta\sin(2\chi), \\
A_\times \sin(\phi_\times-\phi) &=& -A \sin\beta \cos(2\chi).
\end{eqnarray}
From here we see that one can identify unambiguously the amplitude of the signal $A$, calculated as $A=\sqrt{|h_+|^2+|h_\times^2|}=\sqrt{A_+^2+A_\times^2}$. Therefore, this is the appropriate definition of GW amplitude that we will be using in this paper. 
In the context of the main text, changes after propagation in the angle $\chi$ describes a general rotation of the polarization axes, and is equivalent to a shift of the  orientation angle of the LIGO-type antenna pattern function: $\psi\rightarrow \psi+\chi$. Equivalently, with the shift $\chi$ the polarization components of the signal get rotated according to for $+$ and $\times$:
\begin{align}
    h_+\rightarrow & \; h_+\cos 2\Delta\chi-h_\times \sin 2\Delta\chi,\\
    h_\times\rightarrow &\; h_\times\cos 2\Delta\chi+ h_+ \sin 2\Delta\chi.
\end{align}
Finally, the angle $\beta$ describes the relative amplitude between the $+$ and $\times$ polarization components, and therefore whether the wave is linearly, elliptically, or circularly polarized. In general, in analogy with an electromagnetic wave, in a transformed coordinate frame where the `semi-major' and `semi-minor' axes of the polarization are aligned with the a linear basis $e_{+,ij}'$ and $e_{\times,ij}'$, the polarization can be written as \cite{Jackson:1998nia}:
\begin{equation}
h_{ij}' = A \cos\beta e^{-i(\omega \eta - \phi)} e_{+,ij}'  - iA \sin\beta e^{-i(\omega \eta - \phi)} e_{\times,ij}',
\end{equation}
where $\tan\beta$ gives the ratio between the semi-major and minor axes. 
Equivalently, in the circularly-polarized basis, we can generically describe the difference between $h_{L}$ and $h_R$ with two real free parameters as:
\begin{equation}
    h_R=h_L\,re^{4i\chi}; \; h_L=A_Le^{-i(\omega \eta -\phi_L)},
\end{equation}
where $h_{L,R} = (h_+ \pm i h_\times)/\sqrt{2} $. Here, $\chi$ describes a general rotation of the basis, and $r$ describes the ratio between the semi-major and semi-minor axes of a general elliptical polarization, and is related to $\beta$ by $(1-r)/(1+r)=-\tan\beta$ or, equivalently, $r=(\cos\beta+\sin\beta)/(\cos\beta-\sin\beta)$. Here we assume $r$ to be positive definite, and hence $\beta\in [-\pi/4,\pi/4]$, and the phase shift $\exp\{4\chi\}$ can take any value, and thus we assume that $\chi\in [-\pi/4,\pi/4]$. Linear polarization corresponds to $\beta=0$, with special cases 
where with only $h_+$ if $\chi=0$ or only $h_\times$ if $\chi=\pm \pi/4$. On the other hand, a right-handed circular polarization has $\beta=+\pi/4$ and a left-handed has $\beta=-\pi/4$.

\bibliographystyle{JHEP}
\bibliography{GWsOscillations_refs}
\end{document}